\documentclass[letter,11pt]{article}
\usepackage{jheppub}
\usepackage{tikz}
\usetikzlibrary{decorations.pathmorphing}
\usetikzlibrary{decorations.markings}
\usepackage{xspace}
\usepackage{amsmath}
\usepackage{amssymb}
\usepackage{mathtools}
\usepackage[utf8]{inputenc}
\usepackage{physics}
\usepackage{slashed}
\usepackage{hyperref}
\usepackage[capitalize]{cleveref}

\usepackage{stackengine}

\newcommand{\ee}{{e^+e^-}}
\newcommand{\ttbar}{{t\bar t}}
\newcommand{\DNP}{{D^{\rm NP}_{b\to B}}}
\newcommand\brabar{\raisebox{-0.6pt}{\scalebox{.2}{(}}\raisebox{-1.9pt}{\scalebox{.7}{$-$}}\raisebox{-0.6pt}{\scalebox{.2}{)}}}

\title{NNLO B-fragmentation fits and their application to $t\bar t$ production and decay at the LHC}

\author[a]{Micha\l{} Czakon,}
\author[a]{Terry Generet,}
\author[b]{Alexander Mitov,}
\author[b]{Rene Poncelet}

\affiliation[a]{Institut f\"ur Theoretische Teilchenphysik und Kosmologie, RWTH Aachen University, D-52056 Aachen, Germany}
\affiliation[b]{Cavendish Laboratory, University of Cambridge, Cambridge CB3 0HE, United Kingdom}

\emailAdd{mczakon@physik.rwth-aachen.de}
\emailAdd{terry.generet@rwth-aachen.de}
\emailAdd{adm74@cam.ac.uk}
\emailAdd{poncelet@hep.phy.cam.ac.uk}

\note{Preprint: Cavendish-HEP-22/08, P3H-22-100, TTK-22-31}

\abstract{In this work we derive three sets of non-perturbative fragmentation functions, with uncertainties, for $B$-hadrons, $J/\psi$'s and muons resulting from semileptonic $B$ decays. All three sets are with next-to-next-to leading order accuracy and include next-to-next-to leading logarithmic soft gluon resummation. The novel feature of these new sets is that they are fully consistent with our formalism for next-to-next-to leading order (NNLO) calculations for final states with identified $B$, $J/\psi$ or a $\mu$. We employ the fragmentation functions derived in this work to make state of the art predictions for such final states in $t\bar t$ events at the LHC. A special emphasis is placed on observables sensitive to the top quark mass. The present work opens the door for many LHC applications, like, open $B$ production or $B$ production in association with bosons.}

\begin{document} 
\maketitle
\flushbottom

\section{Introduction}\label{sec:intro}

Studies of identified hadrons containing a heavy flavor (i.e.~a valence $b$ or $c$ quark) have a long history. Indeed, calculations with next to leading order (NLO) QCD accuracy, including resummation of collinear and soft logarithms at next-to-leading logarithmic accuracy, have been compared to data from lepton and/or hadron colliders for almost three decades \cite{Mele:1990cw,Binnewies:1997xq,Cacciari:1998it,Cacciari:2001td,Cacciari:2002pa,Cacciari:2003uh,Cacciari:2005rk,Cacciari:2005ry,Kniehl:2005de,Cacciari:2005uk,Kniehl:2006mw,Kneesch:2007ey,Cacciari:2012ny,Cacciari:2015fta,Kniehl:2020szu}. Increasing the accuracy of theoretical predictions beyond NLO in QCD has been a longstanding goal, see for example ref.~\cite{Cacciari:2002pa}. A first step in this direction was the calculation of the massless coefficient functions for $\ee$ colliders \cite{Rijken:1996vr,Rijken:1996npa,Rijken:1996ns,Mitov:2006wy}. The resummation of soft and collinear logarithms has been possible for at least a decade with the derivation of certain process-independent ingredients like the heavy quark perturbative fragmentation function (PFF) \cite{Melnikov:2004bm,Mitov:2004du}, the time-like splitting functions \cite{Mitov:2006ic,Moch:2007tx,Almasy:2011eq} and soft anomalous dimensions \cite{Cacciari:2001cw,Cacciari:2002xb,Gardi:2005yi}. Public codes implementing these results exist \cite{Bertone:2013vaa,Bertone:2015cwa}. Combined, these results have made it possible to extract the fragmentation function (FF) for heavy quarks from $\ee$ collisions with next-to-next-to leading order (NNLO) accuracy and with resummed soft and collinear logs to next-to-next-to leading logarithmic (NNLL) accuracy \cite{Salajegheh:2019ach,Fickinger:2016rfd,Ridolfi:2019bch}. Ideas about heavy flavor fragmentation that do not involve data fits have also been put forward \cite{Braaten:1994bz,Aglietti:2006yf}. Heavy flavor production in a jet has also received a lot of attention \cite{Anderle:2017cgl,Dai:2018ywt,Dai:2021mxb,Caletti:2022glq,Caletti:2022hnc}.

Progress towards applying these results at other processes (like the LHC) have been hampered by the lack of process-dependent coefficient functions with NNLO accuracy. First step in this direction was undertaken only very recently \cite{Czakon:2021ohs} where a numerical method for the calculation of the coefficient functions at NNLO in {\it any} process was developed and applied to $b$-fragmentation in $\ttbar$ production and decay at the LHC.

Interesting parallels can be made with the field of light-flavor fragmentation which historically has been developed largely independently from that of heavy flavor fragmentation, despite the many similarities, common features and ingredients between the two. A number of sets of light flavor fragmentation functions exist, mostly at NLO but also more recently at NNLO QCD \cite{Binnewies:1994ju,Kniehl:2000cr,Kniehl:2000fe,Kniehl:2000hk,Albino:2005me,deFlorian:2007aj,deFlorian:2014xna,Anderle:2015lqa,Sato:2016wqj,Bertone:2017tyb,Moffat:2021dji,Khalek:2021gxf,Abdolmaleki:2021yjf,Borsa:2021ran,Borsa:2022vvp,Khalek:2022vgy}. Just like in the case of heavy flavor fragmentation, progress towards their application to other processes has been rather restricted due to the lack of NNLO-accurate process-dependent coefficient functions. The development of ref.~\cite{Czakon:2021ohs} should allow for the calculation of coefficient functions with full flexibility, broadening the past practice of exclusively using analytical calculations which, despite their many useful features, are restricted to a small class of very inclusive observables. 

The calculation of ref.~\cite{Czakon:2021ohs} has utilized non-perturbative fragmentation functions (NPFF) extracted previously. While this made it possible to present predictions for specific realistic observables at the LHC, the predictions were not ideal since, as explained there, the NPFFs used in ref.~\cite{Czakon:2021ohs} were not fully consistent with the calculation of the perturbative part. Although such a mismatch is numerically small, a consistent program for NNLO predictions for processes with identified $B$ hadrons requires the use of a consistently extracted set of NPFFs. 

The main goal of this article is to extract such a set from the available $\ee$ data. As a first application, and building upon the large body of literature \cite{Corcella:2001hz,Cacciari:2002re,Corcella:2005dk,Biswas:2010sa,Kniehl:2012mn,Agashe:2012bn,MoosaviNejad:2013xqs,MoosaviNejad:2016gcd,Agashe:2016bok,Brancaccio:2021gcz}, in this work we also provide NNLO QCD accurate predictions for a set of {\it realistic} $b$-fragmentation observables in $\ttbar$ events at the LHC. A particular emphasis is placed on observables sensitive to the value of the top quark mass. Indeed, the current work allows for the first time to perform extractions of $m_t$ from this class of observables with full NNLO QCD accuracy. A second novel feature of this work is that we consider observables constructed not only at the level of $b$-flavored hadrons but also for their decay products: $J/\psi$ and muons from semileptonic $B$ decays.

This work is organized as follows: in sec.~\ref{sec:BFF} we present the main result of this work, namely, the extraction of NNLO-accurate NPFFs for $B$-hadrons and $J/\psi$ and soft muons resulting from $B$-decays. Theoretical aspects of this extraction related to the treatment of coefficient functions, perturbative fragmentation functions and soft-gluon resummation are given in, respectively, sections~\ref{sec:coeff-func}, \ref{sec:pert-frag} and \ref{sec:resummation}; the global fit of NPFF for $B$-hadrons from $e^+e^-$ data is presented in sec.~\ref{sec:datafits} while sec.~\ref{sec:NPFF-JPsi-muon} presents the derivation of NNLO-accurate NPFFs for $J/\psi$ and $\mu$. Sec.~\ref{sec:pheno} is devoted to phenomenological applications of the newly derived NPFFs, in particular, for state-of-the-art predictions for precision $m_t$ determination at the LHC. Some lengthier expressions related to soft-gluon resummation are collected in appendix~\ref{app:soft-gluon} while appendix~\ref{sec:appendix-one-dataset} presents extraction of $B$ NPFF from one dataset at a time. Our conclusions are summarized in sec.~\ref{sec:conclusions}.

\section{Extraction of NNLO $B$-fragmentation functions from $e^+e^-$ data}\label{sec:BFF}

The main source of data on $b$-fragmentation at present are measurements of $b$-flavored hadrons in $\ee$ collisions at $E_{\rm c.m. }=m_Z$. What is measured is the energy distribution of $b$-flavored hadrons $B$. Throughout this work we denote this energy through the dimensionless variable $z=E/E_{max}$. The relevant cross-section can be written as
\begin{equation}
\frac{d\sigma_B(m_b,z)}{dz} = \sum_i \Bigg\{\frac{d\sigma_i(\mu_{Fr},z)}{dz}\otimes D_{i\to B}(\mu_{Fr},m_b,z)\Bigg\}(z) + {\cal O}(m_b^2)\,,
\label{eq:dsigmaB}
\end{equation}
where $\otimes$ denotes the usual integral convolution, $\mu_{Fr}$ is the factorization scale associated with the fragmentation function, $m_b$ is the pole mass of the $b$ quark and the index $i$ runs over all partons that can contribute. Typically, one includes the heavy flavor ($b$ and $\bar b$ in this case), all lighter flavors and antiflavors as well as the gluon. Heavier quarks (top in this case) are not included. Further details can be found in ref.~\cite{Czakon:2021ohs}.

The process-independent set of fragmentation functions $D_{i\to B}$ can be further factorized into perturbative and non-perturbative components
\begin{equation}
D_{i\to B}(\mu_{Fr},m_b,z) = \Big\{ D_{i\to b}(\mu_{Fr},m_b,z)\otimes \DNP(z)\Big\}(z)\,.
\label{eq:DiB}
\end{equation}

The scale $\mu_{Fr}$ appearing in the above equations sets the scale at which the perturbative fragmentation process $i\to b$ takes place. The only other scale on which $D_{i\to b}$ depends is the mass $m_b$ of the heavy quark, suggesting $\mu_{Fr} \sim m_b$ as the natural choice. On the other hand, the partonic cross sections also depend on the scale $\mu_{Fr}$, suggesting it should be of the order of the hard scale in the process i.e.~$\mu_{Fr}\sim m_Z\gg m_b$. This hierarchy of scales introduces large collinear logarithms to all orders of the strong coupling constant. These logs are resummed with the help of the DGLAP evolution equations which the functions $D_{i\to b}$ satisfy. 

One of the main goals of this work is to provide a fit from $\ee$ data of the non-perturbative fragmentation function $\DNP$ for the fragmentation of a $b$-quark into a $B$-hadron. The function $\DNP$ is fully defined only once the perturbative cross-section $d\sigma_i/dz$ and the perturbative fragmentation function $D_{i\to b}$ have been fully specified, to which we turn next.

\subsection{Treatment of the coefficient functions}\label{sec:coeff-func}

In this work we have utilized the analytically known through NNLO massless partonic cross-sections $d\sigma_i/dz$ \cite{Rijken:1996vr,Rijken:1996npa,Rijken:1996ns,Mitov:2006wy}. As indicated in eq.~(\ref{eq:dsigmaB}), the use of massless coefficient functions leads to missing power corrections ${\cal O}(m_b^2)$ which are negligible for all kinematics except for very small $z$. On the other hand, the mass-independent term and associated logarithms $\ln^n(m_b^2/m_Z^2)$ are correctly accounted for by eq.~(\ref{eq:dsigmaB}). We have resummed soft-gluon logarithms in the coefficient functions through NNLL following the approach of ref.~\cite{Cacciari:2001cw}. The resummation follows the standard Mellin space approach with $N$ being the Mellin variable conjugated to $z$. In the soft limit $z\to 1$ (i.e.~$N\to\infty$) the cross-section further factorizes as
\begin{equation}
\sigma_b(N)= H(\alpha_S) \exp\left[ G(N,\alpha_S) \right] + {\cal O}(1/N)\,.
\label{eq:sigma_N}
\end{equation}
The explicit expressions for the function $G$ (through NNLL) and for the hard function $H$ (through NNLO) can be found in appendix~\ref{app:ee}. The complete resummed expression requires the matching of the above result with the exact fixed order result. Usually this is done through the so-called additive matching procedure where the resummed result takes the form
\begin{equation}
\sigma_b(N)= H(\alpha_S) \exp\left[ G(N,\alpha_S) \right] + M(N,\alpha_S)\,,
\label{eq:sigma_N-matched-add}
\end{equation}
where the function $M$ contains only ${\cal O}(1/N)$ terms and is defined in such a way that the expression in eq.~(\ref{eq:sigma_N-matched-add}), when expanded through NNLO, coincides with the Mellin transform of the coefficient function $d\sigma_b/dz$.

In this work we use a different, multiplicative, version of the above matching procedure, where the resummed and matched cross-section reads
\begin{equation}
\sigma_b(N)= \left(H(\alpha_S)+\tilde M(N,\alpha_S)\right) \exp\left[ G(N,\alpha_S) \right]\,.
\label{eq:sigma_N-matched-mult}
\end{equation}
The function $\tilde M$ above is defined from the same requirement as the function $M$ appearing in the additive approach. The multiplicative and additive versions differ only beyond NNLO. The main reason we choose to work with the multiplicatively matched resummed version is that for $z\to 1$ the resummed cross-section does not become negative, see sec.~\ref{sec:resummation} for details.

Finally, any $N$-space result needs to be inverted back to $z$-space. To this end we utilize the Minimal Prescription of ref.~\cite{Catani:1996yz}.

\subsection{Treatment of the perturbative fragmentation function}\label{sec:pert-frag}

The perturbative fragmentation function $D_{i\to b}$ can be further factorized as follows 
\begin{equation}
D_{i\to b}(\mu_{Fr},m,z) = \Bigg\{\sum_j E_{i\to j}(\mu_{Fr},\mu_{Fr,0},z) \otimes D^{\rm ini}_{j\to b}(\mu_{Fr,0},m,z)\Bigg\}(z)\,,
\label{eq:D-fact}
\end{equation}
where the index $j$ runs over all partons that contribute above the new arbitrary scale $\mu_{Fr,0}$. The standard choice for this scale, which we also adopt in this work, is $\mu_{Fr,0}=m_b$. This avoids large logarithms in $D^{\rm ini}_{j\to b}$. Instead, because $\mu_{Fr,0} \ll \mu_{Fr}$, $E_{i\to j}$ contains large logarithms which need to be resummed.

The evolution factor $E_{i\to j}$ is standard; it satisfies the DGLAP evolution equation with a trivial boundary condition. In particular, it does not require soft gluon resummation. In our implementation we take this factor from the public code {\tt APFEL} \cite{Bertone:2013vaa} where timelike evolution is implemented through NNLO. We do not restrict our implementation to the flavor non-singlet combination $(b-\bar b)$ but include all partonic contributions.

The non-trivial information about the PFF at the scale $\mu_{Fr,0}\sim m_b$ is contained within the initial condition $D^{\rm ini}_{j\to b}$ \cite{Mele:1990cw}. In the case $j=b$ (i.e.~the transition  $b\to b$), the function $D^{\rm ini}_{b\to b}$ contains terms that are enhanced in the soft-limit and thus require soft-gluon resummation. This has been first discussed in ref.~\cite{Cacciari:2001cw} through NLL and extended to NNLL in ref.~\cite{Gardi:2005yi}. Similarly to the case of the coefficient function discussed above, we implement the soft-gluon resummation of the initial condition through a multiplicative matching
\begin{equation}
D^{\rm ini}_{b\to b}(m_b,N)= \left(H^{\rm ini}(m_b,\alpha_S)+\tilde M^{\rm ini}(m_b,N,\alpha_S)\right) \exp\left[ G^{\rm ini}(m_b,N,\alpha_S) \right]\,.
\label{eq:D-ini-matched-mult}
\end{equation}

For brevity, we have not shown the dependence on $\mu_{Fr,0}$ in the above equation. Further discussion of the soft gluon resummation is given in sec.~\ref{sec:resummation}. The explicit expressions for the functions $G^{\rm ini}$ and $H^{\rm ini}$, through NNLL and NNLO respectively, can be found in appendix~\ref{app:Dini}.

\subsection{Implementation of the soft gluon resummation}\label{sec:resummation}

There are subtleties in the implementation of the soft gluon resummation of $D^{\rm ini}$ related to the decoupling of the heavy flavor $b$, which require a more detailed discussion. Let us start by recalling the ``master" resummation formula for $D^{\rm ini}$ \cite{Cacciari:2001cw}
\begin{eqnarray}
D^{\rm ini}_{b\to b}(\mu_{Fr,0},m_b,N) &\sim& \exp\int_0^1dz\frac{z^{N-1}-1}{1-z}\left( \int_{(1-z)^2m_b^2}^{\mu_{Fr,0}^2} \frac{dk^2}{k^2} A\left(\alpha_S(k^2)\right)+D\left(\alpha_S((1-z)^2m_b^2)\right)\right)\nonumber\\
&& +{\cal O}(1/N)\,.
\label{eq:master-res}
\end{eqnarray}
The function $D$ is often denoted as $H$ in the literature; we refrain from using this notation here since it can lead to confusion with the generic notation $H$ for the hard matching function.

Fixing $\mu_{Fr,0}\ge m_b$ for convenience, the scale $k^2$ of the running coupling in the function $A$ above takes values both below and above the mass $m_b$ of the heavy quark $b$. The scale $(1-z)^2m_b^2$ that enters the wide-angle emission function $D$ never exceeds $m_b^2$. With this in mind, and following eq.~(\ref{eq:D-ini-matched-mult}), we write the above master expression in the following way
\begin{eqnarray}
D^{\rm ini}_{b\to b}(\mu_{Fr,0},m_b,N) &=& H^{\rm ini} \exp\left[ G^{\rm ini} \right] C_{\rm dec}(\mu_{\rm th}=m_b,m_b) + {\cal O}(1/N)\,.
\label{eq:Dini-soft}
\end{eqnarray}

The LHS of the above equation is evaluated at a scale $\mu_{Fr,0}$ which implies that it is considered in a theory with $n_f=n_l+1$ active flavors where the heavy flavor is considered active. In contrast, some of the terms in the RHS of eq.~(\ref{eq:Dini-soft}) are defined in a theory with $n_f=n_l$ active flavors. Therefore, care must be taken that the final expression for the RHS of eq.~(\ref{eq:Dini-soft}) as a whole is converted to a theory with $n_f=n_l+1$ active flavors. In the following we explain how this is done as well as the meaning of each term. 

The hard matching function contains terms that behave as constant in the limit $N\to\infty$. It has the following expansion
\begin{eqnarray}
H^{\rm ini} &=& \sum_{i=0}\left(\frac{\alpha_S(\mu_{R,0})}{2\pi}\right)^i H_i^{\rm ini}\,.
\label{eq:H-expand}
\end{eqnarray}
The coefficients $H_i^{\rm ini}, i=0,1,2$ can be found in appendix~\ref{app:Dini}.

The Sudakov exponent $G^{\rm ini}$ can be written as
\begin{eqnarray}
G^{\rm ini} &=& \int_0^1dz\frac{z^{N-1}-1}{1-z}\int_{m_b^2}^{\mu_{Fr,0}^2} \frac{dk^2}{k^2} A'\left(\alpha_S(k^2)\right) \label{eq:S}\\
&&+ \int_0^1dz\frac{z^{N-1}-1}{1-z}\left( \int_{(1-z)^2m_b^2}^{m_b^2} \frac{dk^2}{k^2} A\left(\alpha_S^{(n_l)}(k^2)\right)+D\left(\alpha_S^{(n_l)}((1-z)^2m_b^2)\right)\right)\,,\nonumber
\end{eqnarray}
with anomalous dimensions $A$ and $D$ defined through the following expansions
\begin{eqnarray}
A^{(')}(a) &=& \sum_{i=1}\left(\frac{a}{\pi}\right)^i A^{(')}_i \,,\label{eq:A-expand}\\
D(a) &=& \sum_{i=1}\left(\frac{a}{\pi}\right)^i D_i \,. \label{eq:D-expand}
\end{eqnarray}
Throughout this work, all quantities denoted with a prime are defined in a theory with $n_f=n_l+1$ active flavors, while quantities without primes are in a theory with $n_f=n_l$ active flavors. Similarly, $\alpha_S^{(n_l)}$ is the coupling in a theory with $n_f=n_l$ active flavors while $\alpha_S$ is the one in a theory with $n_f=n_l+1$ active flavors. The explicit expressions can be found in appendix~\ref{app:Dini}. 

The term in the first line of eq.~(\ref{eq:S}) is defined above the flavor threshold $\mu_{\rm th}=m_b$, i.e.~in a theory with $n_f=n_l+1$ active flavors. In contrast, the term in the second line of that equation is defined below $\mu_{\rm th}=m_b$, i.e.~in a theory with $n_f=n_l$ active flavors. Upon its evaluation (discussed below) it needs to be converted to a theory with $n_l+1$ active flavors. This is achieved with the help of the decoupling relations for the strong coupling constant \cite{Larin:1994va,Chetyrkin:1997un,Schroder:2005hy,Chetyrkin:2005ia} {\it and} the decoupling constant $C_{\rm dec}(\mu_{\rm th},m_b)$ relating the fragmentation function $D^{\rm ini}_{b\to b}$ between theories with $n_l$ and $n_l+1$ active flavors. Schematically, the latter relation takes the following form
\begin{equation}
D^{{\rm ini}\, (n_f=n_l+1)}_{b\to b}(\mu_{Fr,0}=\mu_{\rm th},m_b) = C_{\rm dec}(\mu_{\rm th},m_b)\, D^{{\rm ini}\, (n_f=n_l)}_{b\to b}(\mu_{Fr,0}=\mu_{\rm th},m_b)\,.
\end{equation}

The decoupling coefficient $C_{\rm dec}$ depends on the scales $\mu_{\rm th}$ and $m_b$. Since, as indicated in eq.~(\ref{eq:Dini-soft}), we have chosen the threshold to be at $\mu_{\rm th}=m_b$ in this work we only need $C_{\rm dec}$ for this value of $\mu_{\rm th}$. The matching coefficient is known through NNLO \cite{Neubert:2007je}. In this work we are only interested in its $z\to 1$ limit
\begin{eqnarray}
C_{\rm dec}(\mu_{\rm th},m_b) &=& 1+ \bigg(\frac{\alpha_S(\mu_{R,0})}{2\pi}\bigg)^2 C_F T_F  \nonumber\\
&& \times \ln(N)\bigg[-\frac{2}{3}\ln^2\bigg(\frac{m_b^2}{\mu_\text{th}^2}\bigg)-\frac{20}{9}\ln\bigg(\frac{m_b^2}{\mu_\text{th}^2}\bigg)-\frac{56}{27}\bigg] + {\cal O}(N^0)\,.
\label{eq:C-dec}
\end{eqnarray}
We will not be concerned with the explicit expression for the $N$-independent constant term above since we have absorbed it into the hard matching coefficient $H_2^{\rm ini}$, see eq.~(\ref{eq:Dini-soft}).

The agreement between the NNLO expansion of the resummation formula eq.~(\ref{eq:Dini-soft}) and the direct fixed order calculation of $D^{\rm ini}_{b\to b}$ represents a non-trivial consistency check between the soft-gluon resummation formalism, the anomalous dimension $D_2$ (which is specific to $D^{\rm ini}$) computed in ref.~\cite{Gardi:2005yi} with the help of two independent approaches and the decoupling coefficient $C_{\rm dec}$ derived in ref.~\cite{Neubert:2007je}. On the other hand, given $C_{\rm dec}$ has been extracted from $D^{\rm ini}_{b\to b}$ itself, one may wonder how strong this consistency check is. To that end, in the following we invoke another {\it independent} argument confirming the result (\ref{eq:C-dec}) for $C_{\rm dec}$.

Since we are only interested in the strict soft limit $z\to 1$, one may exploit the relationship between parton distribution functions (PDF) and FF in this limit \cite{Gardi:2005yi} to confirm the decoupling relation (\ref{eq:C-dec}) from the known decoupling relations for PDF \cite{Buza:1996wv}. One difference between the case of heavy flavor FF we study in this work and PDF for heavy flavor production is that the literature on PDF almost exclusively deals with the case when ``intrinsic" heavy flavor is not present (``intrinsic" heavy flavor would be the direct analogue of heavy flavor FF). The results of ref.~\cite{Buza:1996wv} are derived in this context. 

The results of ref.~\cite{Buza:1996wv} are sufficient to address the question of interest here. Recall that we are interested in the double limit $z\to 1$ (which corresponds to the limit $x\to 1$ for PDF) and $m_b\to 0$ where all power corrections ${\cal O}(m_b^n)$ are neglected. In this double limit one expects that the ``valence" heavy and light quarks behave the same way, i.e.~any (cut) loop corrections involving heavy quarks affect them equally. From this one may conclude that the decoupling constant relevant for the heavy quark is the same as the one for the light quarks. From the results of ref.~\cite{Buza:1996wv} one can see that it is the operator matrix element $A^{{\rm NS},(2)}_{qq,H}$ which describes the decoupling of both the non-singlet and the singlet quark densities in the soft limit. One can easily see than in the limit $x\to 1$ the terms $\sim 1/(1-x)_+$ in $A^{{\rm NS},(2)}_{qq,H}$ are identical to $C_{\rm dec}$ in (\ref{eq:C-dec}). In fact, at NNLO, the argument that the ``valence" heavy and light quarks behave the same way can also be extended to a ``valence" gluon upon the replacement of overall color factors. Indeed, the terms $\sim 1/(1-x)_+$ in the diagonal matrix element $A^{{\rm S},(2)}_{gg,H}$ coincide with the ones in $A^{{\rm NS},(2)}_{qq,H}$ after the color factor replacement $C_A\to C_F$. 

Regarding our numerical implementation, we do not resum the $\ln(N)$ term in $C_{\rm dec}$ since, as it has been pointed out in ref.~\cite{Neubert:2007je}, it is of UV origin. Similar approach to this term has been taken in ref.~\cite{Fickinger:2016rfd}. We have checked that the numerical contribution from this term is very small, less than 1\% of the hard matching function $H^{\rm ini}$ for $N<450$.

The evaluation of the integrals in eq.~(\ref{eq:S}) is standard: one first performs the integration over $k^2$ via a change of variables from $k^2$ to $\alpha_S(k^2)$ (first line) or $\alpha_S^{(n_l)}(k^2)$ (second line), perturbatively expanding the integrand in $\alpha_S$. This transforms the integrands into sums of powers of $\alpha_S$, which can be integrated trivially. The result of the integral over $k^2$ on the first line of eq.~(\ref{eq:S}) is independent of $z$, so the integral over $z$ on the same line can also be performed trivially. The result of the integral over $k^2$ on the second line, however, is expressed in terms of $\alpha_S^{(n_l)}((1-z)^2m_b^2)$. The integral over $z$ is performed by another change of variables from $z$ to $\alpha_S^{(n_l)}((1-z)^2m_b^2)$, perturbatively expanding the integrand in $\alpha_S^{(n_l)}((1-z)^2m_b^2)$. The handling of the term $z^{N-1}-1$ through NNLL accuracy can be found in appendix A of ref.~\cite{Catani:2003zt}. This procedure yields a result for eq.~(\ref{eq:S}) which is expressed in terms of $\alpha_S(m_b^2)$, $\alpha_S(\mu_{Fr,0}^2)$, $\alpha_S^{(n_l)}(m_b^2/N^2)$ and $\alpha_S^{(n_l)}(m_b^2)$, all of which are re-expressed perturbatively in terms of $\alpha_S(\mu_{R,0}^2)$, using the decoupling relations for the strong coupling constant \cite{Larin:1994va,Chetyrkin:1997un,Schroder:2005hy,Chetyrkin:2005ia} to replace $\alpha_S^{(n_l)}(m_b^2)$ with $\alpha_S(m_b^2)$. As a result, one arrives at the following explicit expression for the function $G^{\rm ini}$ defined in eq.~(\ref{eq:S})
\begin{equation}
G^{\rm ini}(\mu_{Fr,0},\mu_{R,0},m_b,N,\alpha_S(\mu_{R,0})) = \sum_{i =1}^\infty\alpha_S^{i-2}(\mu_{R,0}) g_{i}(\lambda)\,,
\label{eq:Gini-exact}
\end{equation}
with $\lambda=\alpha_S(\mu_{R,0}) b_0\ln(N)$. The QCD $\beta$-function coefficients relevant for this work can be found in appendix~\ref{app:soft-gluon}. Only the terms with $i=1,2,3$ in eq.~(\ref{eq:Gini-exact}) contribute through NNLL. The explicit expressions for the functions $g_1, g_2$ and $g_3$ can be found in appendix~\ref{app:Dini}.

It is a well-known fact that the functions $g_i$ contain terms of the form $\ln^j(1-2\lambda)/(1-2\lambda)^k$ which become singular for $\lambda=1/2$. In terms of the Mellin variable $N$, this corresponds to a singularity at $N=N_L\approx 34$. While such singularities can be consistently treated within the Minimal Prescription approach \cite{Catani:1996yz} for inverting resummed expressions from $N$ to $z$ space, it is known that such resummed expressions tend to become negative for large values of $z$. This, in turn, may reduce the usefulness of the soft gluon resummation. Ref.~\cite{Cacciari:2005uk} has introduced a way around this issue, based on remapping the Mellin variable $N$ in the soft-gluon resummed exponent
\begin{equation}
N\to N\frac{1+f/N_L}{1+f N/N_L}\,.
\label{eq:N-rescale}
\end{equation}
The parameter $f$ is chosen to have a value $f=1.25$.

Given the importance of the large $z$ region and the NNLO+NNLL precision of our work, we have found it important to reanalyse the above issues. In fact, we propose another prescription, where the function $G^{\rm ini}$ is truncated in an expansion around $\lambda=0$
\begin{align}
G^{\rm ini}(N,\alpha_S) =\phantom{g_{1,2}\alpha_S\ln^2(N)+g_{1,3}\alpha_S^2\ln^3(N)+g_{1,4}\alpha_S^3\ln^4(N)+g_{1,5}\alpha_S^4\ln^5(N)+g_{1,6}}&
\notag\\ g_{1,2}\alpha_S\ln^2(N)+g_{1,3}\alpha_S^2\ln^3(N)+g_{1,4}\alpha_S^3\ln^4(N)+g_{1,5}\alpha_S^4\ln^5(N)+g_{1,6}\alpha_S^5\ln^6(N)&
\notag\\ +g_{2,1}\alpha_S\ln(N)+g_{2,2}\alpha_S^2\ln^2(N)+g_{2,3}\alpha_S^3\ln^3(N)+g_{2,4}\alpha_S^4\ln^4(N)&
\notag\\
+g_{3,1}\alpha_S^2\ln(N)+g_{3,2}\alpha_S^3\ln^2(N)&\,,
\label{eq:G-expand}
\end{align}
where $g_{i,j}$ is the coefficient of $\ln^j(N)$ in the expansion of the function $g_{i}$ around $\lambda = 0$:
\begin{equation}
  g_{i}(\lambda) = \sum_{j = 1}^\infty g_{i,j} \alpha_S^j \ln^j(N) \; , \qquad g_{1,1} = 0 \; .
\end{equation}
The depth of the expansion of the functions $g_1, g_2$ and $g_3$ in eq.~(\ref{eq:G-expand}) is sufficient for NNLL accuracy as we now explain.

Define the N$^n$LL accurate truncation
\begin{equation}
  G^{\rm ini}(N,\alpha_S) \Big|_{\text{N$^n$LL}} = \sum_{i=1}^{n+1} \alpha_S^{i-2} g_{i}(\lambda) \; ,
\end{equation}
with
\begin{equation}
  \exp[G^{\rm ini}(N,\alpha_S) \Big|_{\text{N$^n$LL}}] = 1 + \sum_{l=1}^\infty \sum_{k = 1}^{2l} c_{l,k} \alpha_S^l \ln^k(N) \; ,
\end{equation}
where $c_{l,k}$ are polynomial in $g_{i,j}$, $i \leq n+1$.

At N$^{n+1}$LL accuracy, the coefficients $c_{l,k}$ with $k < 2(l-n)$ depend on $g_{n+2,j}$. Indeed, write
\begin{equation}
  \exp[G^{\rm ini}(N,\alpha_S) \Big|_{\text{N$^{n+1}$LL}}] = \exp[G^{\rm ini}(N,\alpha_S) \Big|_{\text{N$^n$LL}}] \exp[\alpha_S^n \, g_{n+2}] \; .
\end{equation}
The LL term $\big( g_{1,2} \alpha_S \ln^2(N) \big)^{l-n-1}/(l-n-1)!$ of the expansion of $\exp[G^{\rm ini}(N,\alpha_S) \Big|_{\text{N$^n$LL}}]$ enhances the N$^{n+1}$LL term $\alpha_S^{n} \,  g_{n+2,1} \alpha_S \ln(N)$ of the expansion of $\exp[\alpha_S^n \, g_{n+2}]$ to contribute
\begin{equation} \label{eq:LLenhancement}
  g_{1,2}^{l-n-1} g_{n+2,1} \alpha_S^l \ln^{2(l-n)-1}(N) / (l-n-1)! \; ,
\end{equation}
to $\exp[G^{\rm ini}(N,\alpha_S) \Big|_{\text{N$^{n+1}$LL}}]$. Hence, $c_{l,2(l-n)-1}$ contains the term $g_{1,2}^{l-n-1} g_{n+2,1} / (l-n-1)!$. Similar arguments apply for $c_{l,k}$ with $k < 2(l-n)-1$.

Since $c_{l,k}$ with $k < 2(l-n)$ requires N$^{n+1}$LL resummation, the N$^n$LL resummed expression is correct at most for coefficients $c_{l,k}$ with $k \geq 2(l-n)$. Furthermore, two expressions that only differ in the coefficients $c_{l,k}$ with $k < 2(l-n)$ have the same N$^n$LL accuracy. Hence, any coefficient $g_{i,j}$ that does not contribute to $c_{l,k}$ with $k \geq 2(l-n)$ but contributes to $k < 2(l-n)$ is irrelevant for N$^n$LL accuracy and can be modified at will. The maximal $k$ of $c_{l,k}$ to which $g_{i,j}$ contributes, is determined similarly as above eq.~\eqref{eq:LLenhancement} by considering LL enhancement
\begin{equation}
    \big( g_{1,2} \alpha_S \ln^2(N) \big)^{l-i-j+2} g_{i,j} \alpha_S^{i-2+j} \ln^j(N) = g_{1,2}^{l-i-j+2} g_{i,j} \, \alpha_S^l \ln^{2(l-i+2)-j}(N) \; .
\end{equation}
Thus, if
\begin{equation}
  2(l-i+2)-j < 2(l-n) \qquad \Leftrightarrow \qquad j > 2(n-i+2) \; ,
\end{equation}
then $g_{i,j}$ is irrelevant for N$^n$LL resummation. In our prescription, we simply set such $g_{i,j}$ to zero, which results in eq.~\eqref{eq:G-expand} for $n = 2$. Notice finally that, owing to the negative sign of all coefficients $g_{1,j}$, the numeric inversion of the resummed $D^{\rm ini}$ within the expanded prescription is unproblematic. 

The Landau pole discussed above is not specific to $D^{\rm ini}$. From the results in appendix~\ref{app:ee} one can see that it is also present in the $\ee$ coefficient function, albeit at a different value for $N$, due to terms $\ln^j(1-\lambda)/(1-\lambda)^k$. In order to deal with this pole one can choose to either remap the Mellin variable $N$ or, as we have done in this work, truncate the corresponding function $G$ for the $\ee$ coefficient function in analogy with eq.~(\ref{eq:G-expand}). Note that unlike the case of $D^{\rm ini}$, the leading coefficients $g_{i,j}$ of the $\ee$ coefficient functions are positive which precludes the inversion to $z$ space of the coefficient function. However, the $N$-space product of the resummed coefficient function and $D^{\rm ini}$ is invertible as long as $2\alpha_S(\mu_{R,0}) > \alpha_S(\mu_R)$. This requirement is always satisfied since $\mu_{R,0}\leq \mu_R$. For this reason in our implementation we first multiply in $N$-space the resummed expressions for $D^{\rm ini}$ and the $\ee$ coefficient functions and only invert to $z$-space their product. This combination of $D^\text{ini}$ and the coefficient function is matched to its perturbative expansion in $\alpha_s(\mu_{R,0})$ and  $\alpha_s(\mu_R)$ through $\mathcal{O}(\alpha_s(\mu_{R,0})^2,\alpha_s(\mu_{R,0})\alpha_s(\mu_R),\alpha_s(\mu_R)^2)$.

In addition to preserving the logarithmic accuracy of the resummation, a valid prescription for regulating the Landau pole should not introduce corrections larger than those expected from non-perturbative effects. For the PFF, non-perturbative corrections are expected to be $\mathcal{O}(N\Lambda_\text{QCD}/m_b)$ \cite{Cacciari:2002xb,Nason:1996pk,Jaffe:1993ie,Randall:1994gr} in the simultaneous limit of large $N$ and large mass. For the $e^+e^-$ coefficient function, the non-perturbative corrections are $\mathcal{O}(N\Lambda_\text{QCD}^2/Q^2)$ \cite{Dasgupta:1996ki}. Since $N_L\sim m_b/\Lambda_\text{QCD}$ for the PFF and $N_L\sim Q^2/\Lambda_\text{QCD}^2$ for the coefficient function, the corrections are $\mathcal{O}(N/N_L)$ in both cases. Remapping $N$ according to eq.~(\ref{eq:N-rescale}) clearly leads to corrections which scale like $f\;N/N_L$, thus fulfilling the condition that the regularisation procedure does not change the result by more than can be expected from non-perturbative effects.

\begin{figure}[t]
	\centering
	\includegraphics[width=0.49\textwidth]{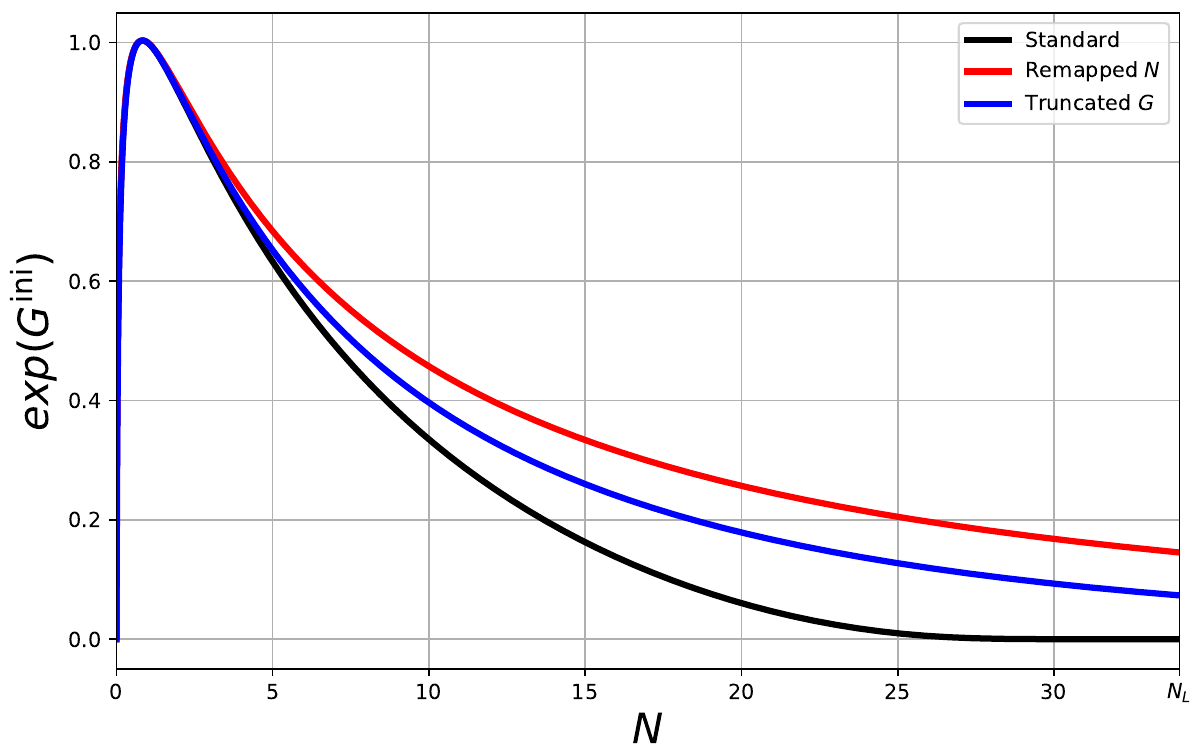}
	\includegraphics[width=0.49\textwidth]{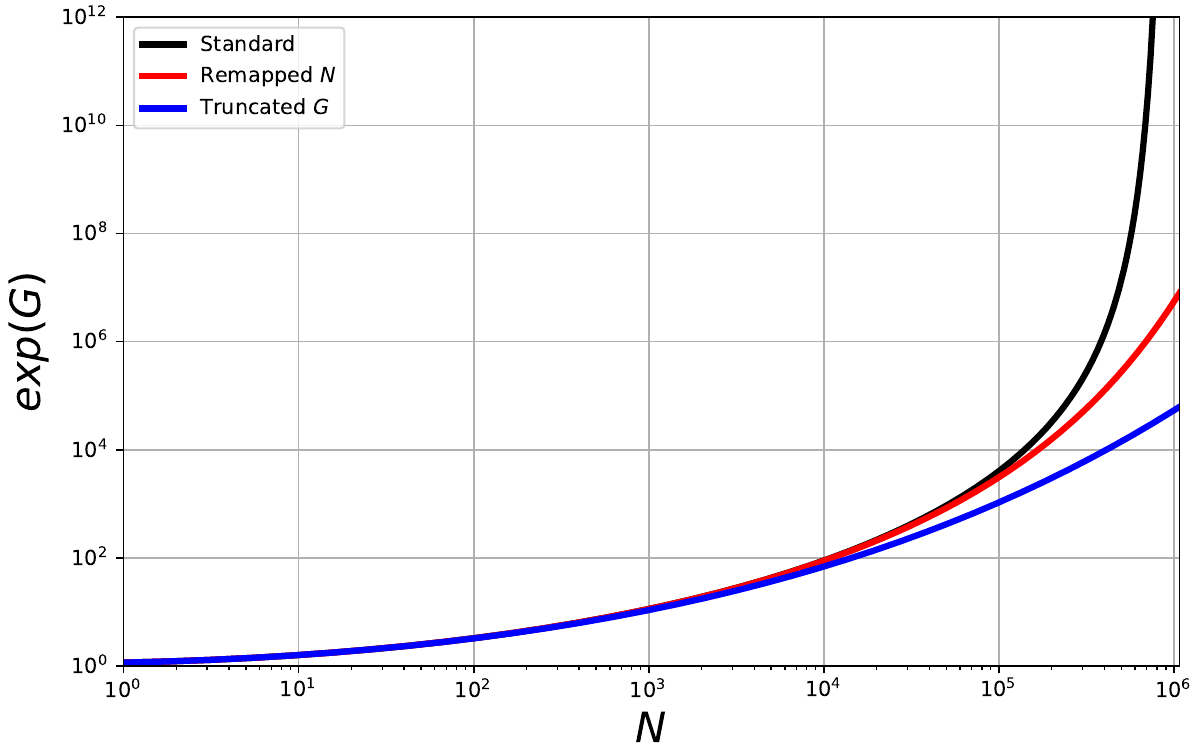}
	\caption{Different ways for dealing with the Landau pole in the NNLL Sudakov factors for the PFF (left) and the coefficient function (right). The PFF Sudakov factor is evaluated for $\mu_{Fr,0} = \mu_{R,0} = m_b$, while the one for the coefficient function is shown for $\mu_{Fr} = \mu_{R} = m_Z$. The various curves are: the Sudakov factor without any regularisation of the Landau pole (black), with $N$ remapped (red) and with a truncated exponent (blue).}
	\label{fig:ComparisonSudakovs}
\end{figure}

For the PFF, one has
\begin{equation}
N_L = \exp\bigg(\frac{1}{2b_0\alpha_S}\bigg) \Leftrightarrow \alpha_S = \frac{1}{2b_0\ln(N_L)}\;,
\end{equation}
as can be trivially shown by requiring $2\lambda = 1$. The leading corrections in the large-mass (i.e.~the large-$N_L$) limit are thus the leading corrections in $\alpha_S$. At N$^{n}$LL, these corrections are $\mathcal{O}(\alpha_S^{n+2}\ln^3(N))$, or equivalently $\mathcal{O}(\ln^3(N)/\ln(m_b/\Lambda_\text{QCD})^{n+2})$. Since $\ln^3(N)$ is $\mathcal{O}(N)$, this scaling is unproblematic. However, one could be concerned that
\begin{equation} 1/\ln(m_b/\Lambda_\text{QCD})^{n+2}\gg\Lambda_\text{QCD}/m_b
\end{equation}
for large $m_b$. This means that the truncation regularisation discussed here cannot lead to an accurate description for arbitrarily large masses. However, the only condition of practical relevance is that the corrections are sufficiently small for the present application, i.e.~$b$-quark fragmentation. At NNLL, we get
\begin{equation}
1/\ln(m_b/\Lambda_\text{QCD})^{2+2} \sim 1/\ln(N_L)^4 \sim 0.006 < 1/N_L \sim 0.03\;.
\end{equation}
Combined with the fact that $\ln^3(N) < 1.4\:N$ for $N \ge 1$, one would thus expect the truncation regularisation to lead to smaller corrections than those resulting from remapping $N$, at least for $b$-quark fragmentation. This is explicitly confirmed in the left panel of fig.~\ref{fig:ComparisonSudakovs}, where the unregulated Sudakov factor for the PFF is compared to the two regulated versions discussed above. Truncating $G^\text{ini}$ consistently yields results closer to the original curve than remapping $N$ does, showing that the new regularisation does not introduce unreasonably large corrections.

The analysis for the coefficient function is completely analogous, but with
\begin{equation}
N_L = \exp\bigg(\frac{1}{b_0\alpha_S}\bigg) \approx 10^6 \Leftrightarrow \alpha_S = \frac{1}{b_0\ln(N_L)}\;.
\end{equation}
This time, the corrections due to the truncation are much larger than those due to remapping $N$, as can be seen in the right panel of fig.~\ref{fig:ComparisonSudakovs}. However, this has no numerical effect in practice: because $N_L$ is so much larger for the coefficient function, the regularisation corrections are tiny for the phenomenologically relevant regime $N < 100$. Additionally, we are only interested in the convolution of the coefficient function with the PFF. The latter receives much larger corrections anyway, making the corrections to the coefficient function completely inconsequential.

Remapping $N$ thus corresponds to adding corrections analytically consistent in size with non-perturbative effects, whereas truncating the Sudakov exponent does not, even if numerically the corrections are actually smaller. However, the newly proposed regularisation has several advantages. First of all, there is no theoretical justification for remapping $N$. It is merely an ad hoc prescription which yields acceptable results. Additionally, it introduces a new free parameter, $f$, which needs to be tuned to data. The new prescription is based on the freedom to change the Sudakov exponent in ways which only affect higher logarithmic orders, which is a theoretically perfectly justifiable approach. It also does not introduce any new free parameters. Additionally, as will be shown below, regulating the Landau pole by remapping $N$ yields a PFF which behaves irregularly as $z\to 1$, whereas truncating $G$ yields a PFF which smoothly approaches zero in this limit.

\begin{figure}[t]
\centering
\includegraphics[width=0.49\textwidth]{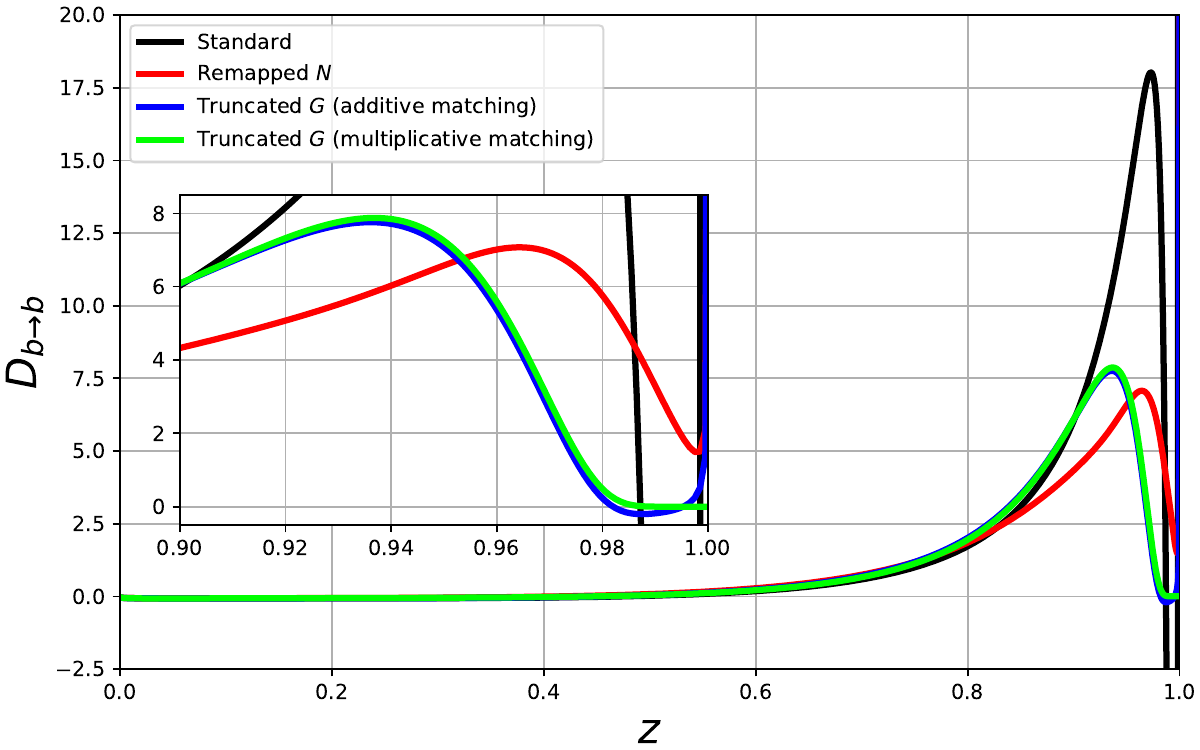}
\includegraphics[width=0.49\textwidth]{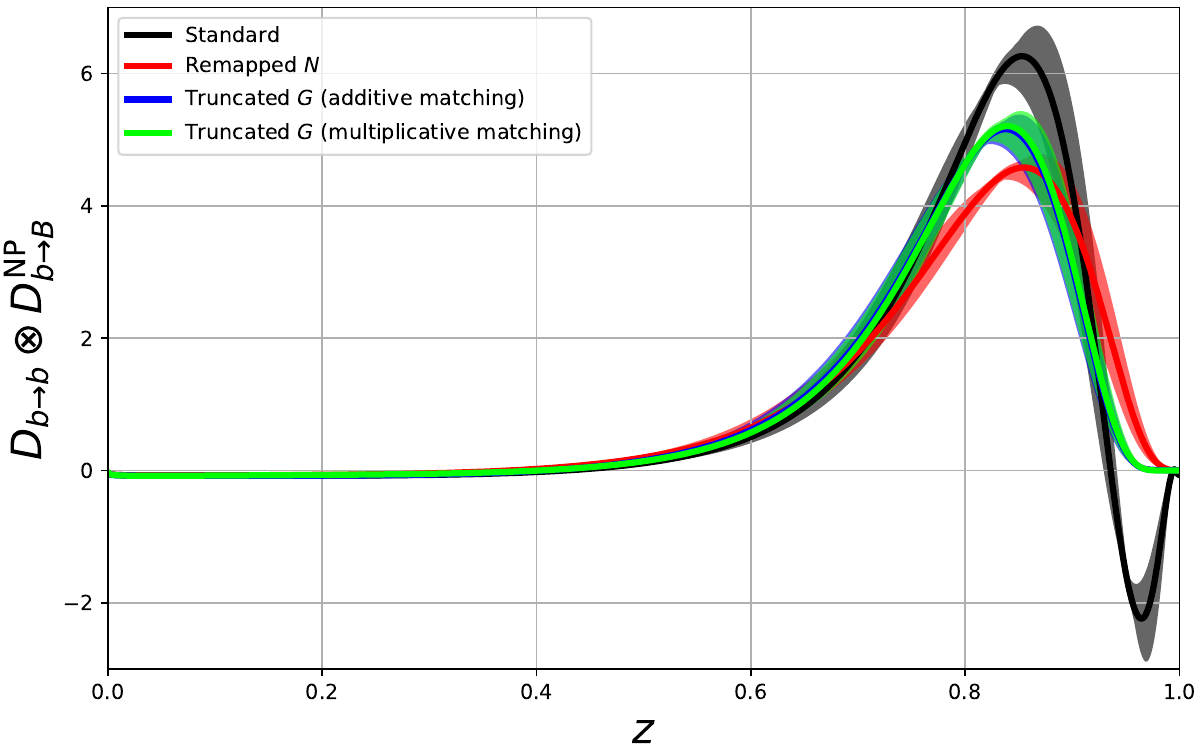}
\caption{Different ways for dealing with the Landau pole in the soft gluon resummed PFF. All curves are at NNLO+NNLL for $\mu_{Fr,0} = \mu_{R,0} = m_b$. Shown are both the PFFs (left) and their convolution with a fixed NPFF eq.~(\ref{eq:NPFF}) (right). The various curves are: the PFF without any regularization of the Landau pole (black), with $N$ remapped (red) and the PFF with a truncated exponent matched to the fixed-order result additively (blue) and multiplicatively (green). The coloured bands in the plot on the right correspond to the uncertainty on the NPFF as shown in eq.~(\ref{eq:NPFF}).}
\label{fig:ComparisonRegularisations}
\end{figure}

The three possible ways of dealing with the Landau pole in the soft gluon resummed result, namely, no prescription, rescaling of $N$ (\ref{eq:N-rescale}) and truncation of $G$ (\ref{eq:G-expand}), are compared in fig.~\ref{fig:ComparisonRegularisations}. Also shown in that figure is the numerical impact of the additive (\ref{eq:sigma_N-matched-add}) versus multiplicative (\ref{eq:sigma_N-matched-mult}) matching. The $N$ variable is only remapped in the resummed part of the PFF, i.e.~in the term $H^{\rm ini}\exp(G^{\rm ini})$ but not in the matching term $M^{\rm ini}$. We show separately the PFFs and their convolution with a fixed NPFF, specifically, the NPFF fitted in ref.~\cite{Cacciari:2005uk}
\begin{equation}
D_\text{NP}(x) = \frac{\Gamma(a+b+2)}{\Gamma(a+1)\Gamma(b+1)}x^a(1-x)^b,\;\;\;\;a = 24\pm 2,\;b = 1.5\pm 0.2\;.
\label{eq:NPFF}
\end{equation}
All PFFs are with NNLO+NNLL accuracy and are evaluated for $\mu_{Fr,0} = \mu_{R,0} = m_b$. The green curve shows the PFF with a truncated Sudakov exponent $G^{\rm ini}$ that is matched multiplicatively. It corresponds to the PFF used in this work. The unregulated (in black) and remapped (in red) PFFs are matched additively.

From fig.~\ref{fig:ComparisonRegularisations} we conclude that the impact of the way the Landau pole in the resummed result for $D^{\rm ini}$ is handled can be significant. In particular, the PFF without any regularization becomes negative and a significant portion of its area is in this region. This negative contribution is compensated by a much larger value in the peak region. On the other hand, the two regularized expressions, with remapped $N$ and with truncated $G^{\rm ini}$, are much closer to each other than the un-regularized result. Still, their behaviors in the large $z$ region show notable differences. The remapped PFF does not vanish at large $z$ while the truncated one rapidly approaches zero as $z\to1$. Given that both curves are normalized, this difference translates into differences for both their peak values and the positions of their maxima. The phenomenological significance of this difference is harder to assess since although the two curves are similar, they are not consistent with each other within the NPFF uncertainty. The un-regularized one (in black) is significantly different from the other two curves. We also note that the difference between the additive (in blue) and multiplicative (in green) matching is negligible everywhere except the extremely large $z$ region where the curve with additive matching becomes slightly negative above $z=0.98$ and diverges in the limit $z\to 1$, while the multiplicative one stays non-negative for all values of $z$ and vanishes in the limit $z\to 1$.

Before closing this section, we would like to make one more remark. When estimating the uncertainty due to scale variation we do not vary the scale $\mu_{R,0}$ but keep it fixed at $m_b$. The reason for this is a subtlety in the soft-gluon resummation of $D^{\rm ini}$. As is well known and as can be seen from eq.~(\ref{eq:master-res}), due to the running of the coupling to scales as low as the Landau pole $\Lambda$, the resummed result breaks down for $z\ge z_{max}$, where $z_{max}$ is defined by the relation $\mu(1-z_{max})\sim \Lambda$. The scale $\mu$ is a fixed hard scale specific to the resummed observable. For the case of $D^{\rm ini}$, one has $\mu=\mu_{R,0}$ which is quite low. Indeed, as can be seen in fig.~\ref{fig:ComparisonRegularisations}, this breakdown of the resummed result is consistent with the behavior of the curve with no regularization which turns negative at large $z$.  Thus, variation of the scale $\mu_{R,0}$ by a factor of two can noticeably change the value of $z_{max}$ and, therefore, the large-$z$ behavior of the resummed PFF. We also note that varying $\mu_{R,0}$ below $m_b$ would be problematic in the context of the decoupling of the heavy flavor. For these reasons in this work we keep the scale $\mu_{R,0}$ fixed at $m_b$. We have verified that the difference between the $b$-quark PFF evaluated at $\mu_{R,0}=m_b$ and at $\mu_{R,0}=2m_b$ is tiny everywhere except for large $z$, as expected, due to the shift in the breakdown point $z_{max}$. We also note that in this context the scales $\mu_{Fr,0}$ and $\mu_{R,0}$ are typically set equal, meaning that the above comments are also applicable for the setting of the scale $\mu_{Fr,0}$. We have checked that the dependence of the PFF on the scale $\mu_{Fr,0}$ is much milder than the one on $\mu_{R,0}$. We have also confirmed that the dependence of the NNLO+NNLL $e^+e^-$ cross section on $\mu_{R}$ and $\mu_{Fr}$ is negligible compared to the size of the experimental uncertainties. For this reason, we fix all scales to their central values $\mu_{R,0} = \mu_{Fr,0} = m_b$ and $\mu_{R} = \mu_{Fr} = m_Z$ when performing the NPFF fits described in the next section.

\subsection{Fits to data}\label{sec:datafits}

Having specified our implementation of the $\ee$ coefficient functions and perturbative fragmentation functions, we are ready to proceed with the fit of the NPFF from $\ee$ data. We perform a combined fit to datasets from ALEPH \cite{ALEPH:2001pfo}, DELPHI \cite{DELPHI:2011aa}, OPAL \cite{OPAL:2002plk} and SLD \cite{SLD:2002poq}. We note that these datasets contain weighted averages of various $B$ mesons, and some of the sets include baryons. This implies that predictions based on NPFF extracted from these data sets will also be for the same combination of $B$ hadrons.

\subsubsection{Treatment of the datasets}

The data -- each dataset, in fact -- correspond to the normalized differential distribution 
\begin{equation}
{\rm Data}_B(z)=\frac{1}{\sigma_B}\frac{d\sigma_B(m_b,z)}{dz}\,.
\label{eq:data(z)}
\end{equation}
In practice, the distribution ${\rm Data}_B(z)$ is not measured directly. The experiments' published data represents the binned version of this differential distribution, supplemented by a covariance matrix representing the correlated experimental uncertainty of this binned measurement. To aid our fits, we would like to represent the binned data, including its uncertainty, in a differential form like in eq.~(\ref{eq:data(z)}). To that end we proceed as follows. We interpret the binned data and its uncertainty as a multidimensional normal random distribution. Since the data is normalized, the dimension of that distribution is given by the number of bins minus one. We then generate 100 binned ``datasets"; one of them corresponds to the measured binned data (i.e.~the ``central value") while the remaining 99 sets are generated randomly, by sampling this multidimensional random distribution. We then treat each one of these 100 datasets as if it is the true data and produce a data fit of the form ${\rm Data}_B(z)$ for each one of those binned sets. As a whole, the 100 data fits represent the true differential data distribution and its uncertainty. 

As a last step, the 100 fitted data distributions are used to fit a set of 100 NPFF's, one being the central member, accompanied by 99 error sets (the detailed description of this procedure is given in sec.~\ref{sec:NPFFfits} below). When predictions based on this set of NPFFs are made, 100 predictions need to be computed. One represents the central prediction for the given observable while the other 99 predictions are used to derive the NPFF uncertainty for this variable. This uncertainty is taken to be the root mean square (r.m.s.) $1\sigma$ deviation with respect to the mean of the full FF set of 100 curves. Throughout this work we present this uncertainty in such a way that it is always centered around the mean and not around the central member.

For our data fits we assume two functional forms. One has 5 parameters (as in ref.~\cite{DELPHI:2011aa}), while the other has 8
\begin{eqnarray}
{\rm Data}_B(z) &=& n\frac{\Gamma(a+b+2)}{\Gamma(a+1)\Gamma(b+1)}z^a(1-z)^b+(1-n)\frac{\Gamma(c+d+2)}{\Gamma(c+1)\Gamma(d+1)}z^c(1-z)^d \,,\nonumber\\
{\rm Data}_B(z) &=& m\bigg[n\frac{\Gamma(a+b+2)}{\Gamma(a+1)\Gamma(b+1)}z^a(1-z)^b+(1-n)\frac{\Gamma(c+d+2)}{\Gamma(c+1)\Gamma(d+1)}z^c(1-z)^d\bigg]\nonumber\\
&&+(1-m)\frac{\Gamma(e+f+2)}{\Gamma(e+1)\Gamma(f+1)}z^e(1-z)^f \,.
\label{eq:data-func-form}
\end{eqnarray}

We perform fits for each individual measurement as well as to their combination. The following procedure was followed for both models. First, the statistical and systematic covariance matrices of the data were added. The OPAL data has asymmetrical systematic uncertainties, meaning four systematic covariance matrices are needed, but only two are made available. Because of this, the two provided systematic covariance matrices were averaged before being added to the statistical one. SLD only provides the diagonal elements of the covariance matrices, so the off-diagonal elements were set to zero. The SLD data also has two systematic uncertainties, which were added in quadrature. All covariance matrices were then diagonalized by determining their eigenvalues and eigenvectors. This procedure is necessary for both the ALEPH and OPAL datasets, since the provided covariance matrices are not positive-definite, likely due to rounding errors. By diagonalizing the covariance matrices it is possible to reject all negative eigenvalues and only use the positive ones in the fit. 

\begin{table}
	\centering	
	\begin{tabular}{||c|c|c|c|c|} 
		\hline
		Dataset & ALEPH & DELPHI & OPAL & SLD \\
		\hline
		\hline
		Number of bins & $19$ & $9$ & $20$ & $22$ \\
		\hline
		Number of positive eigenvalues & $13$ & $9$ & $13$ & $22$ \\
		\hline
		Number of used eigenvalues ($5$ parameters) & $7$ & $8$ & $8$ & $22$ \\
		\hline
		$\chi^2/n_\text{dof}$ ($5$ parameters) & $3.41/2$ & $1.76/3$ & $4.97/3$ & $5.31/17$ \\
		\hline
		Number of used eigenvalues ($8$ parameters) & $13$ & $9$ & $12$ & $22$ \\
		\hline
		$\chi^2/n_\text{dof}$ ($8$ parameters) & $5.39/5$ & $0.574/1$ & $4.09/4$ & $1.06/14$ \\
		\hline
	\end{tabular}
\caption{The number of bins, number of positive eigenvalues of the covariance matrix, number of used eigenvalues for both central fits and the $\chi^2/n_\text{dof}$ for both central fits for all four datasets.}
\label{tab:DataFits}
\end{table}

It is still possible, however, that some of the positive eigenvalues are artificially small due to the rounding errors. Because of this, only the $n$ largest eigenvalues were included in the fit, where $n$ is chosen such that $n+1$ is the smallest number of used eigenvalues that leads to an unacceptably large $\chi^2/n_\text{dof}$ if the data model is fitted to the respective dataset only. Since, typically, the difference in $\chi^2$ between $n$ and $n+1$ is at least a factor of ten, this choice is quite unambiguous in practice, despite the seemingly subjective criterion. The results of the fits to each individual dataset are given in table \ref{tab:DataFits}.

\begin{figure}[t]
\centering
\includegraphics[width=0.49\textwidth]{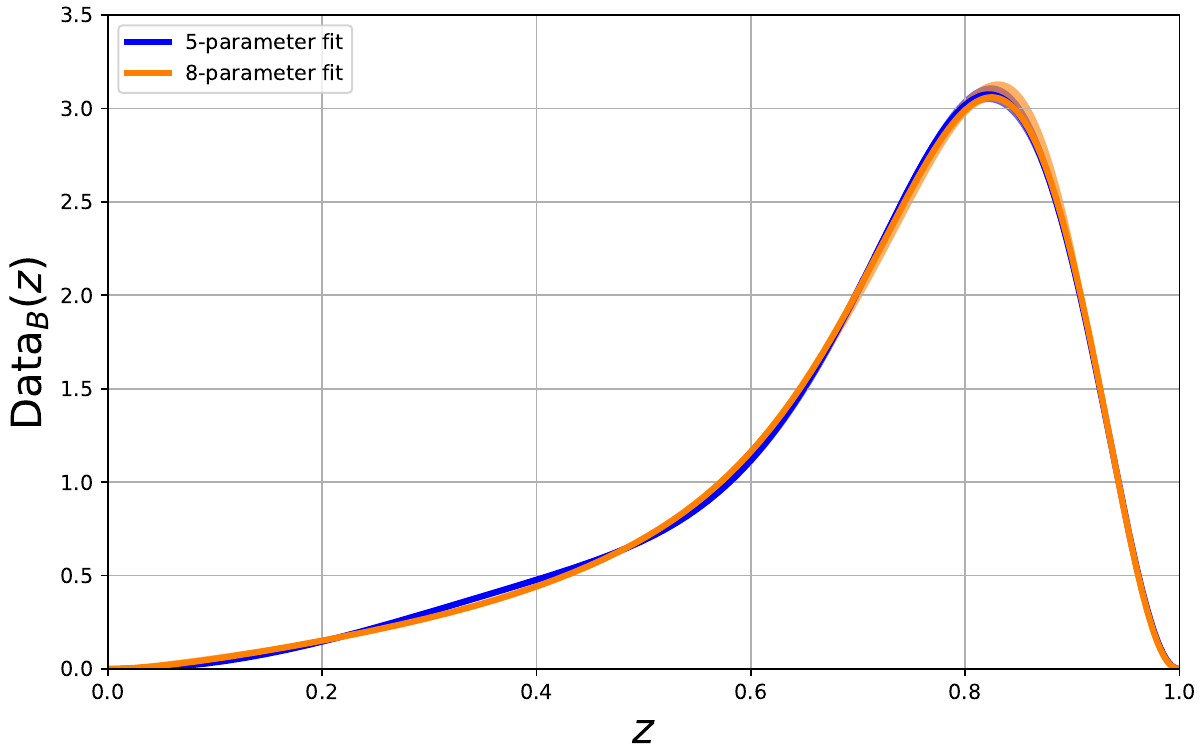}
\includegraphics[width=0.49\textwidth]{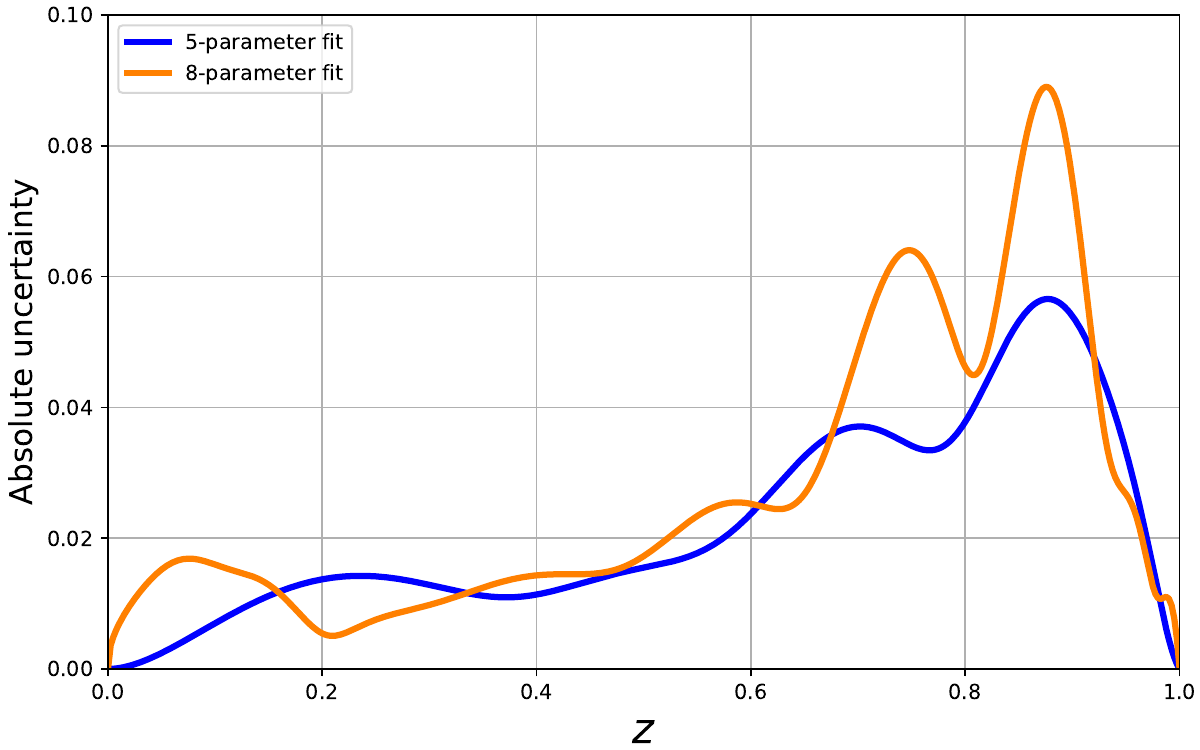}
\includegraphics[width=0.49\textwidth]{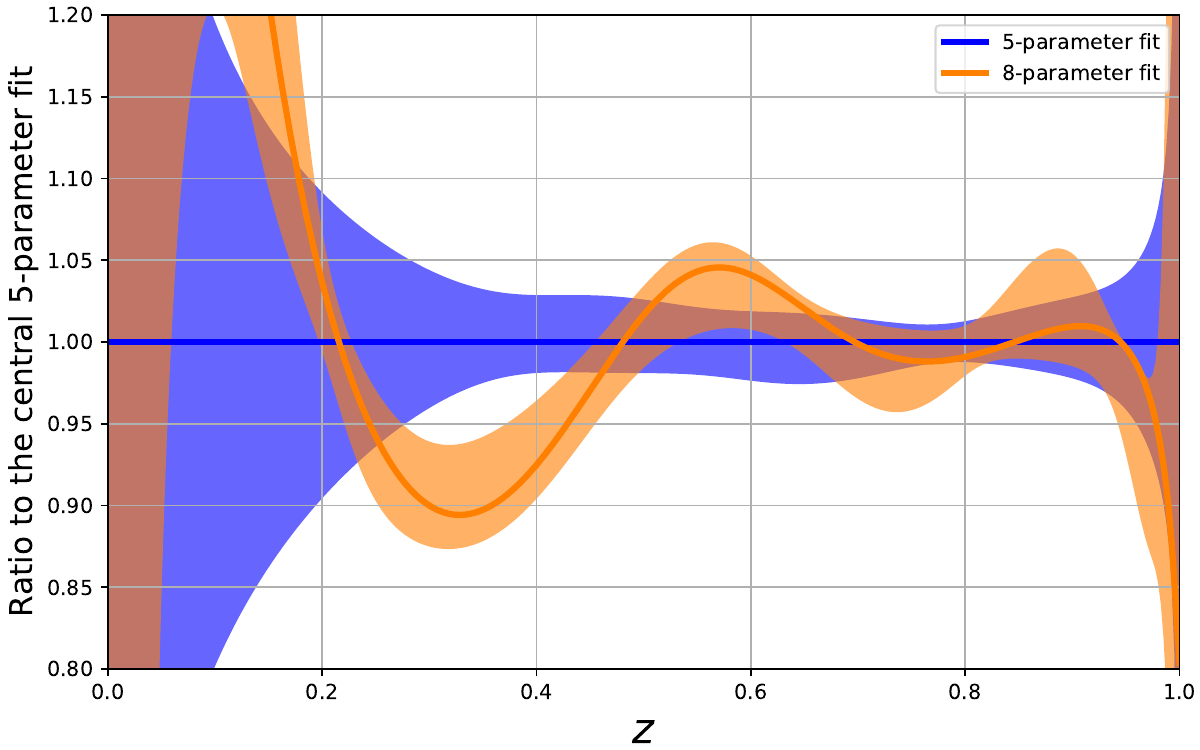}
\includegraphics[width=0.49\textwidth]{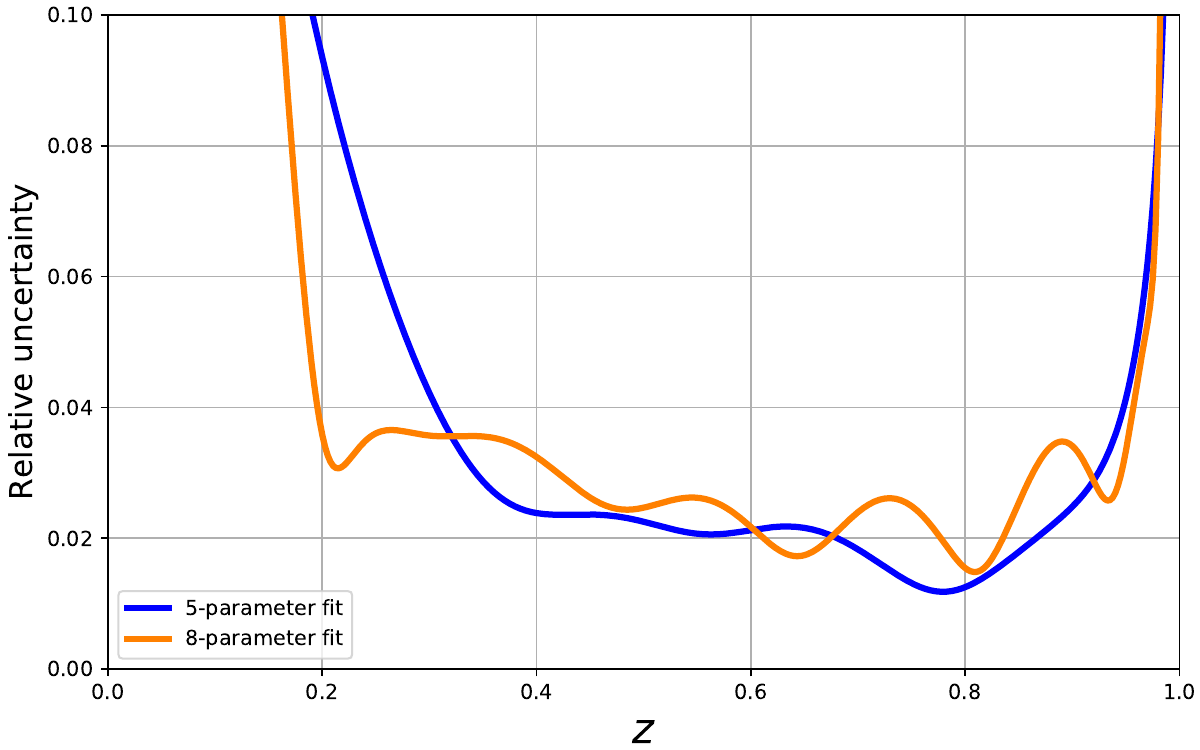}	
\caption{Comparison of the 5-parameter (blue) and 8-parameter (orange) fits for the distribution ${\rm Data}_B(z)$. Shown are the absolute distributions (top left) and their ratio (bottom left) as well as their absolute (top right) and relative (bottom right) uncertainties. The uncertainty bands correspond to the root mean square (r.m.s.) of the set of 100 data fits.}
\label{fig:DataFits}
\end{figure}

Finally, a fit is performed on the combined data. For the $5$-parameter model, this yields $\chi^2/n_\text{dof}=85.4/40$, whereas the $8$-parameter fit has $\chi^2/n_\text{dof}=115/48$. Since both the 5- and 8-parameter models can describe each individual dataset very well by construction, these unacceptably large values of $\chi^2/n_\text{dof}$ must be due to tensions between the datasets. We resolve this by rescaling all covariance matrices by $2.14$ for the $5$-parameter case and by $2.40$ for the $8$-parameter case, thus enforcing $\chi^2/n_\text{dof}=1$ for the combined fit. This procedure is analogous to the one used in ref.~\cite{DELPHI:2011aa}. The additional 99 variations of the datasets are then generated using the rescaled covariance matrices.

Table \ref{tab:DataFits} shows that more data points are used for the $8$-parameter fit than for the $5$-parameter one. Normally, this should make the $8$-parameter fit more precise. However, as discussed above, the tensions between the datasets are larger for the $8$-parameter case, so the errors are rescaled by a larger factor. As a result the final uncertainties of the $8$- and $5$-parameter fits are very similar. This can be seen in fig.~\ref{fig:DataFits} which shows a comparison of both fits, including their uncertainties. The uncertainty bands correspond to the r.m.s.~of the complete set of data fits. 

Due to its more restrictive functional form, the $5$-parameter fits exhibit a somewhat sharp turn near $x\sim0.6$. The fits based on the $8$-parameter model do not have this behavior. Apart from this feature, the two fit models are essentially equivalent.

\begin{figure}[t]
\centering
\includegraphics[width=0.49\textwidth]{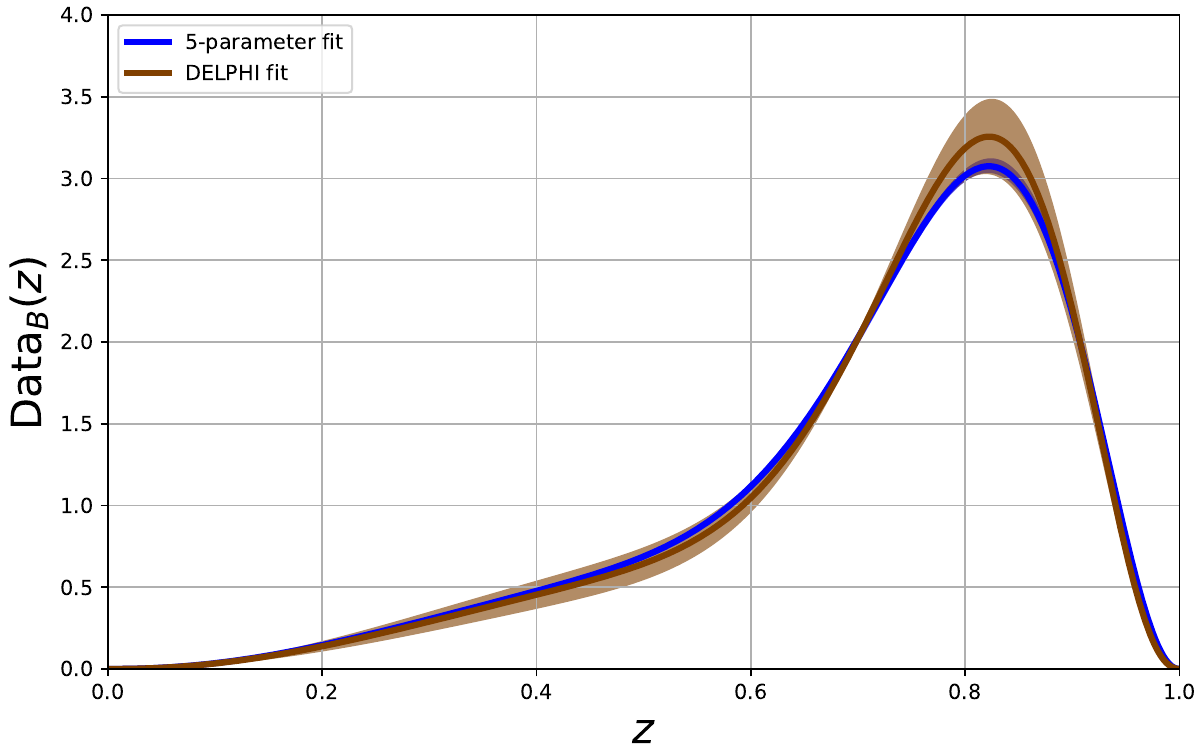}
\includegraphics[width=0.49\textwidth]{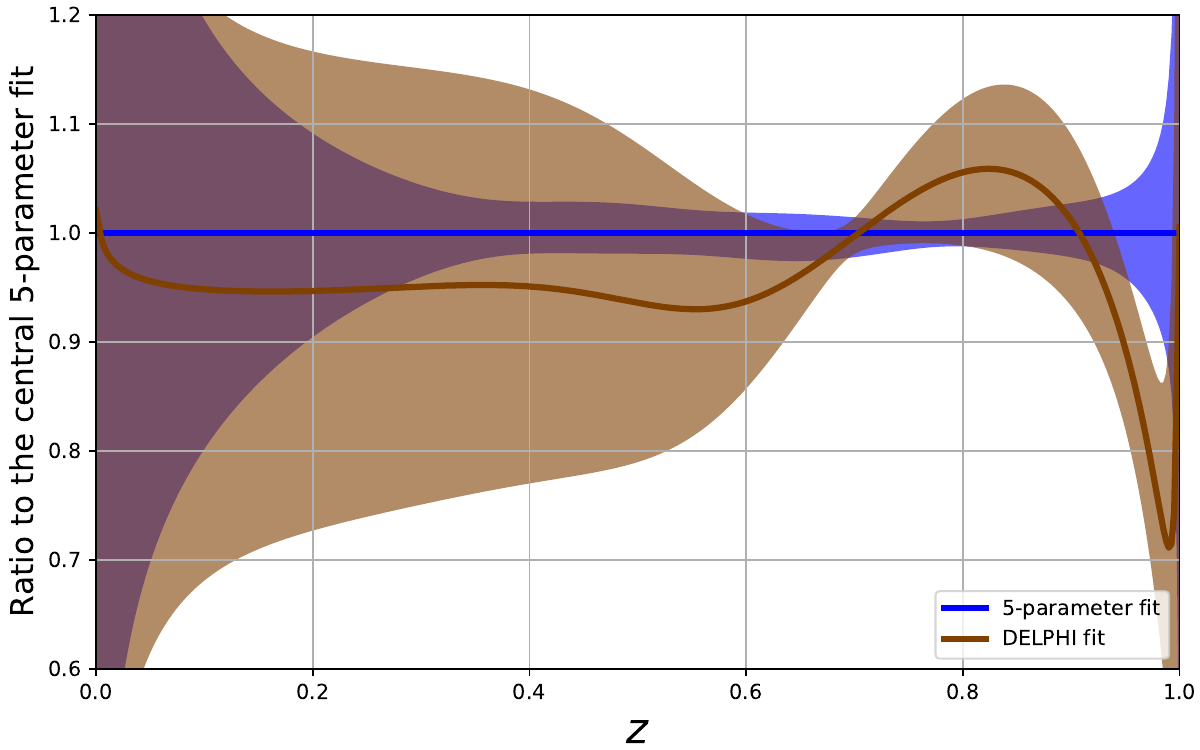}
\caption{Comparison of the $5$-parameter fit (blue) with the fit from ref.~\cite{DELPHI:2011aa} (brown). Absolute distributions (left) and their ratio (right). The blue error band corresponds to the r.m.s. of the 100 data fits while the brown error band corresponds to the propagated errors on the fitted parameters taken from table 12 of ref.~\cite{DELPHI:2011aa}.}
\label{fig:DataFits-comp-DELPHI}
\end{figure}

It is interesting to compare the results from our combined 5-parameter fit to the combined fit performed in ref.~\cite{DELPHI:2011aa}, since both fits are based on the same datasets and utilize the same functional forms. This comparison can be found in fig.~\ref{fig:DataFits-comp-DELPHI}. The two uncertainty bands are modeled differently: our uncertainty band (in blue) corresponds to the r.m.s. of the 100 data fits while the DELPHI error band (in brown) corresponds to the propagated errors on the fitted parameters taken from table 12 of ref.~\cite{DELPHI:2011aa}. The error bands of our new fits are significantly smaller. This is partly due to the fact our new fits retain more eigenvalues of the covariance matrices of the data sets, in particular of the one from OPAL. We have checked that even when using the same eigenvalues as in ref.~\cite{DELPHI:2011aa}, the error bands obtained using our method are still significantly smaller. We suspect the reason for this difference is related to an inconsistency between the published results for the data fit and the first 40 moments of the same fit also given in that reference. Assuming for a moment that the published moments are correct, the size of the uncertainties of the fit parameters must be smaller by a factor of about 5 which would make the uncertainties roughly equal. As an argument that the uncertainty estimate of our data fit is of the correct magnitude, we compare in the following section the size of the uncertainty of the derived NPFF and find it to be of similar size to other extractions in the literature. We have also verified that when using the same number of eigenvalues as ref.~\cite{DELPHI:2011aa} our fit reproduces the first 10 moments given explicitly in table 14 of that reference within 1 permile, while the moments of our fit differ from the moments of the fit published in ref.~\cite{DELPHI:2011aa} by about 4 percent.

\subsubsection{NPFF fitting procedure}\label{sec:NPFFfits}

The NPFF is extracted by comparing theory prediction eqns.~(\ref{eq:dsigmaB}) and (\ref{eq:DiB}) with the combined data as in eq.~(\ref{eq:data(z)}). Since the data fits themselves already encode the experimental uncertainties, the NPFF should not be fitted to the data fits, but rather obtained by exactly deconvolving the data fits using the perturbative prediction. However, since this cannot be achieved analytically, the NPFF is actually fitted to the data fits, but with the ultimate goal to achieve $\chi^2=0$, unlike a proper fit in the usual sense which should have $\chi^2/n_\text{dof}=1$. Because of this, there is no reason to limit the number of degrees of freedom. In fact, it is desirable to use as many degrees of freedom as computationally acceptable, since this will enable the most precise approximation of the exact deconvolution. We have found it best to use a $z$-space parametrization consisting of a high-degree polynomial supplemented with a $\delta$-function
\begin{equation}
\DNP(z) = a_0 \delta(1-z) + \sum_{i = 3}^{31}a_i z^i(1-z)\,.
\label{eq:xSpaceModelV2}
\end{equation}

The coefficients of the polynomial in eq.~(\ref{eq:xSpaceModelV2}) are fixed by minimizing the squared difference between the data fit eq.~(\ref{eq:data(z)}) and the theory prediction, integrated over the range $0.1\le z \le 0.95$. This range is chosen in order to avoid the low $z$ region which is sensitive to missing mass power corrections and the extremely large $z$ region where soft-gluon resummation starts to be important. We repeat this procedure for each one of the 100 data fits, thus producing 100 NPFFs. In this we utilize the 8 parameter data fit discussed in sec.~\ref{sec:datafits}. 

The coefficient of the $\delta$-function is fixed by normalization. Its value for the mean of the NNLO NPFF fits reads 
\begin{equation}
a_0 = 0.159\pm 0.010\,.
\label{eq:a0}
\end{equation}
The reason we have introduced a $\delta$-function in our fit is to increase the hardness of the NPFF since we have not been able to achieve this satisfactorily by means of a pure polynomial fit. Indeed, as can be concluded from $a_0$'s fit value in eq.~(\ref{eq:a0}), the coefficient of the $\delta$-function is incompatible with zero. Note that a $\delta$-function contribution in the NPFF has already been used in the literature in the past \cite{Cacciari:2005uk}.

\begin{figure}[t]
\centering
\includegraphics[width=0.32\textwidth]{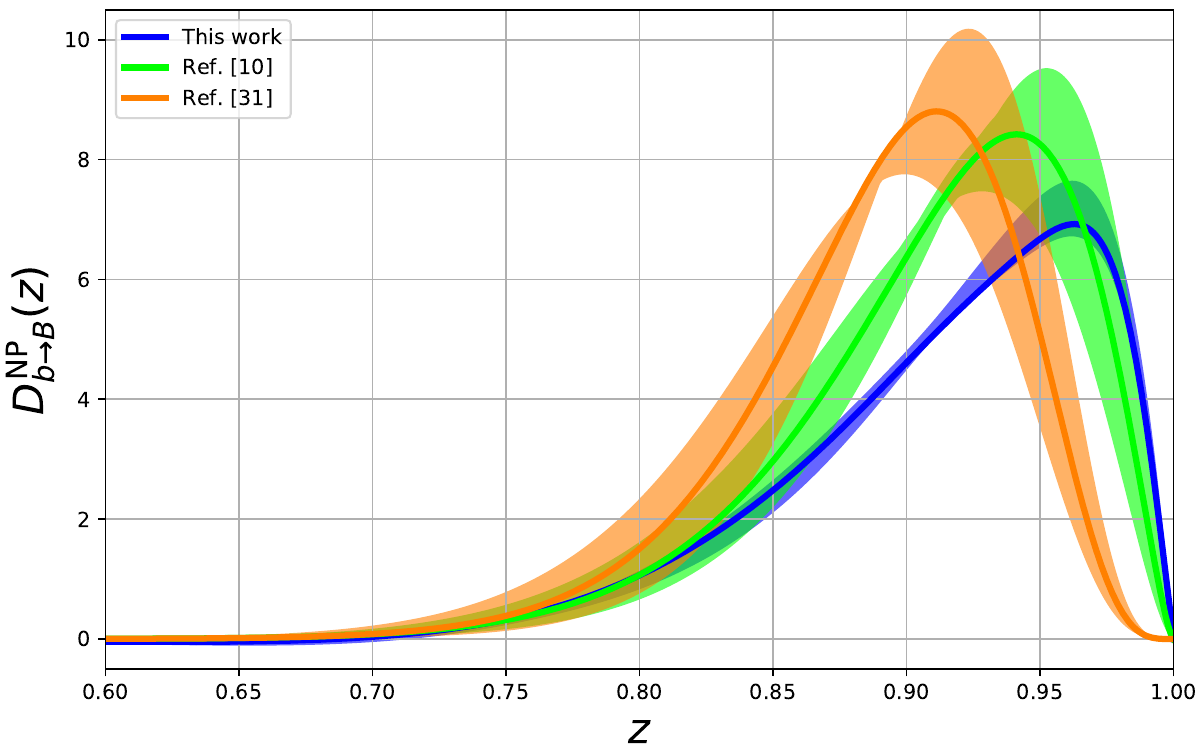}
\includegraphics[width=0.32\textwidth]{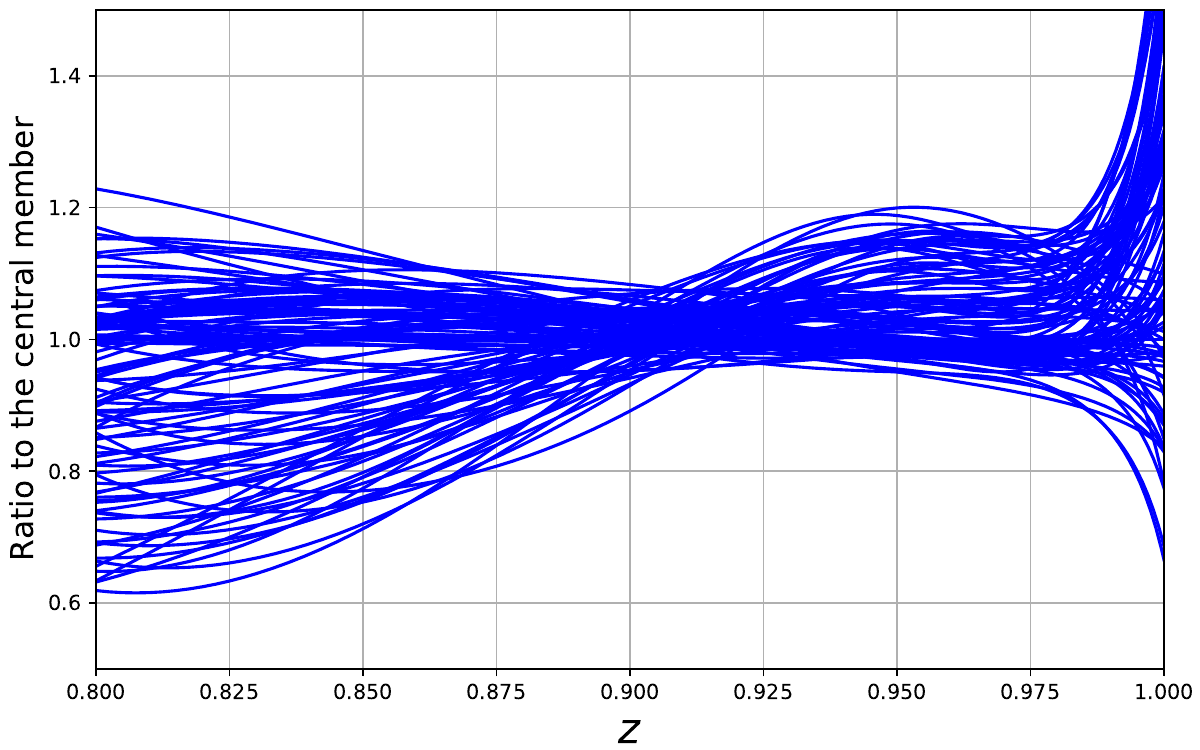}
\includegraphics[width=0.32\textwidth]{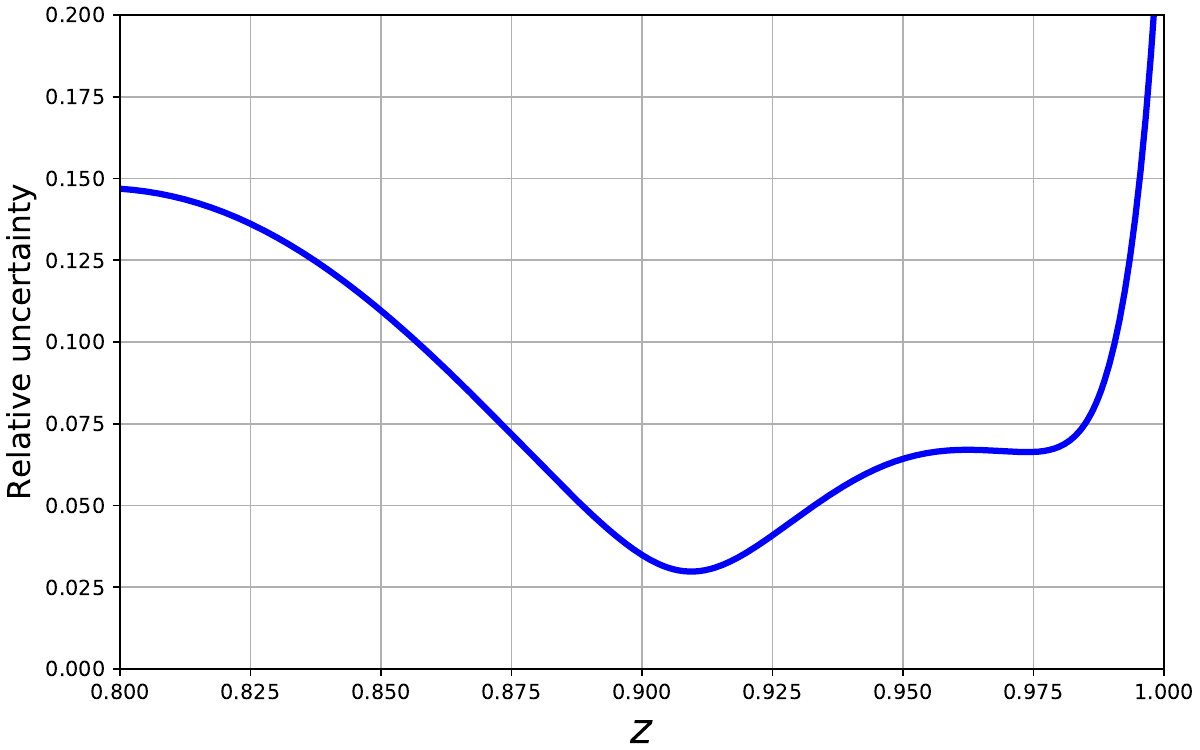}
\caption{Left: a comparison of the new NPFF fit (blue) with the one from ref.~\cite{Fickinger:2016rfd} (orange) and ref.~\cite{Cacciari:2005uk} (green). The error bands correspond to one standard deviation. Center: the ratio plot of the full new NPFF set to its central member, and (right) the relative r.m.s.~of the new NPFF set.}
\label{fig:NPFFfit}
\end{figure}

The new NPFF set derived in this work, including its uncertainty, is shown in fig.~\ref{fig:NPFFfit}. Also shown there is a comparison to existing NPFF sets. One should keep in mind that due to the $\delta$-function term -- which is not plotted -- the normalization of our new set appears different. As can be seen from eq.~(\ref{eq:a0}) the coefficient of this term is about 0.16. Normalization aside, we note that the shape of the new NPFF agrees with the one from ref.~\cite{Cacciari:2005uk} but not with the one from ref.~\cite{Fickinger:2016rfd}. That later set was used previously in ref.~\cite{Czakon:2021ohs}. The difference between the new set derived in this work and the one from ref.~\cite{Fickinger:2016rfd} illustrates the need for a NPFF which is consistently derived within our formalism. 

It is interesting to clarify the reasons behind the differences in shapes and uncertainties between the three sets shown in fig.~\ref{fig:NPFFfit}. First there are differences in the perturbative parts: the set of ref.~\cite{Cacciari:2005uk} is at NLO while the one of ref.~\cite{Fickinger:2016rfd} is with NNLO accuracy but is obtained in a slightly different formalism (a detailed discussion can be found in ref.~\cite{Czakon:2021ohs}). There are differences in the datasets that have been used in the fits: ref.~\cite{Cacciari:2005uk} only fits the ALEPH and SLD datasets. Since the ALEPH data is not very constraining, the error bands of ref.~\cite{Cacciari:2005uk} originate, essentially, from the SLD data. Furthermore, ref.~\cite{Cacciari:2005uk} uses a two-parameter model for the NPFF and does not provide a correlation matrix for the uncertainties on the fit parameters. For this reason the error bands for that set that can be seen in fig.~\ref{fig:NPFFfit} are likely an overestimate. Both of those factors combined could explain why our new NPFF appears to have much smaller uncertainty. The NPFF of ref.~\cite{Fickinger:2016rfd} does not include in their fit the OPAL data, while using less eigenvalues from the other experiments than our new analysis. This explains the reduced uncertainty of our new set relative to the one of ref.~\cite{Fickinger:2016rfd}.

\begin{figure}[t]
	\centering
	\includegraphics[width=0.32\textwidth]{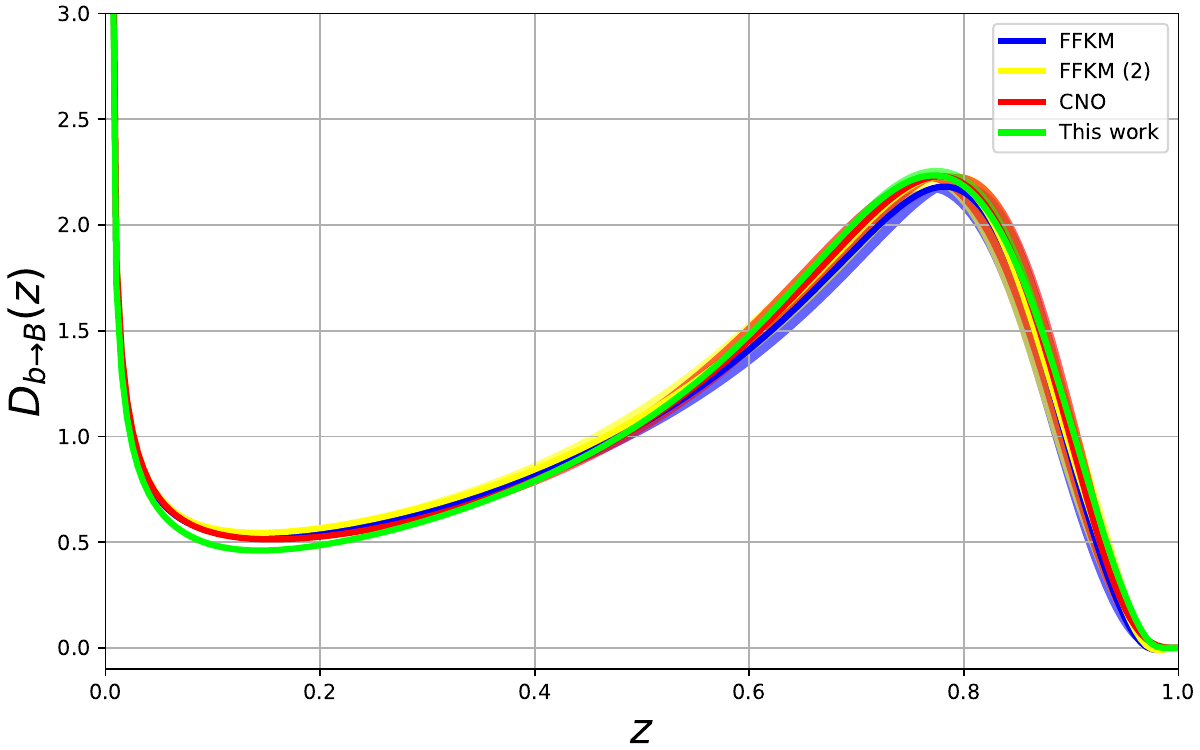}
	\includegraphics[width=0.32\textwidth]{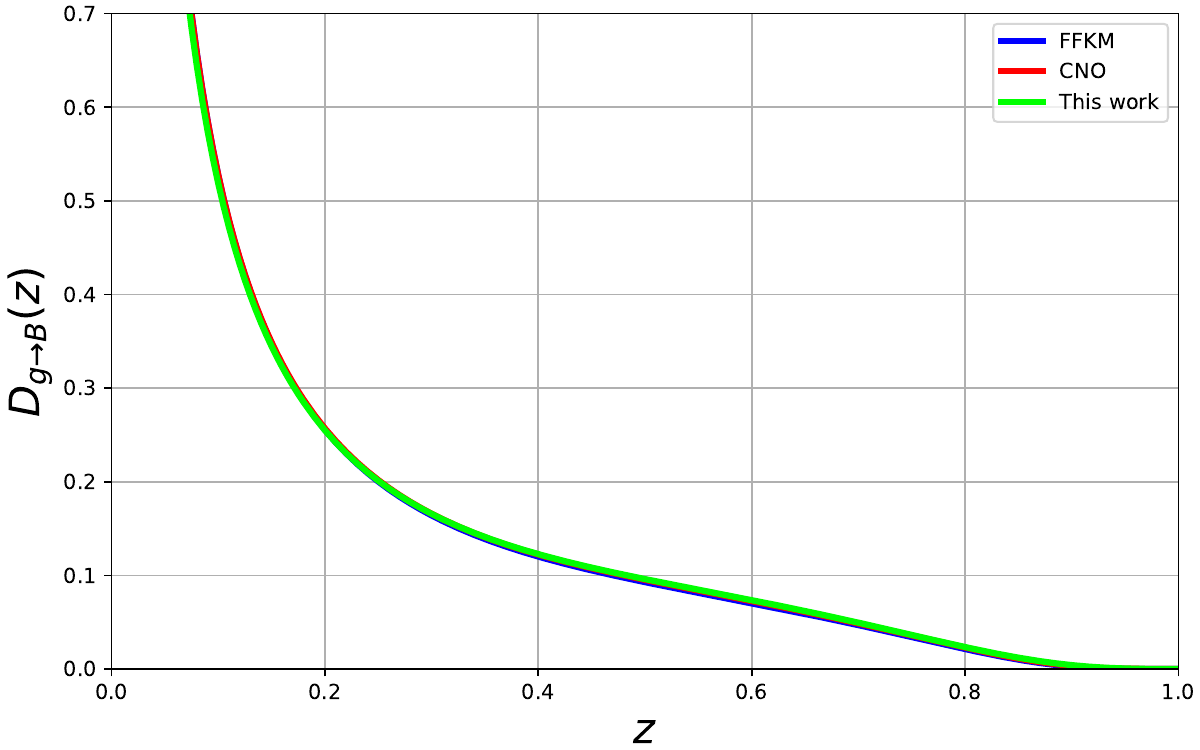}
	\includegraphics[width=0.32\textwidth]{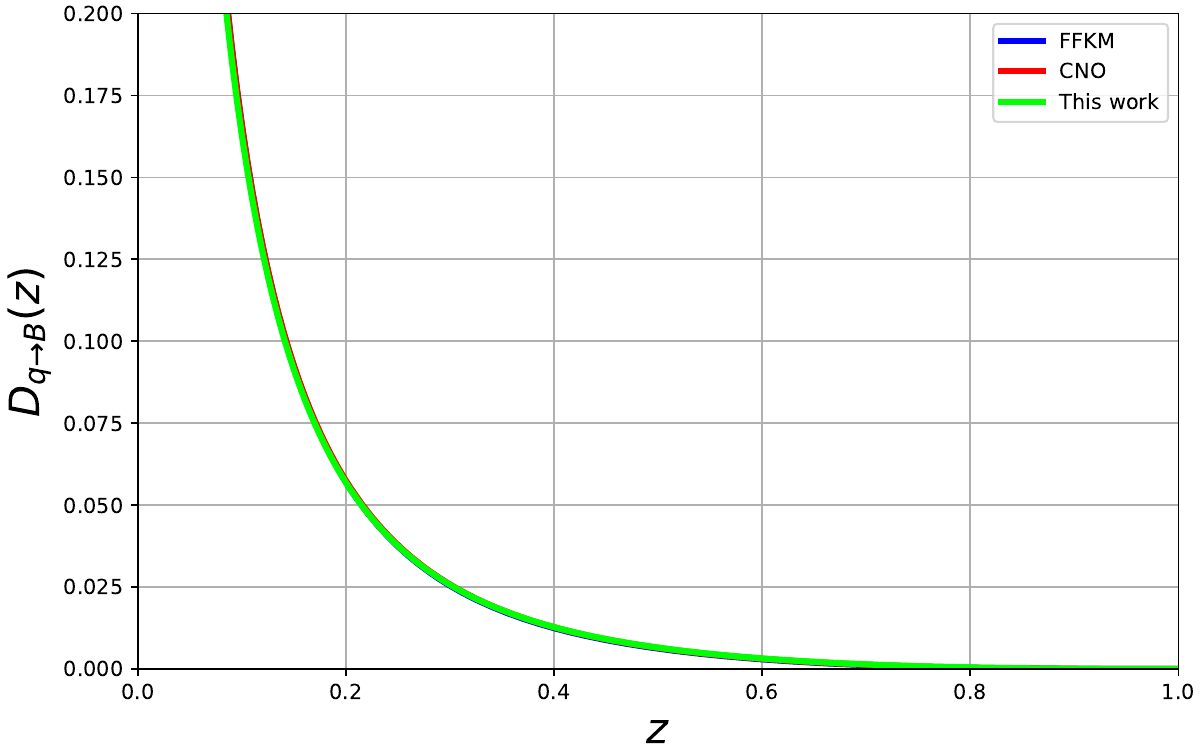}
	\includegraphics[width=0.32\textwidth]{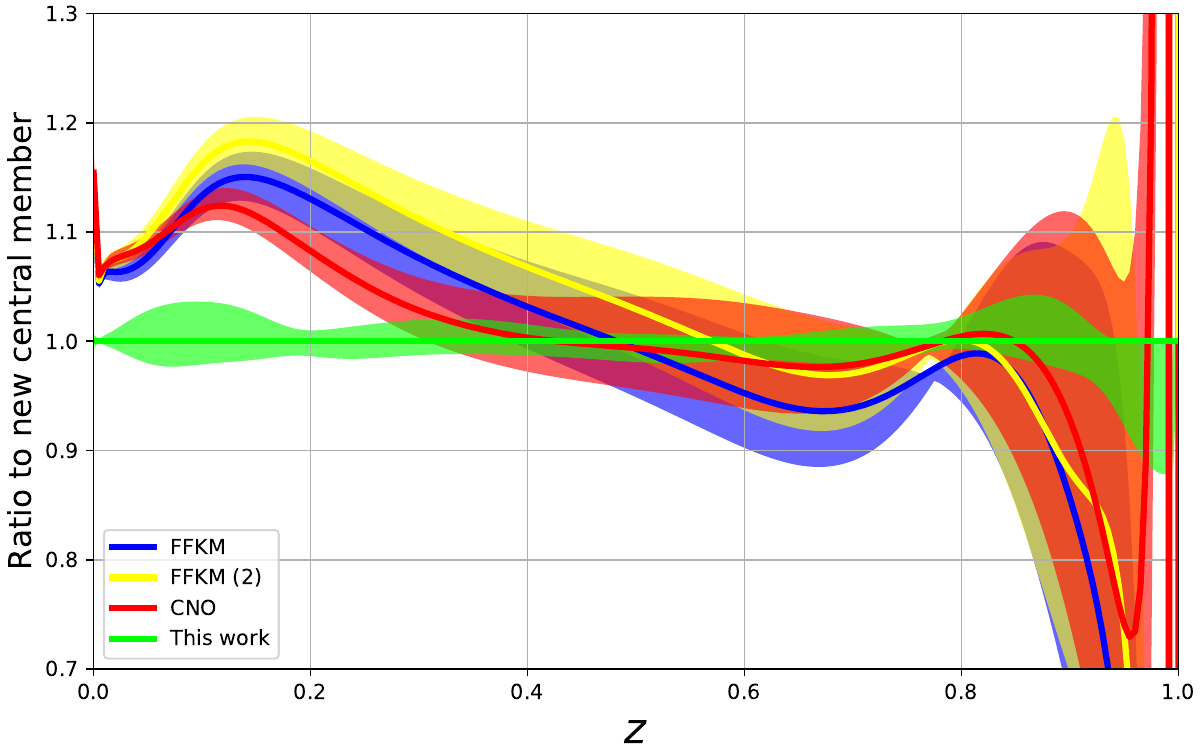}
	\includegraphics[width=0.32\textwidth]{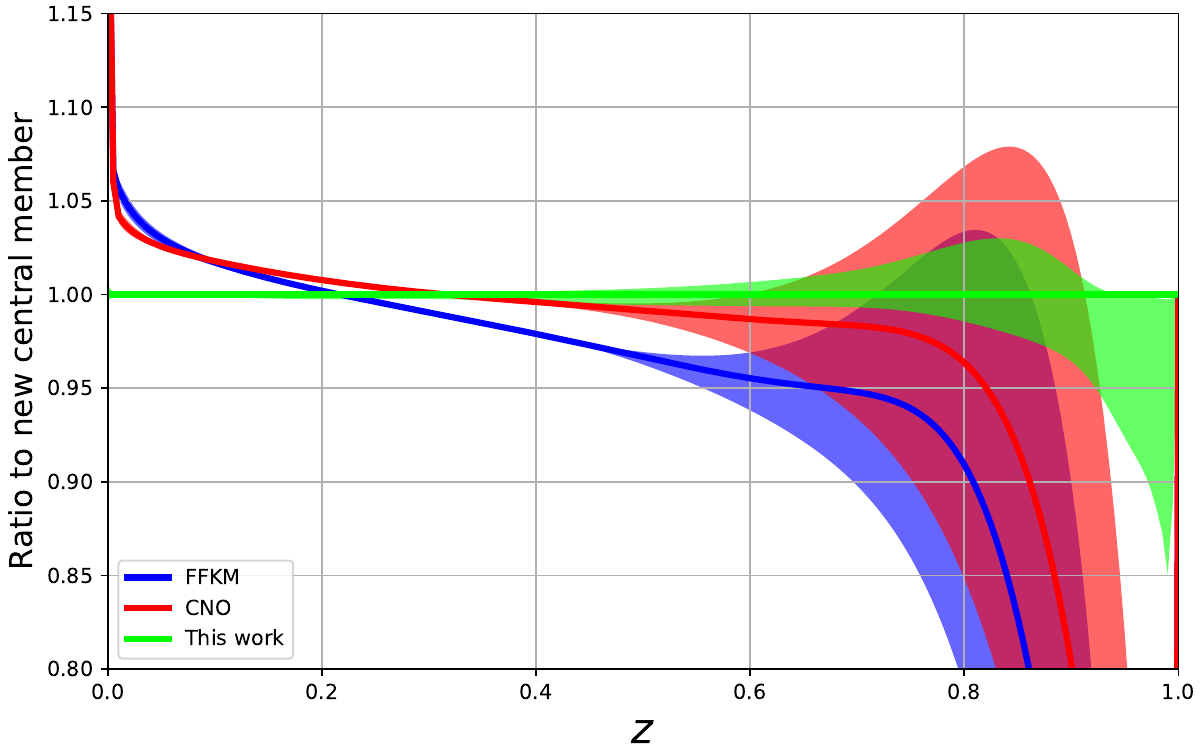}
	\includegraphics[width=0.32\textwidth]{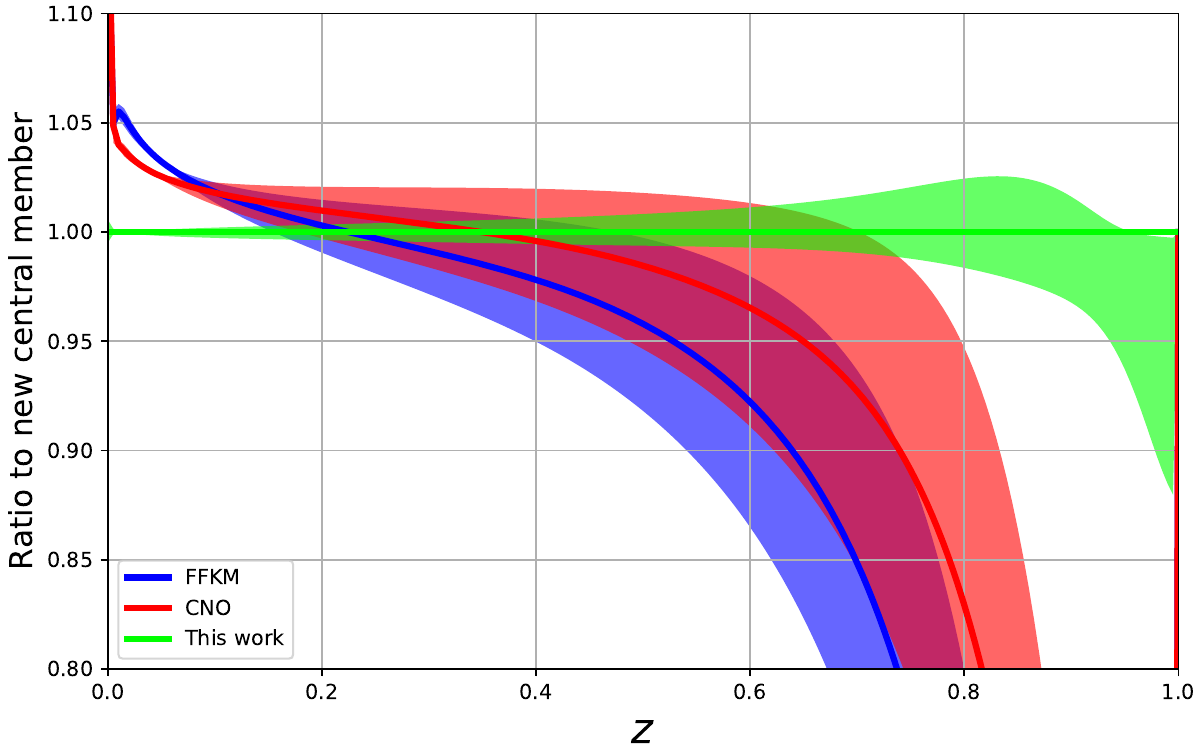}
	\caption{A comparison of the new FF set (green) with the ones constructed in ref.~\cite{Czakon:2021ohs} (yellow, blue and red), together with their uncertainties. From left to right: the $b\to B$ FF,  the $g\to B$ FF and the $q\to B$ FF, where $q$ can be any light quark. The top row shows the absolute plots, while the bottom plots show the ratios to the new FFs. The error bands are as in fig.~\ref{fig:NPFFfit}. The yellow curve is identical to the blue curve for both the gluon and the light-quark FFs, and is not shown in the corresponding plots. All curves are evaluated at $\mu_{Fr} = m_t$.}
	\label{fig:FullFFs}
\end{figure}

A more direct comparison is possible by constructing the full fragmentation function set, i.e.~the convolution of the NPFFs and the PFFs. In ref.~\cite{Czakon:2021ohs}, three full FF sets were constructed from the NPFFs obtained in refs.~\cite{Cacciari:2005uk,Fickinger:2016rfd}. These are compared to the full FF set obtained in this work in fig.~\ref{fig:FullFFs}. Focusing on the $b$-quark FFs, it is clear that there is an excellent agreement within the uncertainties between the different sets for $z > 0.3$. Below $z = 0.3$, our new set deviates from the others. However, this is almost certainly merely an artefact of the fitting procedure used to obtain the old NPFFs. Those NPFFs were fitted using either one or two parameters, whereas our new fit was obtained using 8 parameters. The few parameters of the old fits are almost entirely fixed by the requirement that the position and shape of the peak in the data distribution must be reproduced. Since our fit uses more parameters, it is able to simultaneously describe the peak region and the tail of the data distribution accurately. It was pointed out in ref.~\cite{Fickinger:2016rfd} that the FF obtained there slightly overestimates the tail of the distribution. A similar effect can be observed in figs.~17 and 18 of ref.~\cite{Cacciari:2005uk}. It would thus appear this discrepancy has now been resolved by increasing the number of degrees of freedom of the NPFF fit model.

The gluon and light-quark plots of fig.~\ref{fig:FullFFs} show larger discrepancies. This is to be expected, since these plots correspond, up to mixing with the $b$-quark FF through DGLAP evolution, to the convolution of the NPFFs shown in fig.~\ref{fig:NPFFfit} with the same gluon or light-quark PFF. Since the NPFFs do not agree within their errors, these convolutions cannot be expected to be in agreement. In fact, based on the NPFFs, one would expect our curves to be the hardest, followed by the red curves based on ref.~\cite{Cacciari:2005uk}, which is exactly what can be seen in fig.~\ref{fig:FullFFs}. This highlights the fact that the shape of these FFs is highly sensitive to the perturbative model used to describe the $b$-quark FF. Effectively, the differences between perturbative models for the $b$-quark FF are compensated for by the NPFF fit. This compensation is then passed on to the gluon and light-quark FFs through the convolution. The tensions between the perturbative models for the $b$-quark FF are thus traded for similar tensions between the gluon and light-quark FFs, which are not constrained by the data. Since the model of ref.~\cite{Fickinger:2016rfd} follows a completely different approach, anything beyond a qualitative comparison is perhaps unwarranted. The model of ref.~\cite{Cacciari:2005uk} is identical to the one used here, aside from two differences: the perturbative order, which was limited to NLO in ref.~\cite{Cacciari:2005uk}, and the regularisation of the Landau pole, which ref.~\cite{Cacciari:2005uk} performed by remapping $N$. While the former is a relatively small effect, the latter difference means our NPFF is harder and, in particular, it contains a $\delta$-distribution. This $\delta$-distribution will almost certainly have a large effect on the large-$z$ region, which explains the discrepancy there. As explained above, our new $b$-quark FF features a smaller tail. This effect propagates to the gluon and light-quark FFs, both through the initial condition and through DGLAP evolution, explaining the difference at small $z$ between our new FFs and the red curves of fig.~\ref{fig:FullFFs}. All of this aside, the three sets agree to within a few percent.

The size of the uncertainties on the different fits, as discussed above in the context of fig.~\ref{fig:NPFFfit}, can be more clearly seen in fig.~\ref{fig:FullFFs}, especially in the bottom-left plot. Around the peak, our new FF has an uncertainty of around 2-4\%, whereas the previous FFs all had an uncertainty of roughly 5-10\%. What is especially clear in this plot, is that the procedure for estimating the uncertainty bands of the old sets, i.e.~varying the NPFF parameters by one standard deviation independently, leads to artificially small uncertainties around the peak, whereas the new procedure of taking the r.m.s.~of the full set of 100 members yields very smooth uncertainties without any such artefacts.

One may wonder how the individual datasets that enter our combination affect the determination of the NPFF. To clarify this point in appendix~\ref{sec:appendix-one-dataset} we discuss the extraction of NPFF from each individual dataset and compare the result with the combined extraction. 

Before closing this section, we would like to briefly discuss alternative fitting models. We have also tried to fit the NPFF directly in Mellin space. To that end we have utilized the first 30 or so moments of the $\ee$ coefficient functions and PFF. We have assumed that in $z$-space the NPFF is built out of an exponential (like the one used in ref.~\cite{Fickinger:2016rfd} ) times a polynomial of a high degree and have inverted this expression analytically to $N$-space. We have been able to fit all of the first 30 moments with very high precision, typically better than few permil. Despite that, the predicted $z$-space distribution at the level of $B$ hadrons does not fit well the data it was derived from in the very large $z$ region and close to the peak of the distribution. This seem to suggest that this approach is limited by biases in NPFF's assumed functional form. 

The results for all fits derived in this work are attached as convenient to use LHAPDF \cite{Buckley:2014ana} interpolation grids for the set of fragmentation functions $D_{i\to B}$, evolved to a scale $\mu_{Fr}$. Note that, as in ref.~\cite{Czakon:2021ohs}, these grids have been symmetrized with respect to charge. This means that any results generated with these grids cannot distinguish between hadrons containing a $b$-quark and hadrons containing a $\bar b$-quark; this restriction also applies to the calculations presented in sec.~\ref{sec:pheno} below. The $b\leftrightarrow\bar b$ symmetrization is not an intrinsic limitation of the calculational approach, however. If desirable, future calculations can instead employ unsymmetrized FF sets, allowing a distinction of B-hadrons based on $b$-quark charge to be made.

\subsection{NPFF for $B$ decays to a $J/\psi$ or to a soft muon}\label{sec:NPFF-JPsi-muon}

Not all measurements of $B$-flavored hadrons measure the $B$ directly. What is often measured are $B$ decay products like $J/\psi$ or the $\mu$ originating in semileptonic $B$ decays. Due to their very clean signatures, these final states have been advocated in the context of precise top quark mass determination, see refs.~\cite{Kharchilava:1999yj,CDF:2009mbf,CMS:2016ixg,ATLAS:2022jbw}. In the following we explain how these decays can be incorporated into the $B$-fragmentation formalism discussed in this work.

Consider the decay $B\to d+X$ of a spin-0 $B$-meson at rest, where $d$ is a descendent particle with mass $m_d$. Keeping the decay rate fully differential in the momentum of $d$, the decay rate can be written as
\begin{equation}
d\Gamma(B\to d+X) = \frac{1}{4\pi}f(E_d^\text{rest})dE_d^\text{rest}d\cos(\theta)d\phi\,,
\label{eq:DecayRateDefinition}
\end{equation}
where $E_d^\text{rest}$ is the energy of the descendant in the $B$-hadron rest frame and the normalization is such that
\begin{equation}
\frac{d\Gamma(B\to d+X)}{dE_d^\text{rest}} = f(E_d^\text{rest})\,.
\end{equation}

Introducing the dimensionless variable $y = E_d^\text{rest}/m_B$, with $0\le y\le 1$, eq.~(\ref{eq:DecayRateDefinition}) becomes
\begin{equation}
d\Gamma(B\to d+X) = \frac{m_B}{4\pi}f(y\:m_B)dyd\cos(\theta)d\phi\,.
\end{equation}
This phase space can next be boosted from the $B$-hadron rest frame to a frame with $E_B\gg m_B$. Without loss of generality the boost can be taken to be along the $z$-axis. The goal is to express the energy of $d$ in this boosted frame as a fraction $z$ of $E_B$, i.e.~$E_d = z E_B$. Using
\begin{align}
\delta\bigg(z-\frac{E_d}{E_B}\bigg) &= \delta\bigg(z-\gamma_B\frac{E_d^\text{rest}+\beta_B\cos(\theta)\sqrt{(E_d^\text{rest})^2-m_d^2}}{E_B}\bigg) \notag\\&\approx \delta\bigg(z-\frac{E_d^\text{rest}+\cos(\theta)\sqrt{(E_d^\text{rest})^2-m_d^2}}{m_B}\bigg) \notag\\&= \delta\bigg(z-y-\cos(\theta)\sqrt{y^2-\frac{m_d^2}{m_B^2}}\bigg)\;,
\end{align}
where on the second line $\gamma_B = E_B/m_B$ and $\beta_B \approx 1$ were used, this yields
\begin{align}
\frac{d\Gamma(B\to d+X)}{dz} &= \int_0^{2\pi}\int_{-1}^{1}\int_0^1\frac{m_B}{4\pi}f(y\:m_B)\delta\bigg(z-y-\cos(\theta)\sqrt{y^2-\frac{m_d^2}{m_B^2}}\bigg)dyd\cos(\theta)d\phi \notag\\
&= \int_0^1\frac{m_B}{2\sqrt{y^2-\frac{m_d^2}{m_B^2}}}f(y\:m_B)\theta\bigg(1-\frac{(z-y)^2}{y^2-\frac{m_d^2}{m_B^2}}\bigg)dy \notag\\
&= \int_{\frac{z}{2}+\frac{m_d^2}{2z\:m_B^2}}^1\frac{m_B}{2\sqrt{y^2-\frac{m_d^2}{m_B^2}}}f(y\:m_B)dy \equiv \Gamma_B\:D_{B\to d}(z)\;.
\label{eq:DecayFunctionDerivation}
\end{align}

The function $D_{B\to d}(z)$ can be interpreted as a fragmentation function for the transition $B\to d$. Since the consecutive fragmentations $\DNP$ and $D_{B\to d}$ can be combined via convolution, the decay $B\to d+X$ can be included in the calculation by simply convolving $\DNP$ and $D_{B\to d}$ and then running the calculation as one normally would for the production of $B$-hadrons. The only ingredient necessary in order to incorporate $B$-hadron decays in a fragmentation calculation is $f(E_d^\text{rest})$, i.e.~$d\Gamma(B\to d+X)/dE_d^\text{rest}$. Such energy spectra are highly constrained by very precise measurements at $B$-factories, which means that the total theory error is essentially unaffected by the further inclusion of the $B$ decay. 

The decay energy spectra $B\to J/\psi+X$ and $B\to \mu+X$ needed in this work are obtained from ver.~2.1.1 of the program {\tt EvtGen} \cite{Lange:2001uf}. We study $B$-hadron decays with $J/\psi$ or muons for several different types of $B$-hadrons: $B^+$, $B^0$, $B_s^0$ and $\overline{\Lambda}_b^0$. The mixture of $B$-hadrons produced in high-energy collisions corresponds to $40.8\%$ $B^\pm$, $40.8\%$ $\stackon[.1pt]{$B$}{\brabar}^0$, $10.0\%$ $\stackon[.1pt]{$B$}{\brabar}_s^0$ and $8.4\%$ $B$-baryons \cite{ParticleDataGroup:2020ssz}. In the following, it is assumed that the vast majority of produced B-baryons are $\stackon[.1pt]{$\Lambda$}{\brabar}_b^0$'s, such that the contributions from other $B$-baryons can be neglected.

\subsubsection{Muon spectra}\label{sec:muon-spectra}

In $B\to \mu+X$ decays one needs to distinguish between same-sign decays ($\overline{b}\to\mu^+$) and opposite-sign decays ($\overline{b}\to\mu^-$), as well as between prompt muons ($\overline{b}\to\mu^+$) and non-prompt muons (mostly $\overline{b}\to \overline{c}\to\mu^-$). There are also two types of spectra, corresponding to different ways of treating final states with multiple muons: multiplicities (one counts all muons) and hardest-only (one counts only the hardest muon). Additionally, we could study the spectra in the rest frame or in the boosted one. This results in a total of $144$ different spectra (a few of which are identically equal to zero), all of which have been generated using {\tt EvtGen}. We note that eq.~(\ref{eq:DecayFunctionDerivation}) assumes isotropic decays, which is a valid assumption for spin-0 $B$-mesons, but not for $\overline{\Lambda}_b^0$. This does not appear to be an issue in practice since, we have checked, the analytically boosted spectra and the MC histograms revealed no systematic difference.

We fit the total (the sum of same-sign, opposite-sign, prompt and non-prompt) hardest-only spectrum of each individual $B$-hadron species. This is the type of spectrum needed to describe the measurement presented in ref.~\cite{ATLAS:2022jbw}. The choice to include both same-sign and opposite-sign decays is based on the fact that, as explained in sec.~\ref{sec:NPFFfits}, the $B$-hadron FF sets used in this work have been charge-symmetrized. All spectra are normalized to unity for the fit and then rescaled to their physical normalization according to {\tt EvtGen}. The fit model is a cubic spline, continuous up to the second derivative, which splits the range $[y_\text{min},y_\text{max}]$ into 15 regions, where $y_\text{min} = m_\mu/m_B$ and $y_\text{max} = 1/2(1+m_\mu^2/m_B^2)$. The exact positions of the splits is chosen in such a way that all spectra can be fitted simultaneously. The fit is performed by binning the decay events in $32$ bins of equal width, covering the interval $[y_\text{min},y_\text{max}]$, and then minimizing $\chi^2$ with uncertainties from the Monte Carlo integration only.

\begin{figure}[t]
\centering
\includegraphics[width=0.49\textwidth]{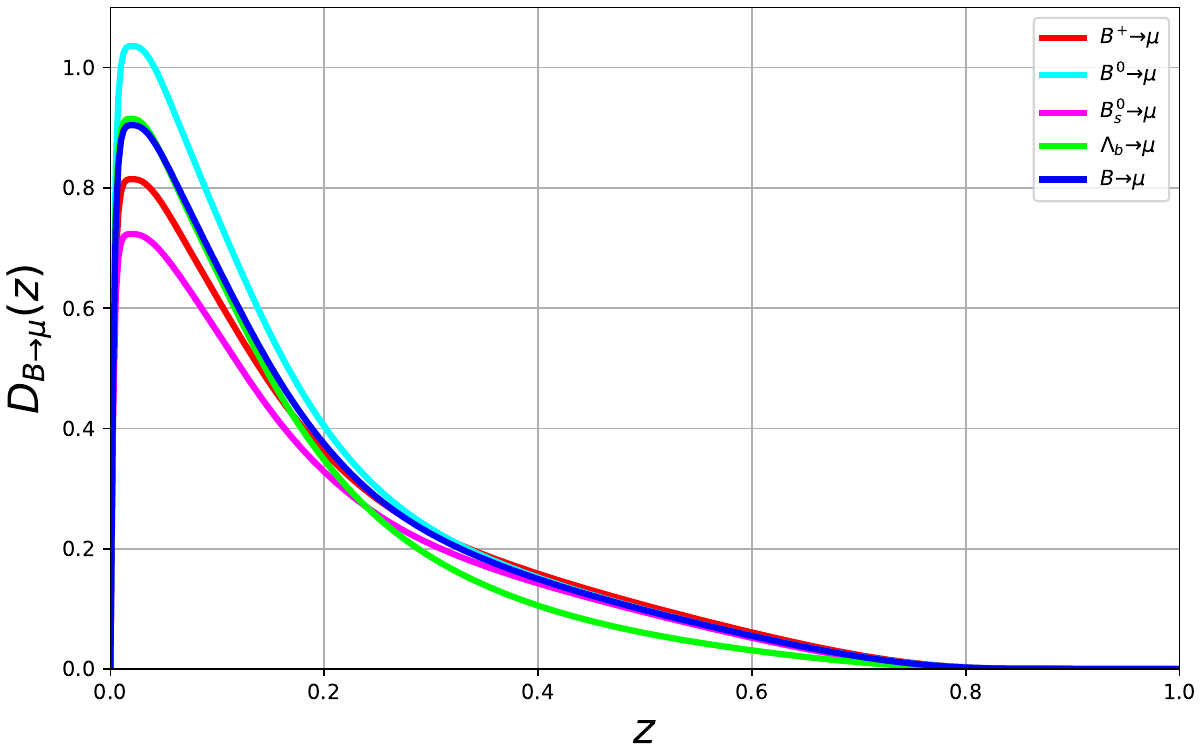}
\includegraphics[width=0.49\textwidth]{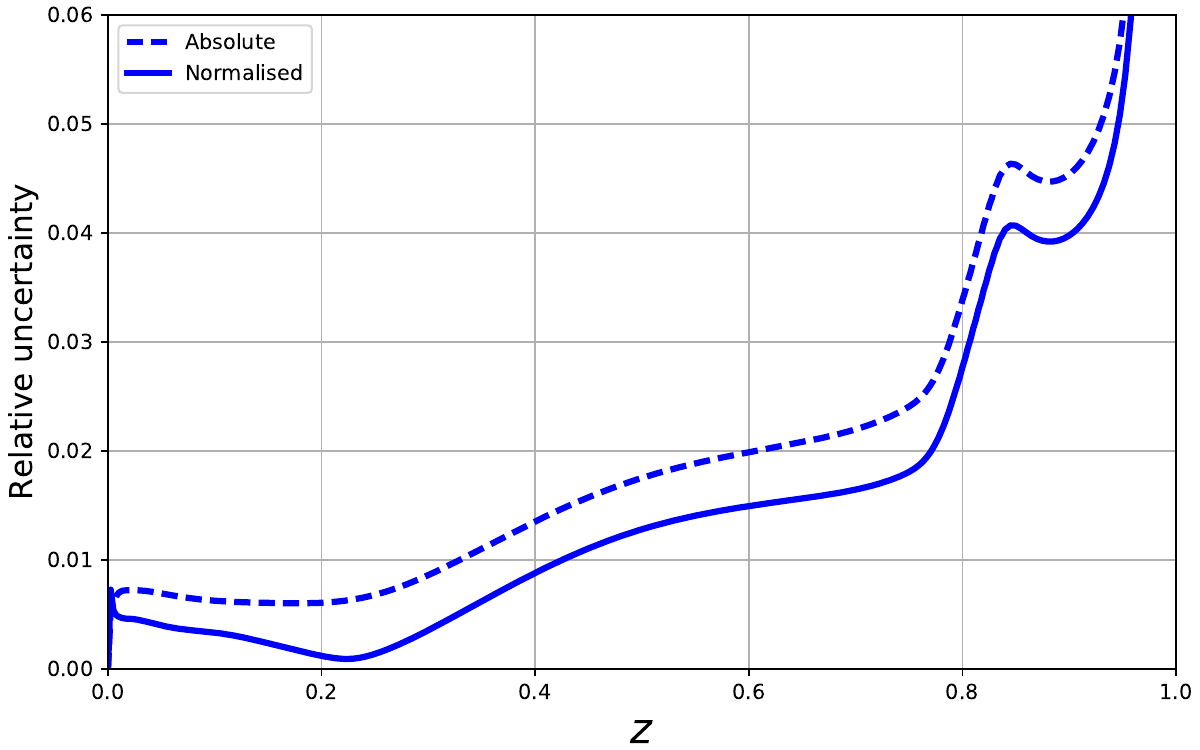}
\caption{Left: the muon decay fragmentation function for the high-energy B-hadron mixture. Right: the corresponding uncertainty plot. The error bands are obtained as described in sec.~\ref{sec:muon-spectra}.}
\label{fig:BDecayFF}
\end{figure}

The final spectrum is the linear combination of the spectra of the four $B$-hadron species according to the hadronization fractions given above. This spectrum is shown in fig.~\ref{fig:BDecayFF}. The error bands are estimated in the following way, in analogy with ref.~\cite{ATLAS:2022jbw}. The total error is given by the combination of the errors on the $B$-hadron hadronization fractions and those on the branching ratios. First, the errors on the $B$-hadron hadronization fractions are propagated. Then, these errors are rescaled by the ratio of the total systematic error coming from $B$-hadron hadronization fractions and branching ratios to the systematic error coming from $B$-hadron hadronization fractions only, as can be extracted from fig.~9 in ref.~\cite{ATLAS:2022jbw}. This method gives a reasonable estimate of the size of the uncertainties, keeping in mind the obtained errors are so small compared to the errors on the NPFF, that they can be safely neglected even if they are underestimated by a factor of 3.

\subsubsection{$J/\psi$ spectra}

We follow a strategy very similar to the one for decays to a muon. There are, however, a few differences. First, since it is kinematically impossible for any of the $B$-hadrons considered here to decay to more than one $J/\psi$ and because the $J/\psi$ is charge-neutral, there are only $24$ different spectra of interest. Only the combination of prompt and non-prompt $J/\psi$'s will be studied here. Note that in {\tt EvtGen 2.1.1} the decay $B^0\to J/\psi\;K^0$ is included twice: once as just $B^0\to J/\psi\;K^0$ and once distinguishing between $B^0\to J/\psi\;K_S^0$ and $B^0\to J/\psi\;K_L^0$. We have made sure to avoid double counting, which otherwise could lead to a visible difference in the spectrum.

Second, each one of the $B$-hadrons considered here has at least one two-body decay to a $J/\psi$, leading to sharp peaks in the rest-frame energy spectrum. Some of these peaks appear delta-function-like (due to decays involving particles with very small decay widths) while others, despite being resolvable, are hard to fit with a pure spline. Since the spectra of interest are the boosted ones, we have fitted such decays directly in the boosted frame using the fact that boosting turns delta-distributions into step functions and sharp Breit-Wigner peaks into smooth steps. The latter can be easily fitted with a cubic spline, whereas the former have been included analytically with the same masses and branching ratios as in {\tt EvtGen}.

The fit model is again a cubic spline, continuous up to the second derivative. However, the fitting of boosted spectra requires extra effort, since they satisfy the following symmetry relation
\begin{equation}
D(x) = D\bigg(\frac{y_\text{min}^2}{x}\bigg)\;,
\label{eq:BoostedSymmetryRelation}
\end{equation}
which is taken into account by covering the region $[y_\text{min},1]$ with the spline and mapping it to $[y_\text{min}^2,y_\text{min}]$ using eq.~(\ref{eq:BoostedSymmetryRelation}). Additionally, a much larger number of bins -- 256 -- was used in order to resolve the smooth steps. The spline splits the range $[y_\text{min},1]$ into $25$ regions, the exact positions of the splits having been manually chosen to ensure a good fit for all spectra simultaneously. The exception is the $\overline{\Lambda}_b^0$ spectrum, which differs considerably from the others in its shape. It was fitted using a spline with $15$ regions.

\begin{figure}[t]
\centering
\includegraphics[width=0.49\textwidth]{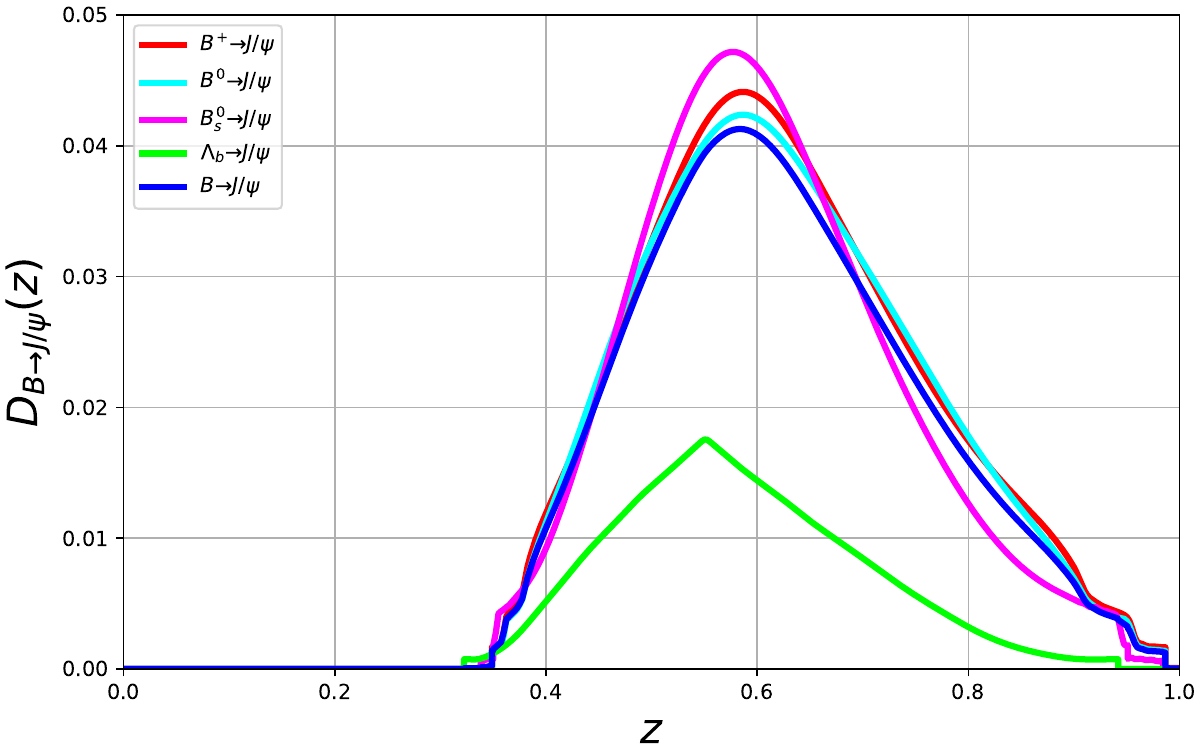}
\includegraphics[width=0.49\textwidth]{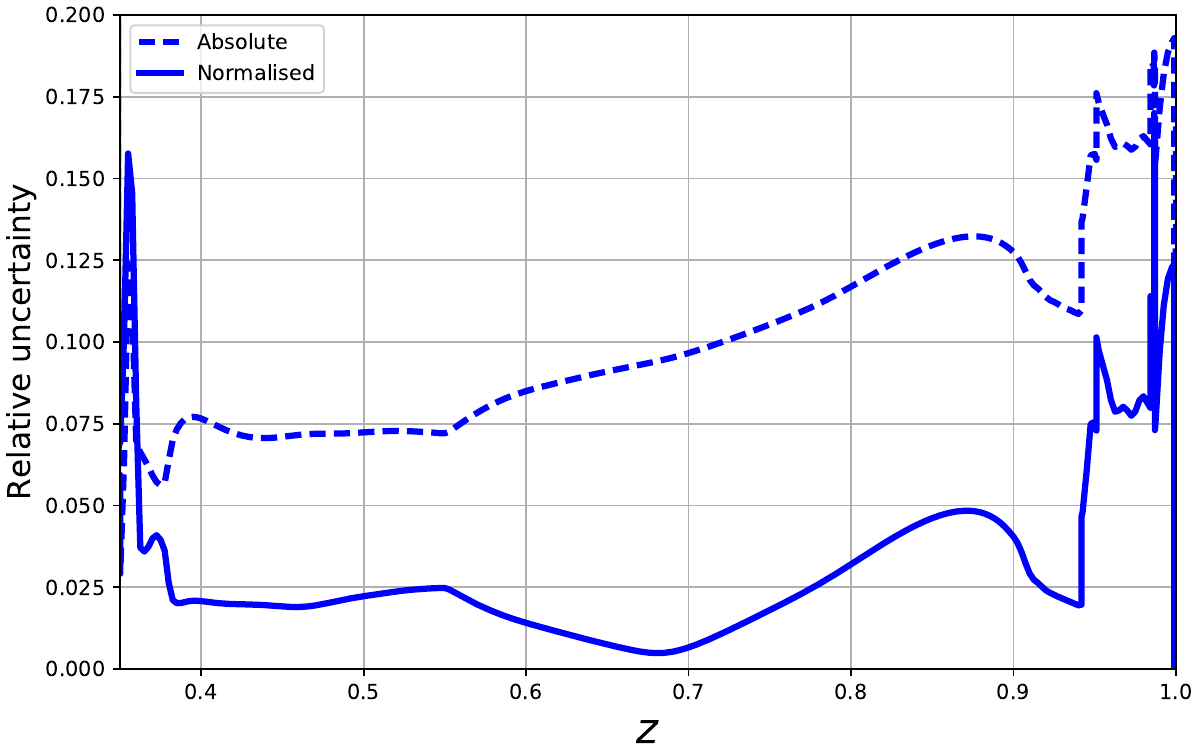}
\caption{The $J/\psi$ decay fragmentation functions for various $B$ decays (left) and corresponding uncertainty plot (right).}
\label{fig:BDecayJpsiFF}
\end{figure}

The spectra of all four $B$-hadron species and the $B$-hadron mixture are shown in fig.~\ref{fig:BDecayJpsiFF}. There is a sizable ($\sim8\%$) uncertainty on the normalization but the uncertainty on the shape is again small enough and can be neglected.

\section{Application: NNLO-accurate $B, J/\psi$ and soft-$\mu$ observables in $t\bar t$ events}\label{sec:pheno}

Top quark decays at the LHC are a major source of $b$ quarks. For this reason $t\bar t$ events at the LHC allow one to investigate $b$-fragmentation in a high-purity, high-statistics environment. Additionally, a number of $b$-fragmentation observables are sensitive to the value of the top quark mass and, since long ago, have been advocated in the context of precision determination of $m_t$. With such applications in mind, we apply the NPFFs extracted in the present work to $t\bar t$ production and decay at the LHC and make NNLO accurate predictions for a number of observables involving an identified $B$-hadron, $J/\psi$ or a soft muon.

Our setup is as follows. We consider the production and decay of a top-quark pair at the LHC at 13 TeV in the Narrow Width Approximation
\begin{equation}
    pp \to t\bar{t} + X \to \ell \bar{\ell} \nu \bar{\nu} b \bar{b} + X\;.
\end{equation}

We consider inclusive final states with a single identified particle $F$, with $F$ being a $B$, a $J/\psi$ or a $\mu$. Unless stated otherwise, the top-quark mass is set to $m_t = 172.5$ GeV. We utilize the {\tt NNPDF3.1} NNLO pdf set \cite{Ball:2017nwa} and the following two combinations for the renormalization ($\mu_R$), factorization ($\mu_F$) and fragmentation ($\mu_{Fr}$) scales
\begin{eqnarray}
&& \mu_R=\mu_F=\mu_{Fr} = m_t/2\;, \label{eq:scale-mt}\\
&& \mu_R = \mu_F = H_T/4~,~ \mu_{Fr} = m_t/2\;. \label{eq:scale-HT}
\end{eqnarray}
The baseline scale variation for the fixed scale choice eq.~(\ref{eq:scale-mt}) is the 15-point scale variation where all three scales are independently varied by a factor of 2 around their central values, subject to the constraints 
\begin{eqnarray}
&& 1/2 \le \mu_{R}/\mu_{F} \le 2\,, \nonumber\\
&& 1/2 \le \mu_{R}/\mu_{Fr} \le 2\,, \label{eq:scales-var}\\
&& 1/2 \le \mu_{F}/\mu_{Fr} \le 2\,.\nonumber
\end{eqnarray}
For the dynamic scale choice eq.~(\ref{eq:scale-HT}) we use the standard 7-point variation of $\mu_R$ and $\mu_F$ with fixed $\mu_{Fr} = m_t/2$.

We do not match the order of the PDF and FF sets to the order of the calculation, choosing to always use the NNLO sets instead. The effect of this choice on the LO and NLO predictions presented below is a higher order effect (NLO for LO predictions and NNLO for NLO predictions) and can be safely ignored compared to the scale uncertainties at those orders. Indeed, the differences between e.g.~the NLO and NNLO FFs would be similar to the differences between the new NNLO set and the set constructed in ref.~\cite{Czakon:2021ohs} based on ref.~\cite{Cacciari:2005uk}. These differences are not much larger than the uncertainties on the new FF set, which, as will be shown below, are barely visible when compared to the (N)LO scale uncertainties.

As in ref.~\cite{Czakon:2021ohs}, the electroweak parameters are defined within the $G_F$ scheme
\begin{eqnarray}
  m_W &=& 80.385 \; \text{GeV} \,,\nonumber\\ 
  \Gamma_W &=& 2.0928\;\text{GeV}\,,\nonumber\\
  m_Z &=& 91.1876 \; \text{GeV} \,,\nonumber\\ 
  \Gamma_Z &=& 2.4952\;\text{GeV}\,,\nonumber\\
  G_F &=& 1.166379\cdot10^{-5} \;\text{GeV}^{-2} \,,\nonumber\\ 
  \alpha &=&\frac{\sqrt{2} G_F}{\pi} m_W^2\left(1-(m_W/m_Z)^2\right)\,.
  \label{eq:GF-scheme}
\end{eqnarray}
The top quark width at leading order reads
\begin{equation}
\Gamma_t^{(0)} = G_F \frac{m_t^3}{8\pi\sqrt{2}}(1-\xi)^2(1+2\xi) = 1.48063 \;\text{GeV}~ ({\rm for}~ m_t = 172.5\;\text{GeV})\,,
\end{equation}
where $\xi=(m_W/m_t)^2$.

\subsection{Differential distributions of a fragmenting $B$, $J/\psi$ or $\mu$}\label{sec:diff-dist}

We consider the following event selection requirements
\begin{itemize}
    \item $p_T(\ell) > 25$ GeV, $|\eta(\ell)| < 2.5$\,,
    \item at least 2 anti-kT jets ($R = 0.4$) with $p_T(j) > 25$ GeV and $|\eta(j)| < 2.5$\,,
    \item $\Delta R (\ell,j) > 0.4$\,,
    \item $p_T(F) > 8$ GeV and $|\eta(F)| < 2.5$,~ $F$ {\rm must be part of one jet}\,.
\end{itemize}
The following differential observables have been considered: $p_T(F), |\eta(F)|, m(F\ell), E(F)$ and $p_T(F)/p_T(J_F)$ where $J_F$ denotes the jet containing $F$. These observables are computed for all values of $F=\{B,J/\psi,\mu\}$. 

\begin{figure}[t]
\centering
\includegraphics[width=0.49\textwidth]{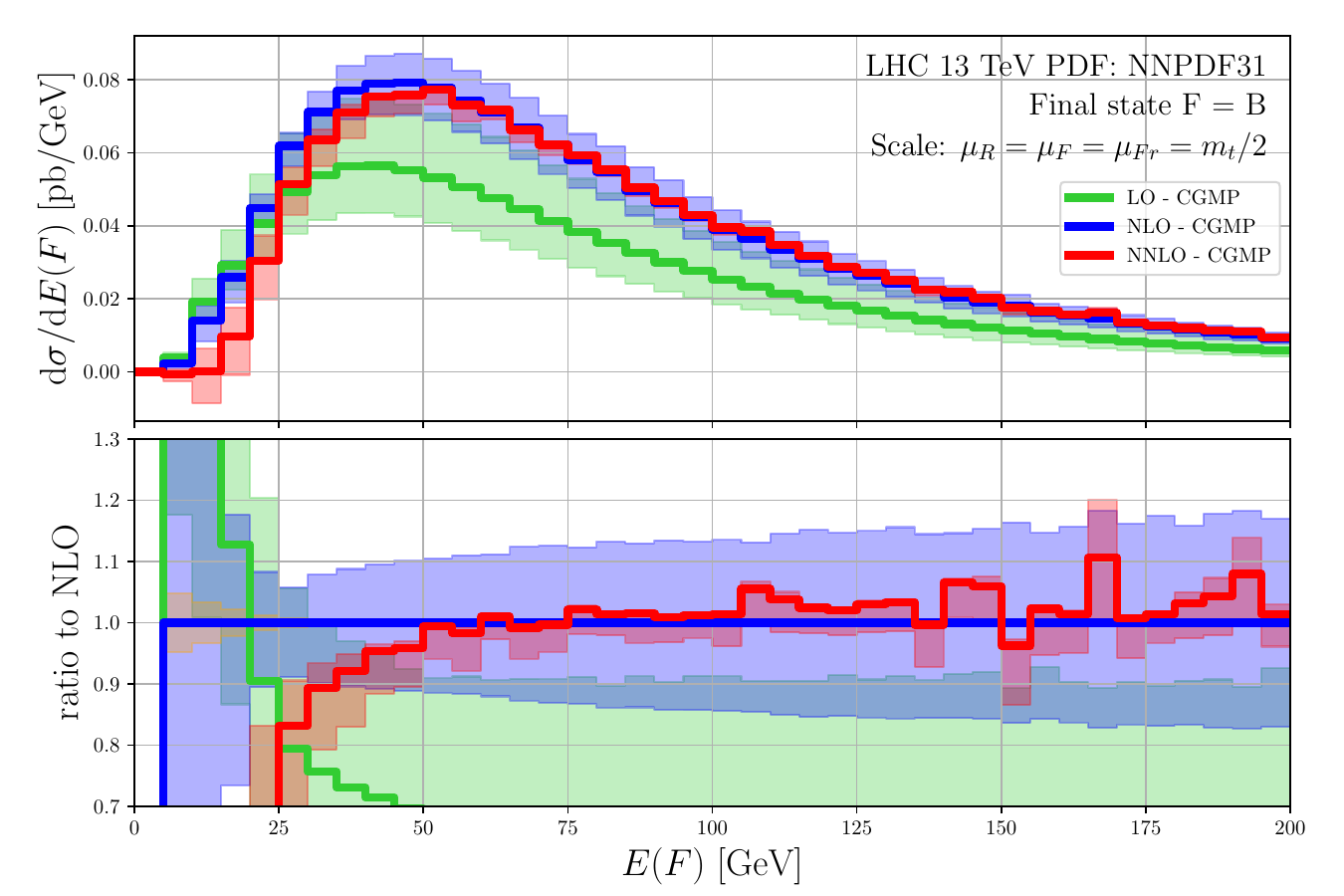}
\includegraphics[width=0.49\textwidth]{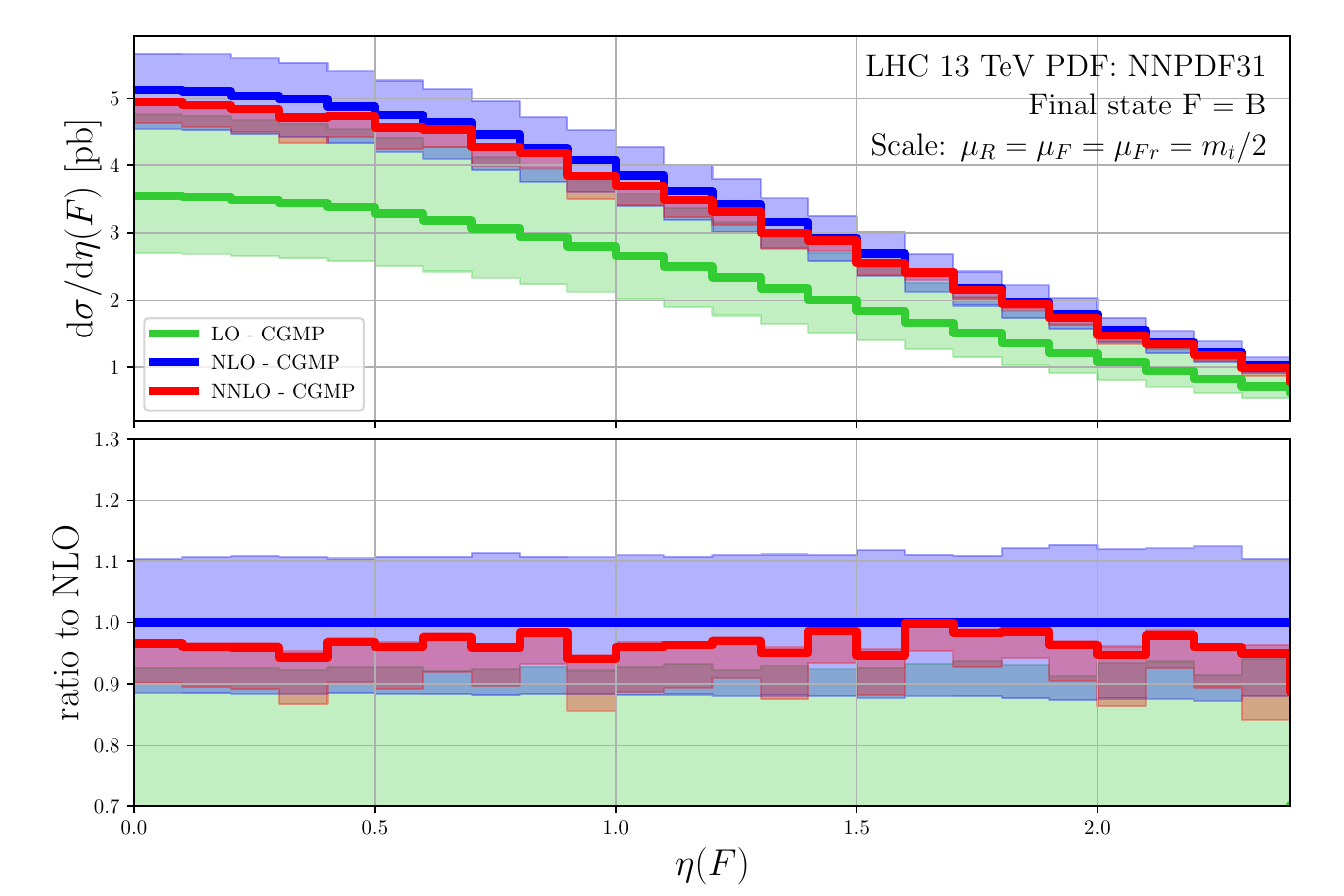}
\includegraphics[width=0.49\textwidth]{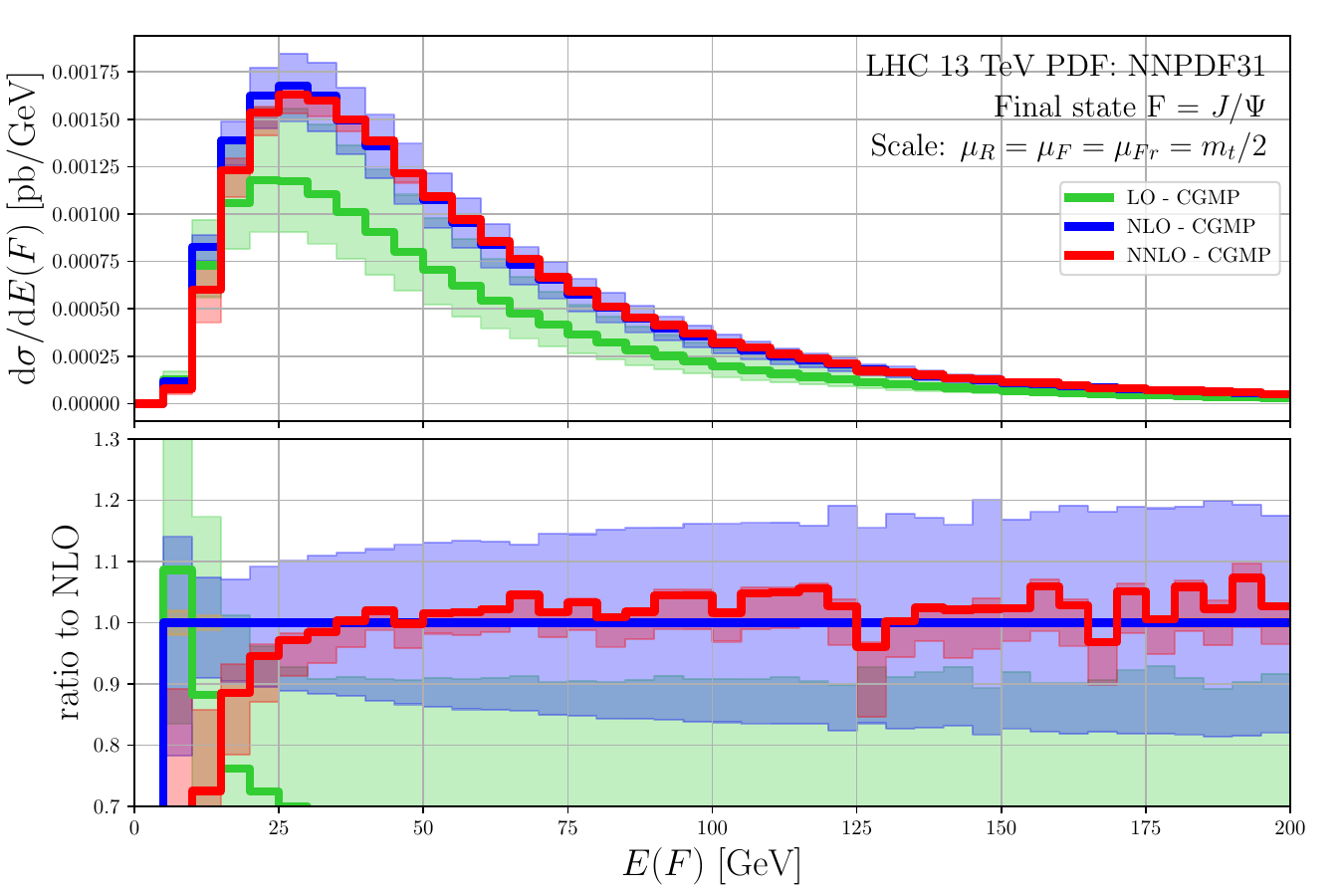}
\includegraphics[width=0.49\textwidth]{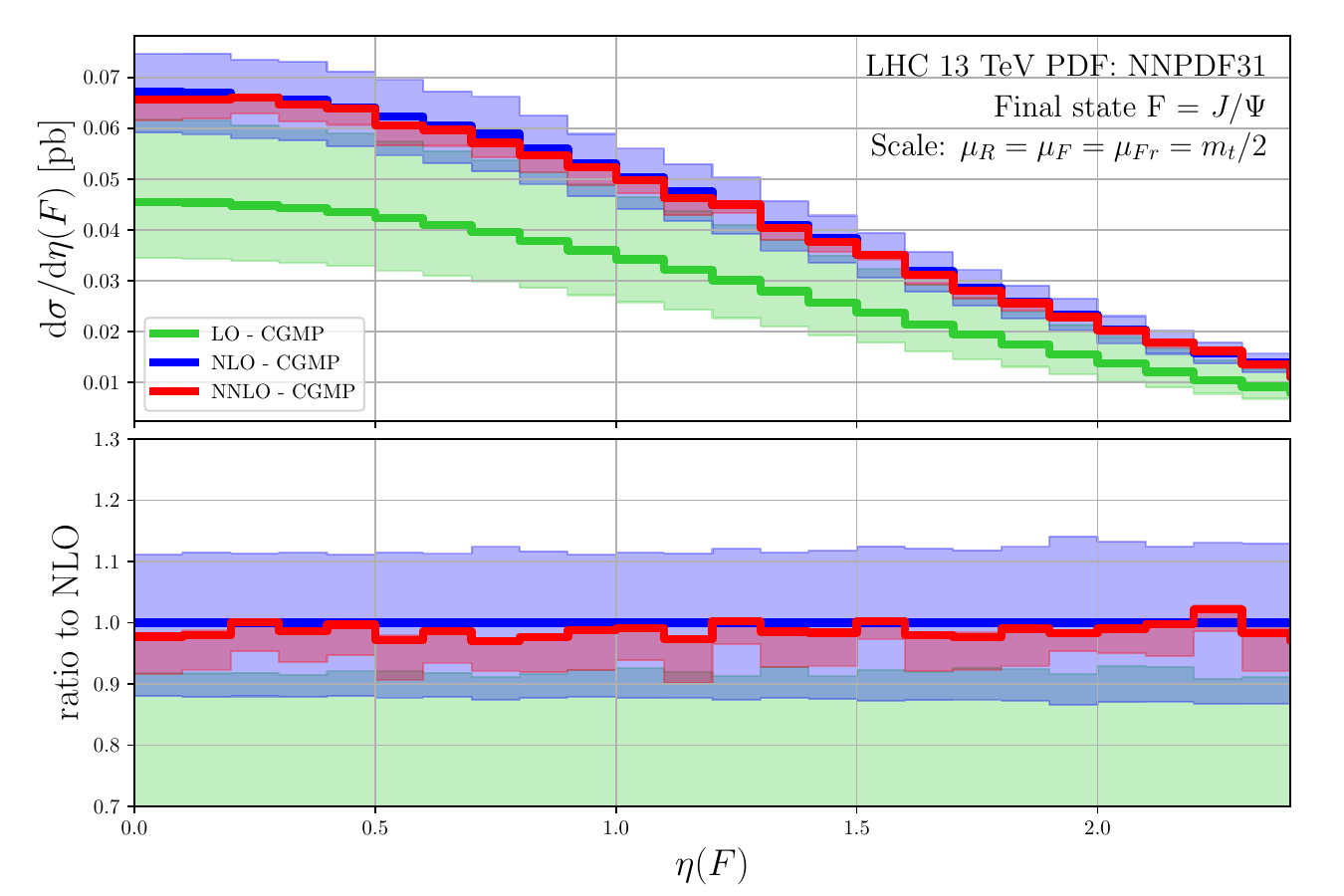}
\includegraphics[width=0.49\textwidth]{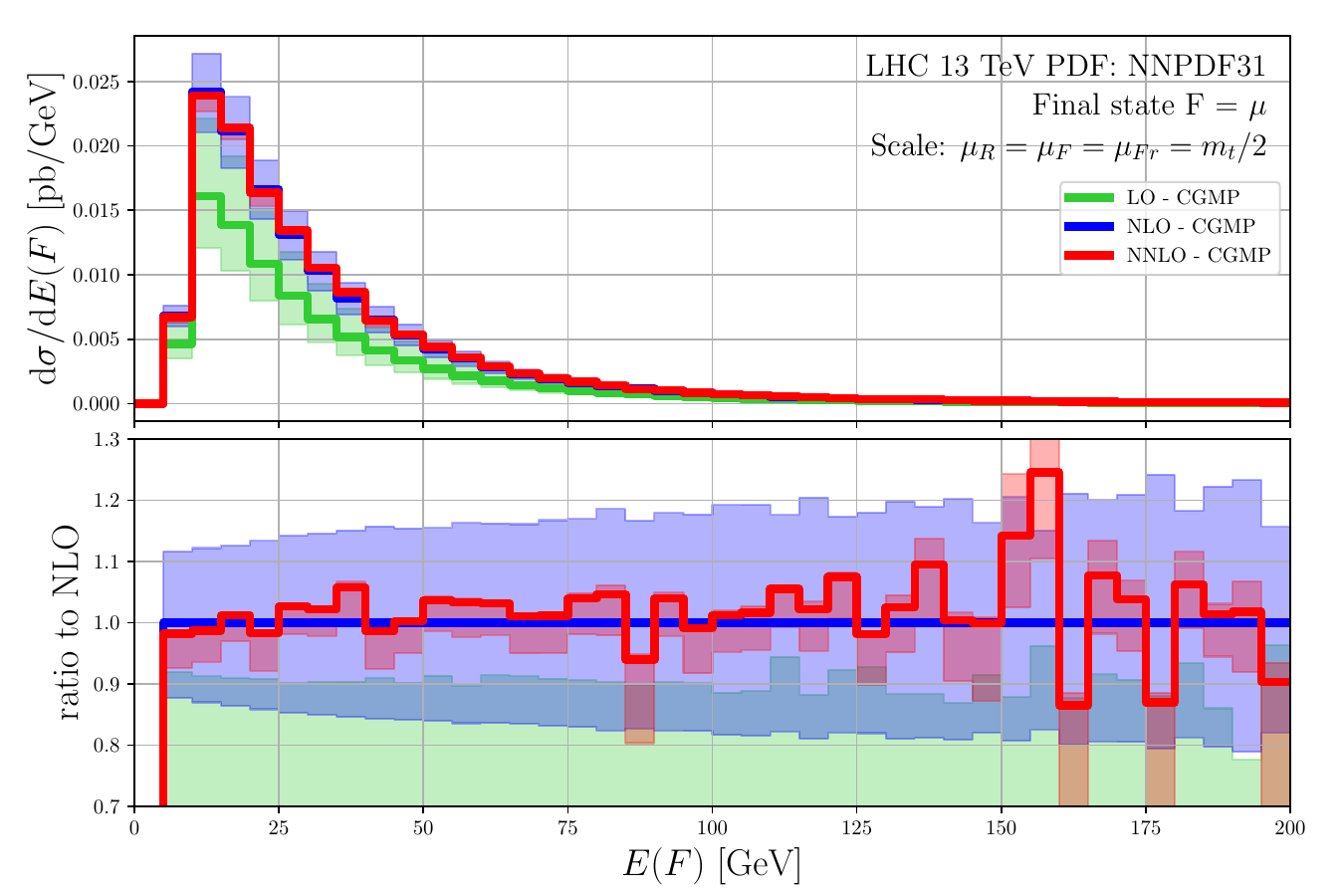}
\includegraphics[width=0.49\textwidth]{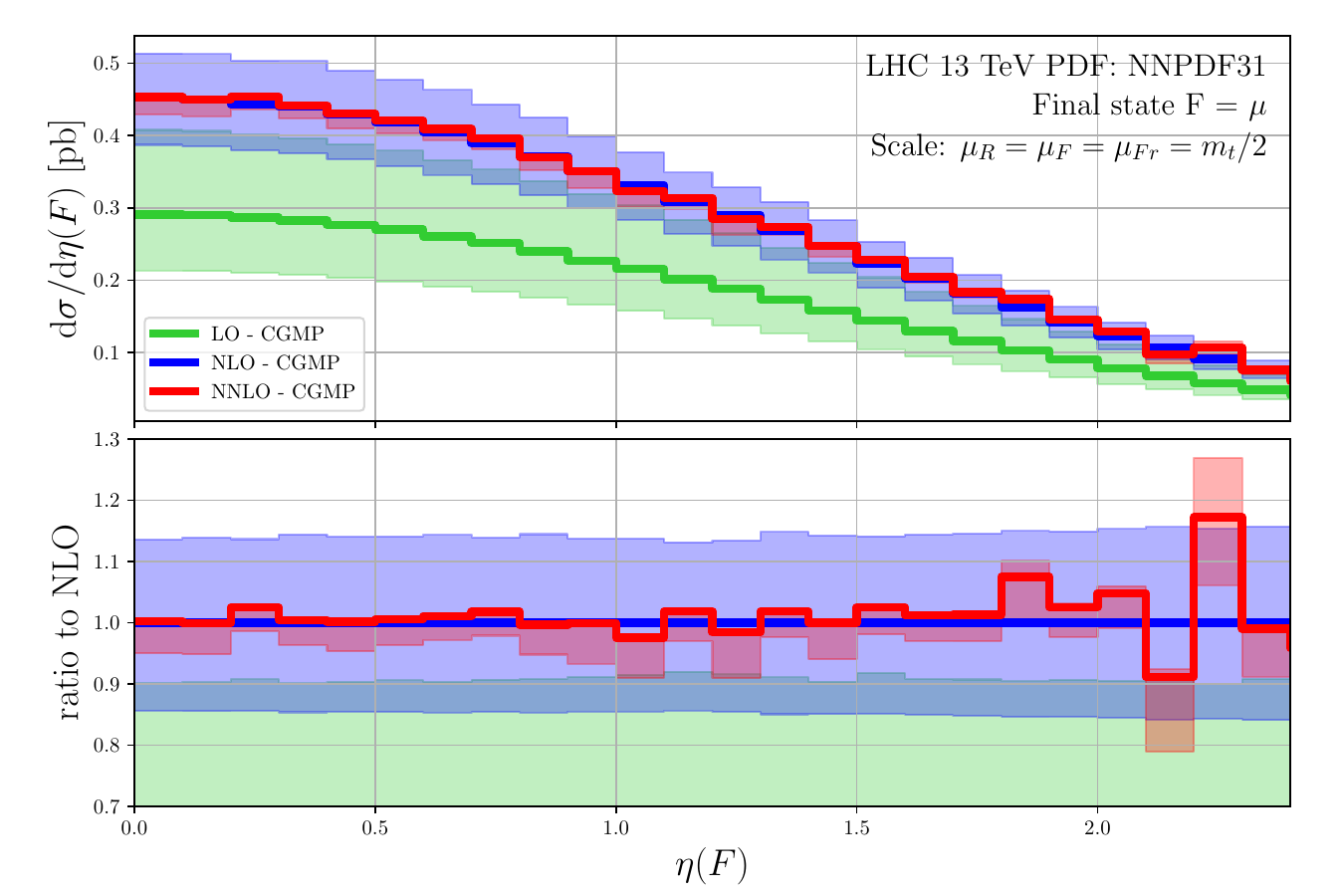}
\caption{The absolute $E(F)$ (left column) and pseudorapidity $|\eta(F)|$ (right column) distributions for $F=B$ (top), $F=J/\psi$ (middle) and $F=\mu$ (bottom). The prediction uses the fixed scale (\ref{eq:scale-mt}). Shown are the 15 point scale variation bands for LO, NLO and NNLO as well as the NPFF r.m.s.~uncertainty band (in yellow, shown with respect to the NLO).}
\label{fig:E_F}
\end{figure}
\begin{figure}[t]
\centering
\includegraphics[width=0.49\textwidth]{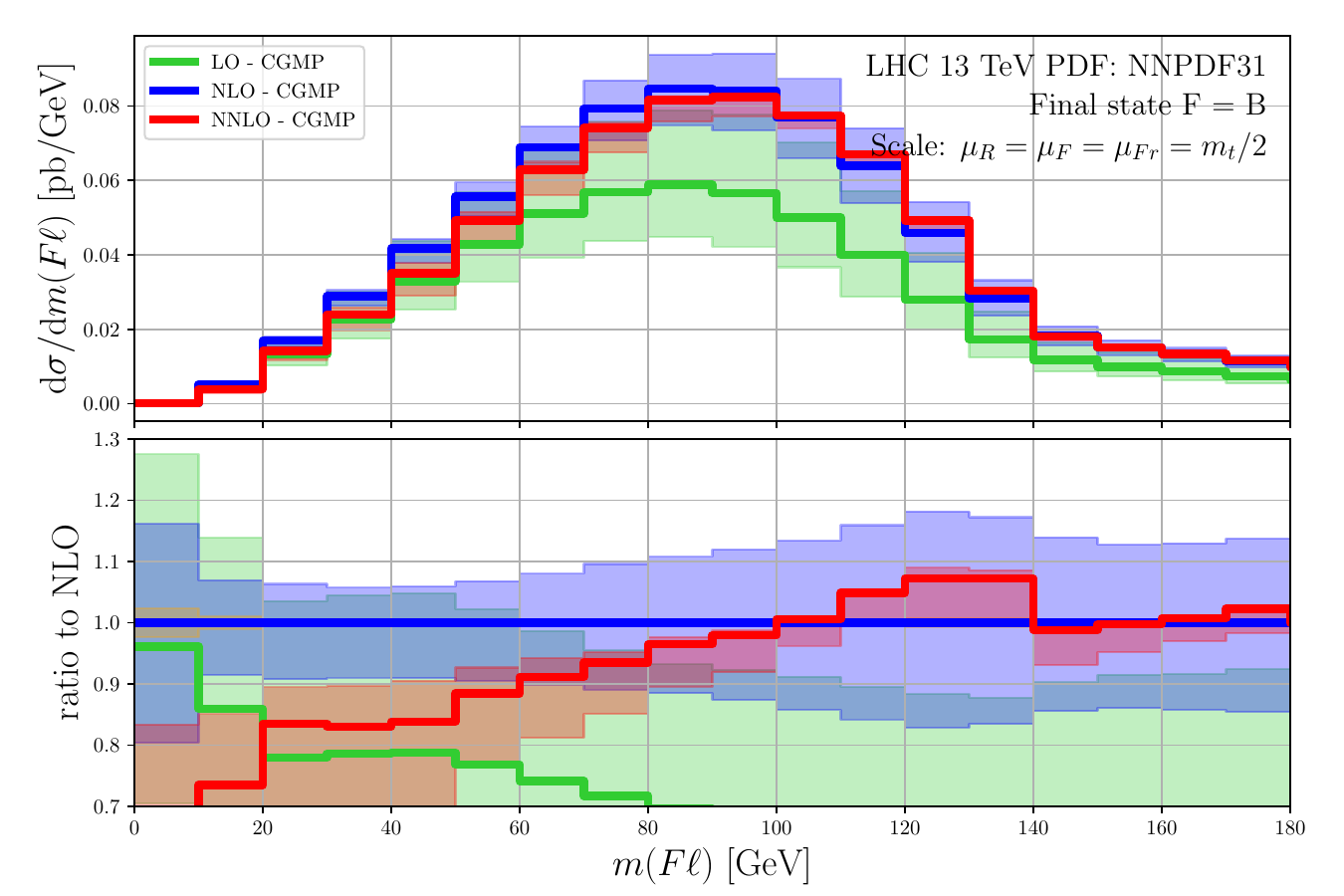}
\includegraphics[width=0.49\textwidth]{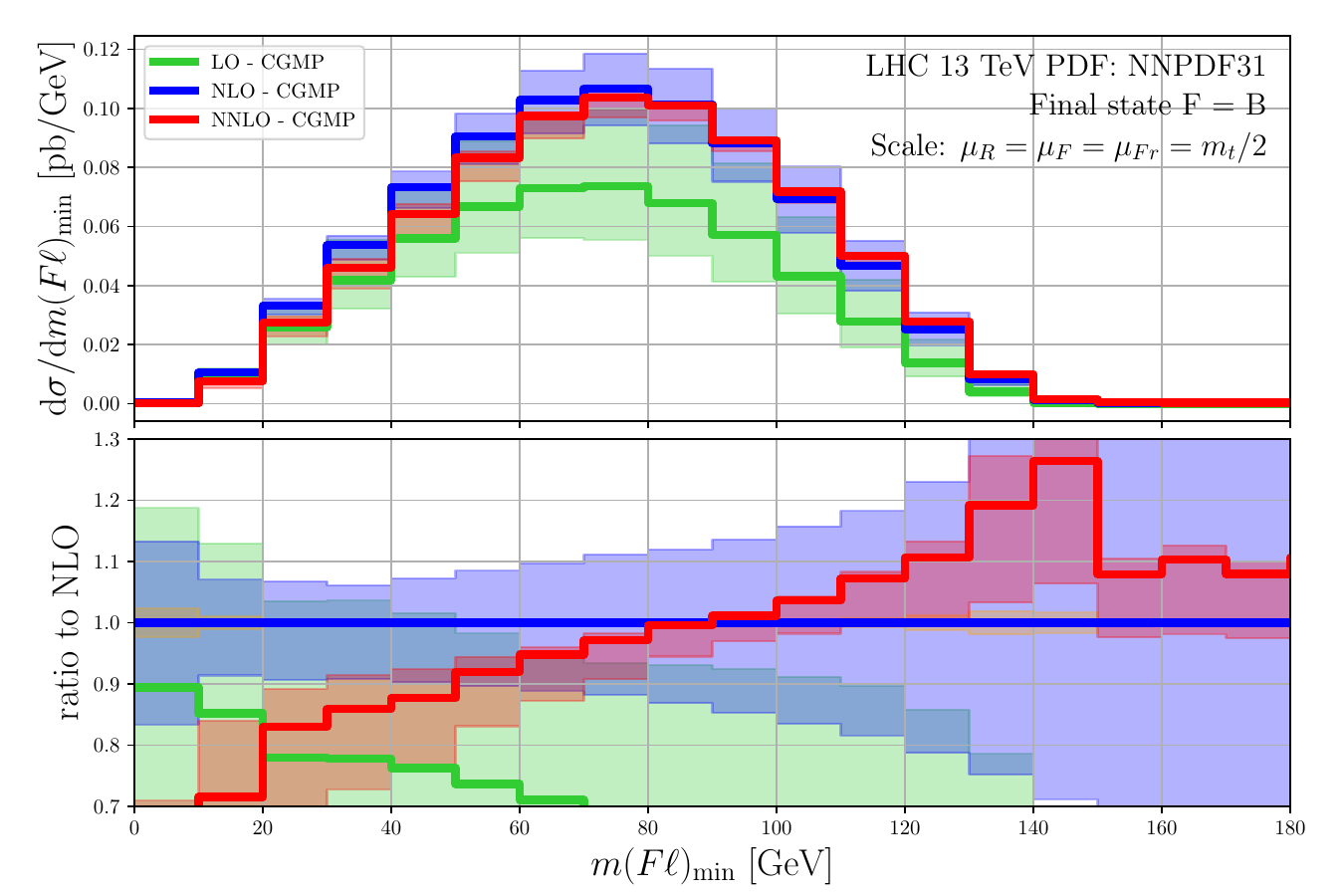}
\includegraphics[width=0.49\textwidth]{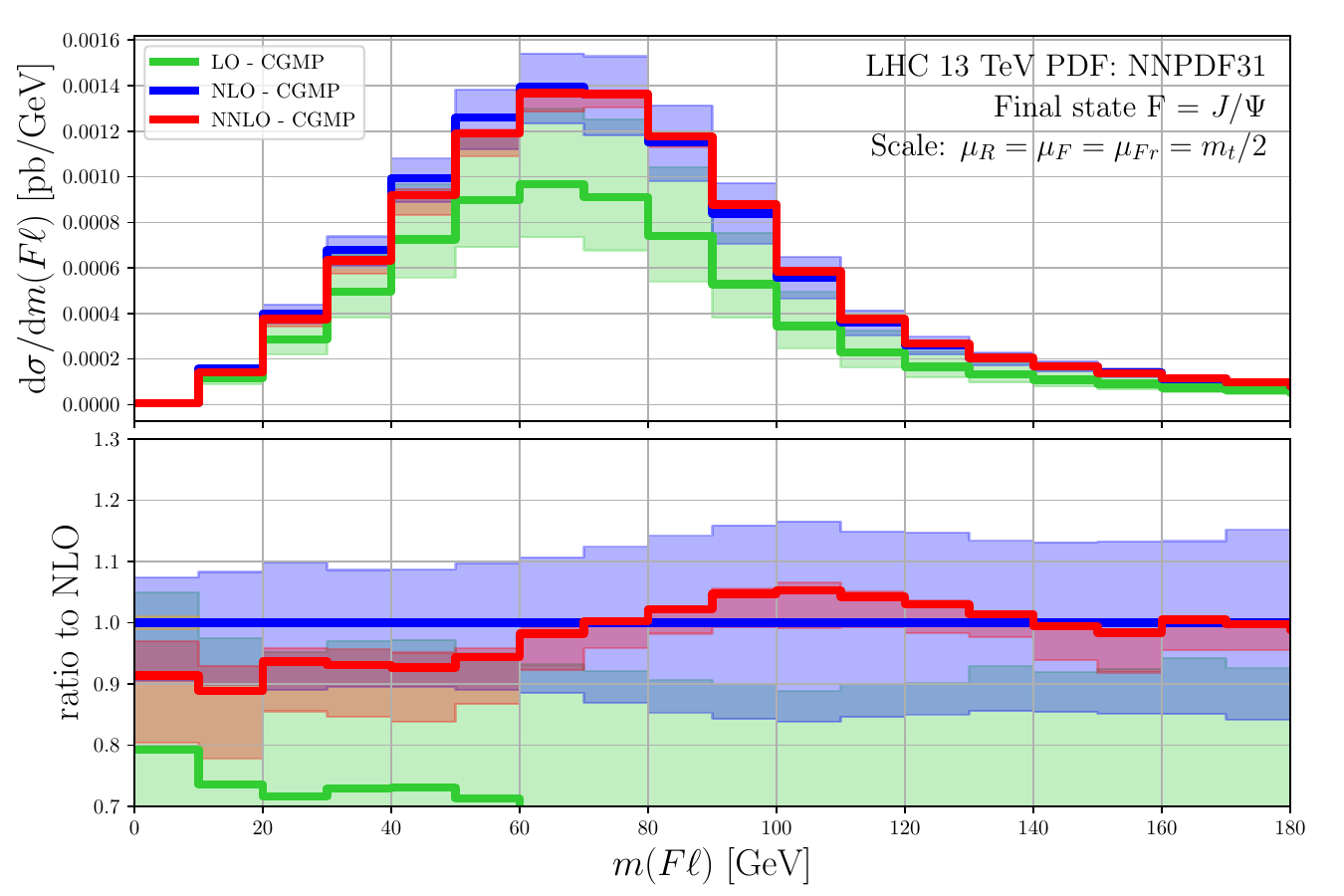}
\includegraphics[width=0.49\textwidth]{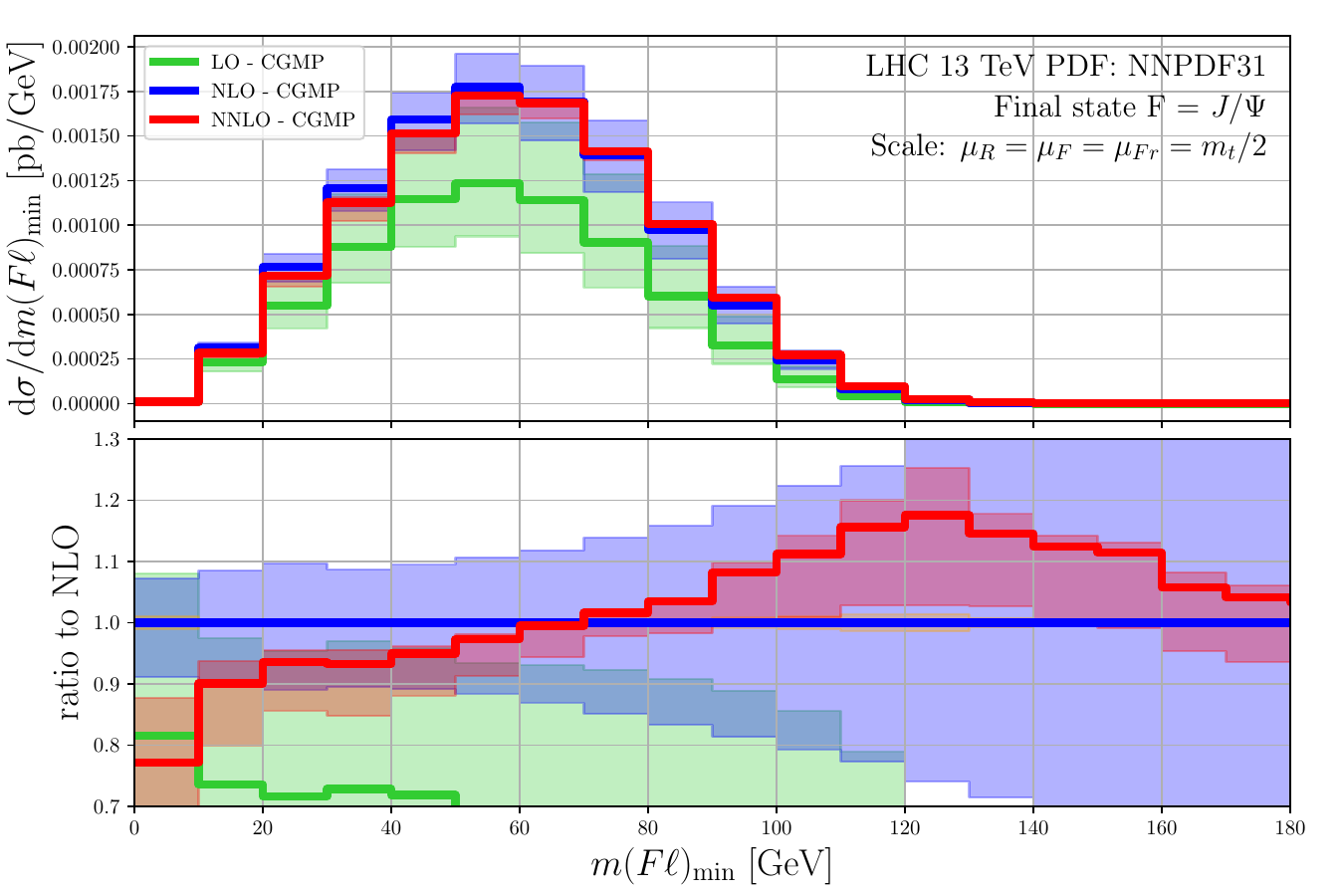}
\includegraphics[width=0.49\textwidth]{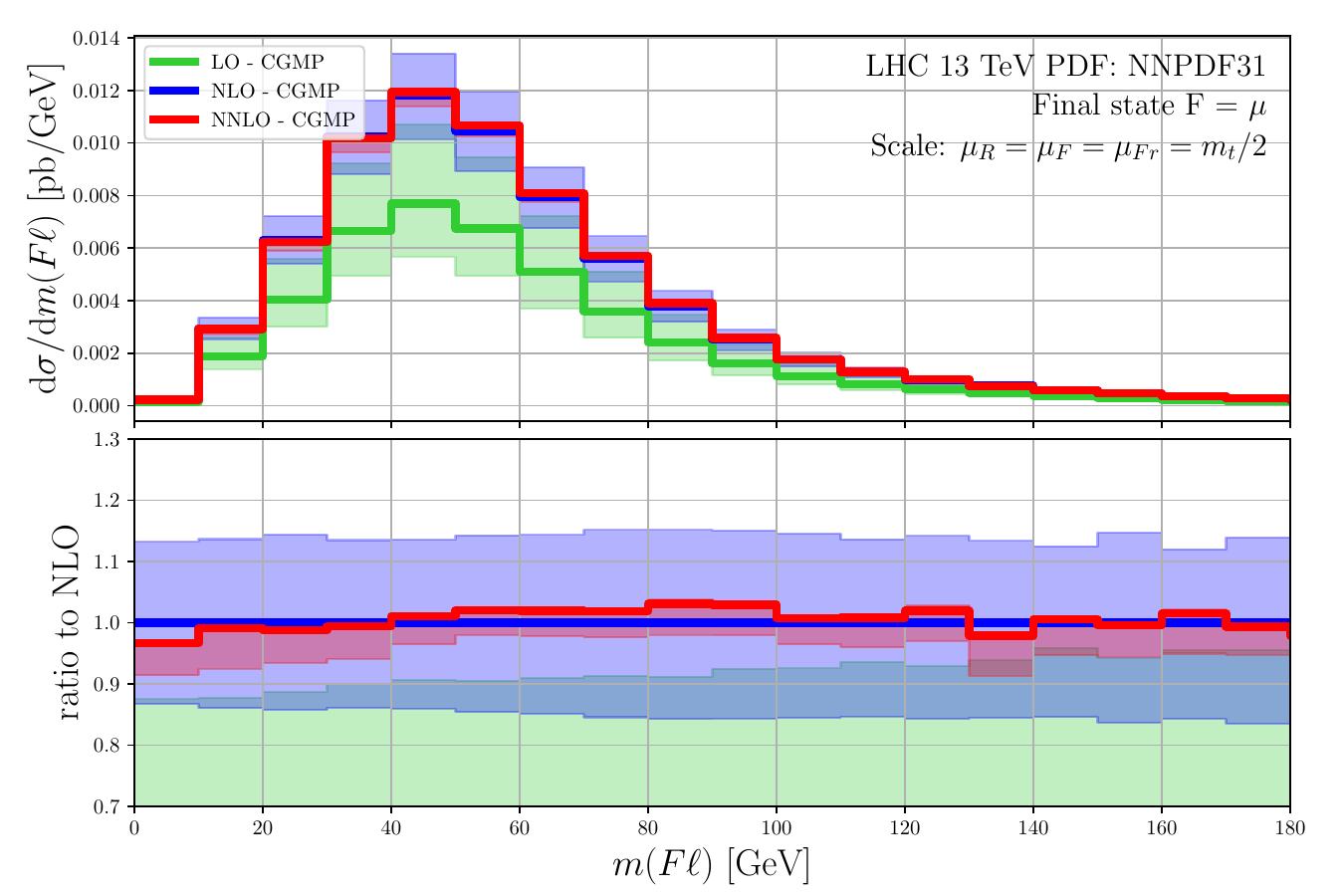}
\includegraphics[width=0.49\textwidth]{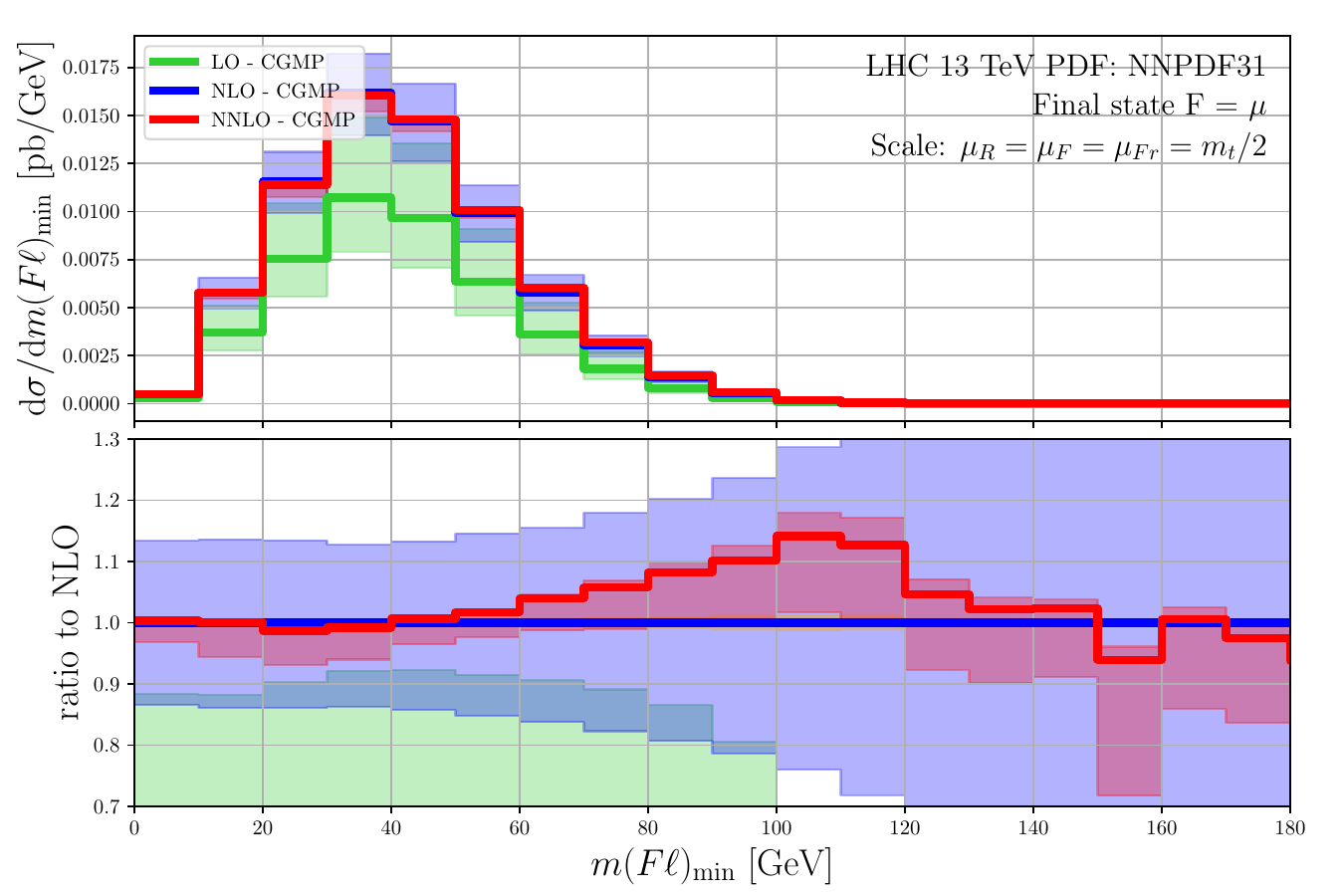}
\caption{As in fig.~\ref{fig:E_F} but for $m(F\ell^-)$ (left column) and $m(F\ell)_{\rm min}$ (right column).}
\label{fig:mFl}
\end{figure}

First we consider the $E(F)$ distribution which can be found in fig.~\ref{fig:E_F}. This distribution is of particular interest in the context of $m_t$ determination and we discuss it extensively in sec.~\ref{sec:mtop}. By comparing the $E(B)$ distribution presented in this section with the one in sec.~\ref{sec:mtop} one can immediately note the potentially large impact of selection cuts. The calculation of sec.~\ref{sec:mtop} is very inclusive, with minimal selection cuts, unlike the observable computed in this section which is strongly impacted by cuts. As can be seen in fig.~\ref{fig:E_F} the region of low $E(F)$ is not reliably described and higher order corrections are large. However, once $E(F)$ increases sufficiently (above about 50 GeV for $E(B)$, 30 GeV for $E(J/\psi)$ and 10 GeV for $E(\mu)$) the behavior of higher order corrections changes dramatically. The perturbative convergence becomes evident and the improvement due to the inclusion of NNLO corrections leads to a decrease of the theoretical uncertainty from scale variation by a factor of about 4 relative to NLO. Furthermore, in this region, the size of NNLO corrections is only few percent above NLO further illustrating the reliability of the theory prediction once NNLO QCD corrections have been included. We also note that the FF uncertainty is tiny relative to the scale uncertainty even at NNLO.

The pseudorapidity distributions $|\eta(F)|$ for the three final states are shown in fig.~\ref{fig:E_F}. As it may be expected, all three are very well behaved in the full kinematic range and show no noticeable shape difference between NLO and NNLO. The NNLO uncertainty bands are much smaller than the NLO ones and the NNLO/NLO $K$-factor is very small. We conclude that this observable is very precisely and reliably predicted at NNLO in QCD.

We next turn our attention to $m(F\ell)$, shown in fig.~\ref{fig:mFl}, which is another kinematic variable studied in the context of $m_t$ in sec.~\ref{sec:mtop}. Specifically we show $m(F\ell^-)$ and $m(F\ell)_{\rm min}$, the latter being defined by the requirement that $\ell$ is chosen in such a way that its invariant mass with $F$ is minimized. We observe a pattern of higher order corrections roughly in line with $E(F)$ discussed above. The most prominent feature is the non-overlap of NLO and NNLO uncertainty bands for $m(B\ell)$ below, roughly, 50 GeV. This feature is likely driven by the selection cuts since, as can be seen in fig.~\ref{fig:mt-mBl}, once more inclusive selection cuts are applied the normalized distribution does not exhibit significant shape change (in this region) between NLO and NNLO. Outside of this range, the NNLO prediction for $m(F\ell)$ is consistent with the NLO one. The $m(B\ell)$ and $m(J/\psi\ell)$ distributions at NNLO are affected by a shape difference, while the $m(\mu\ell)$ one is not, within the NNLO scale uncertainty. In general, the NNLO uncertainty band is much smaller than the NLO one, especially for $m(\mu\ell)$, which demonstrates the important role of NNLO QCD corrections in the precision description of this observable.

\begin{figure}[t]
\centering
\includegraphics[width=0.49\textwidth]{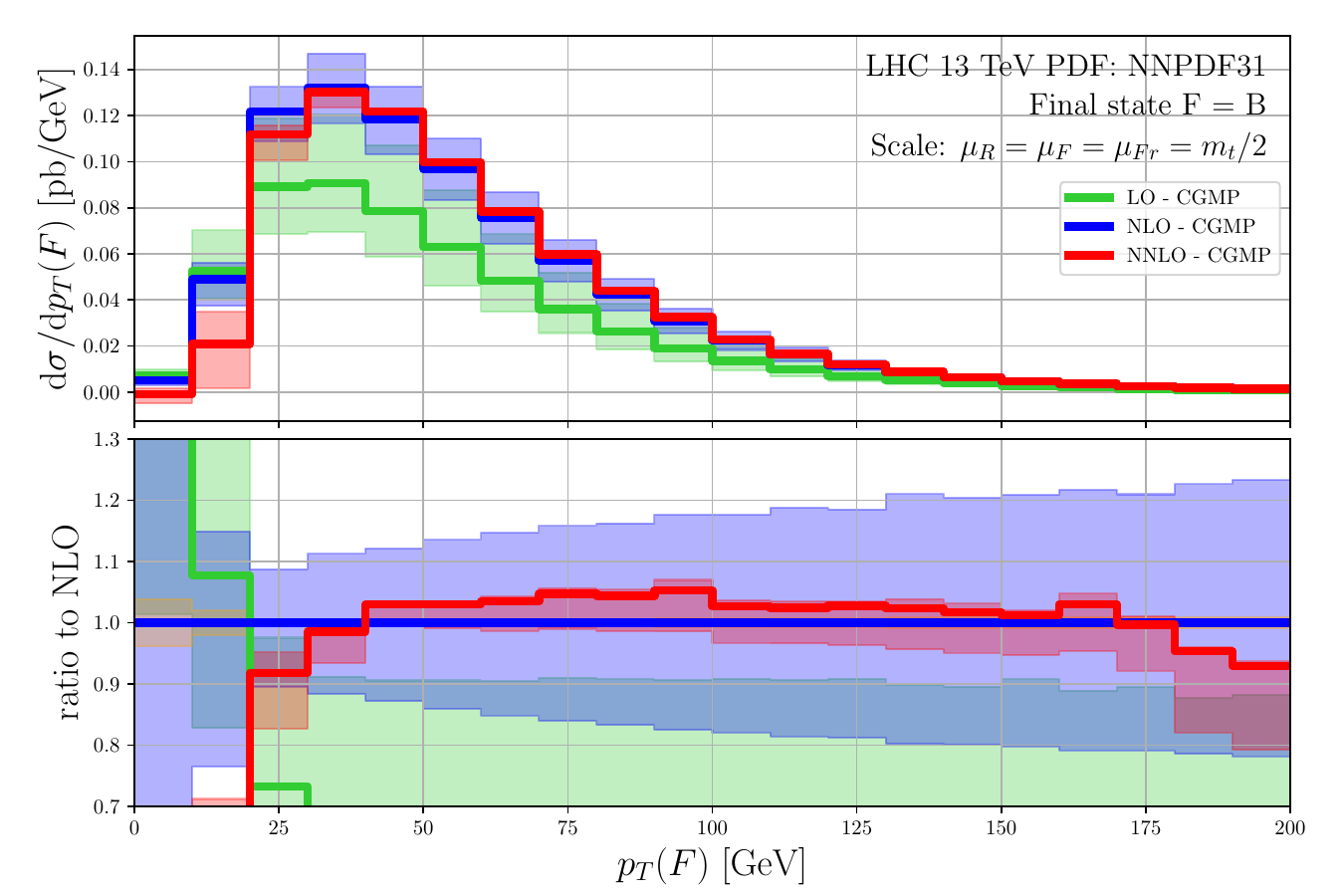}
\includegraphics[width=0.49\textwidth]{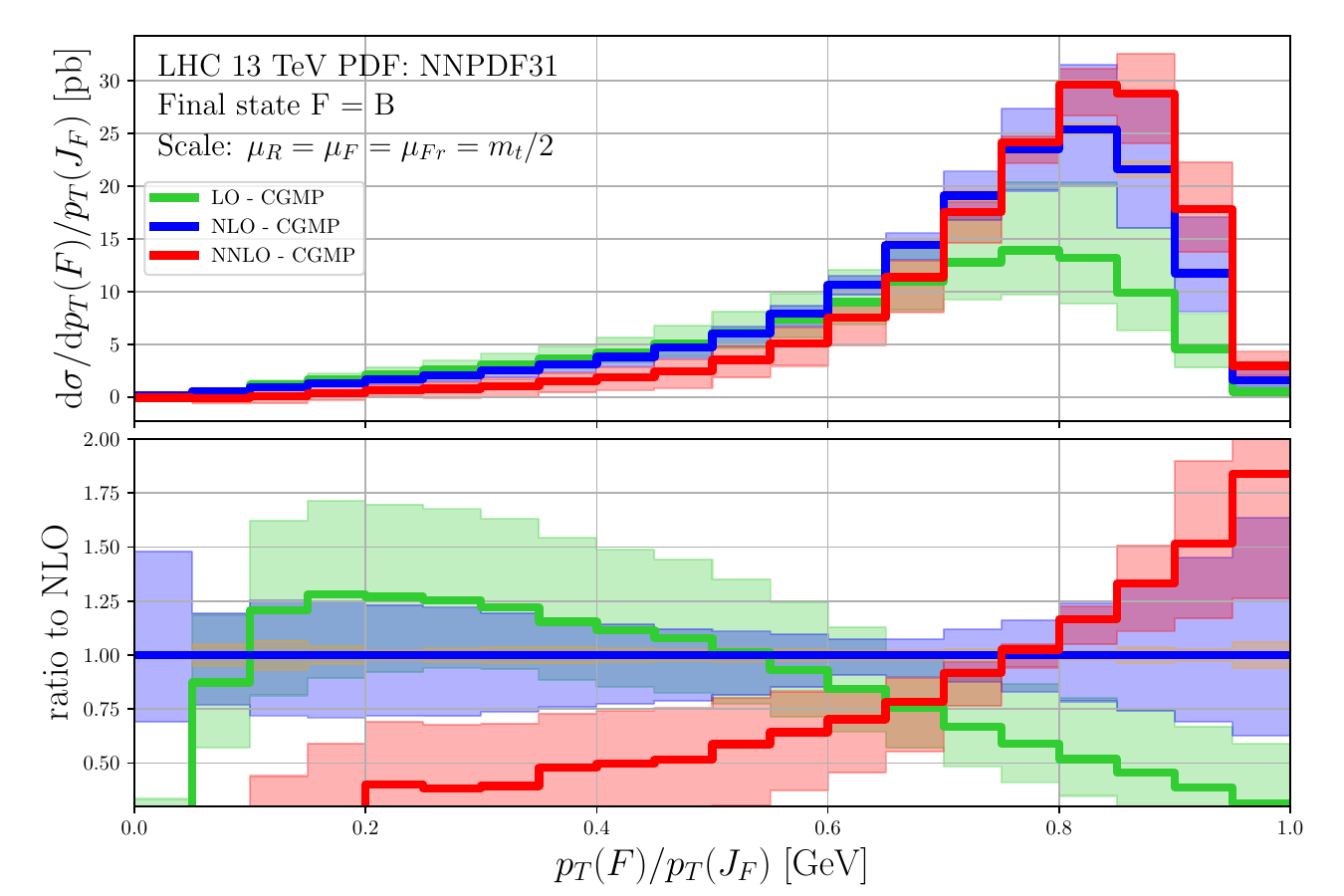}
\includegraphics[width=0.49\textwidth]{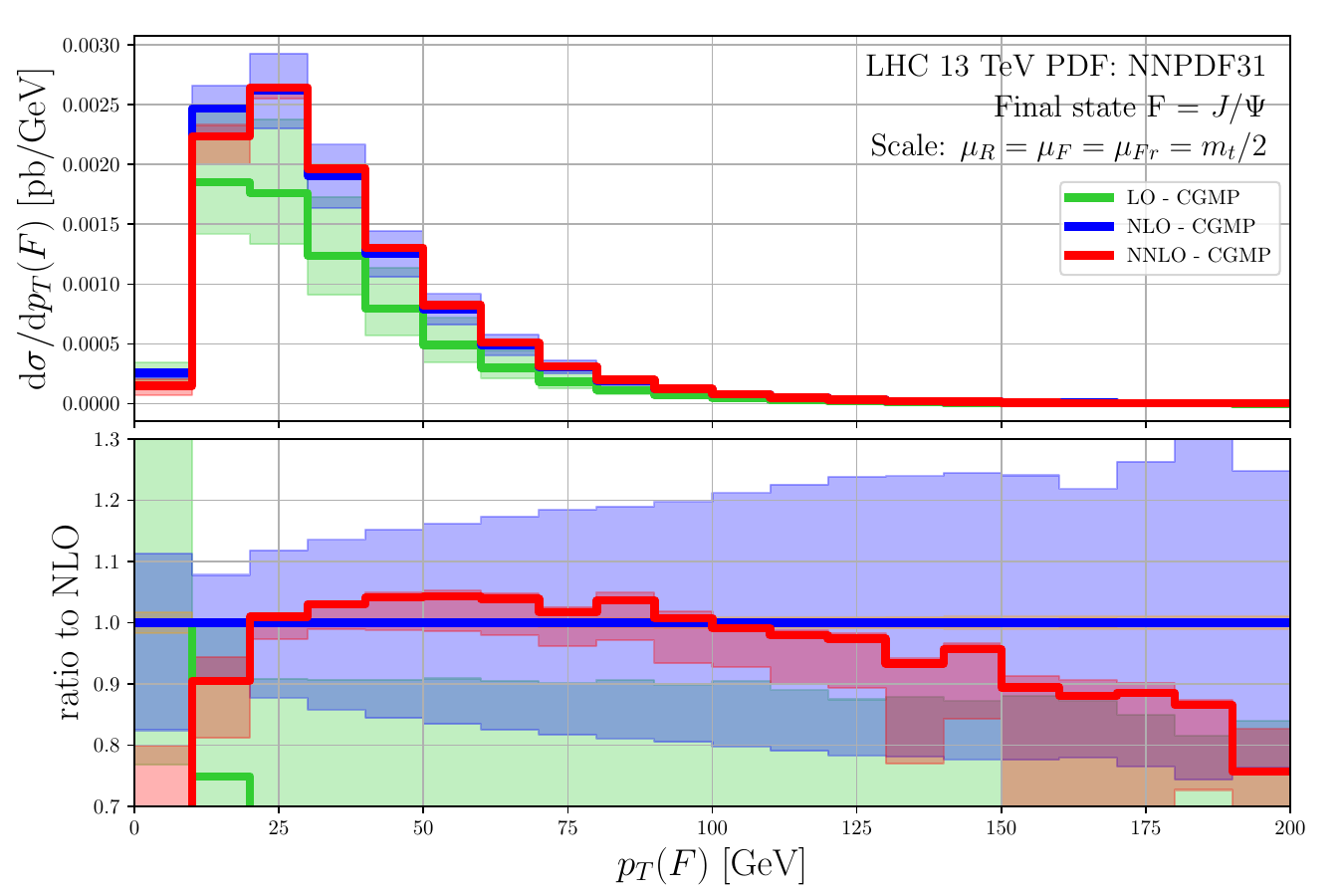}
\includegraphics[width=0.49\textwidth]{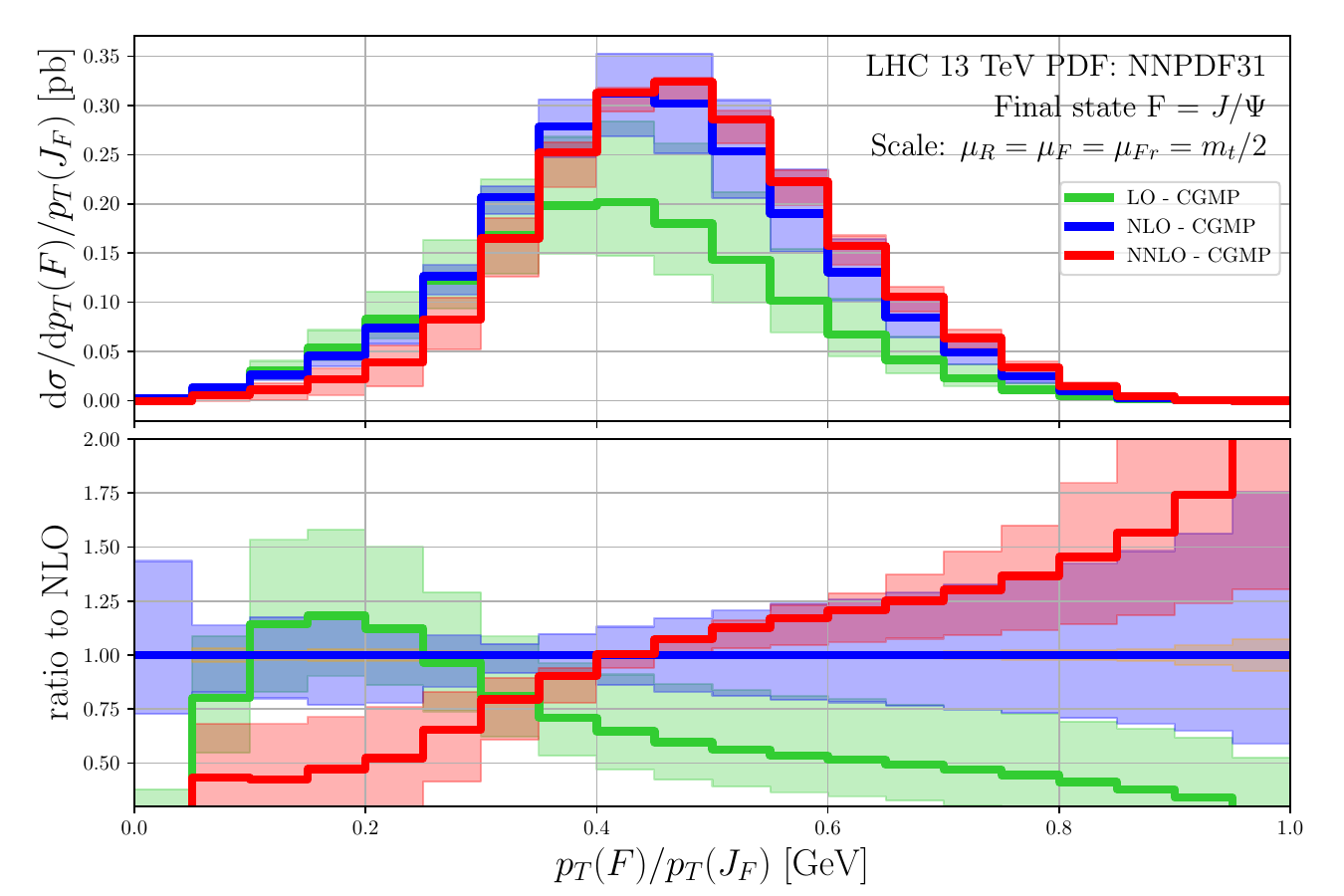}
\includegraphics[width=0.49\textwidth]{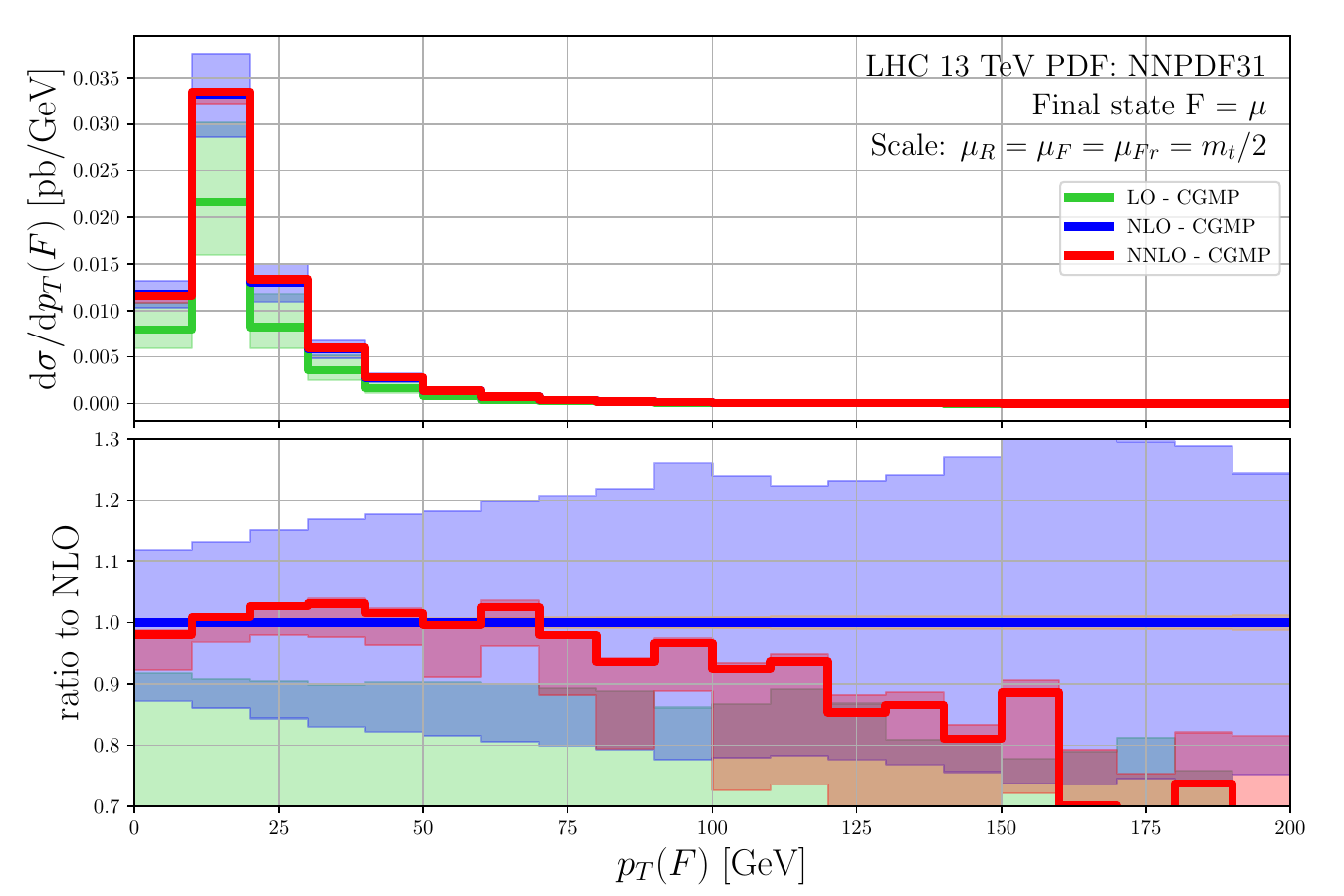}
\includegraphics[width=0.49\textwidth]{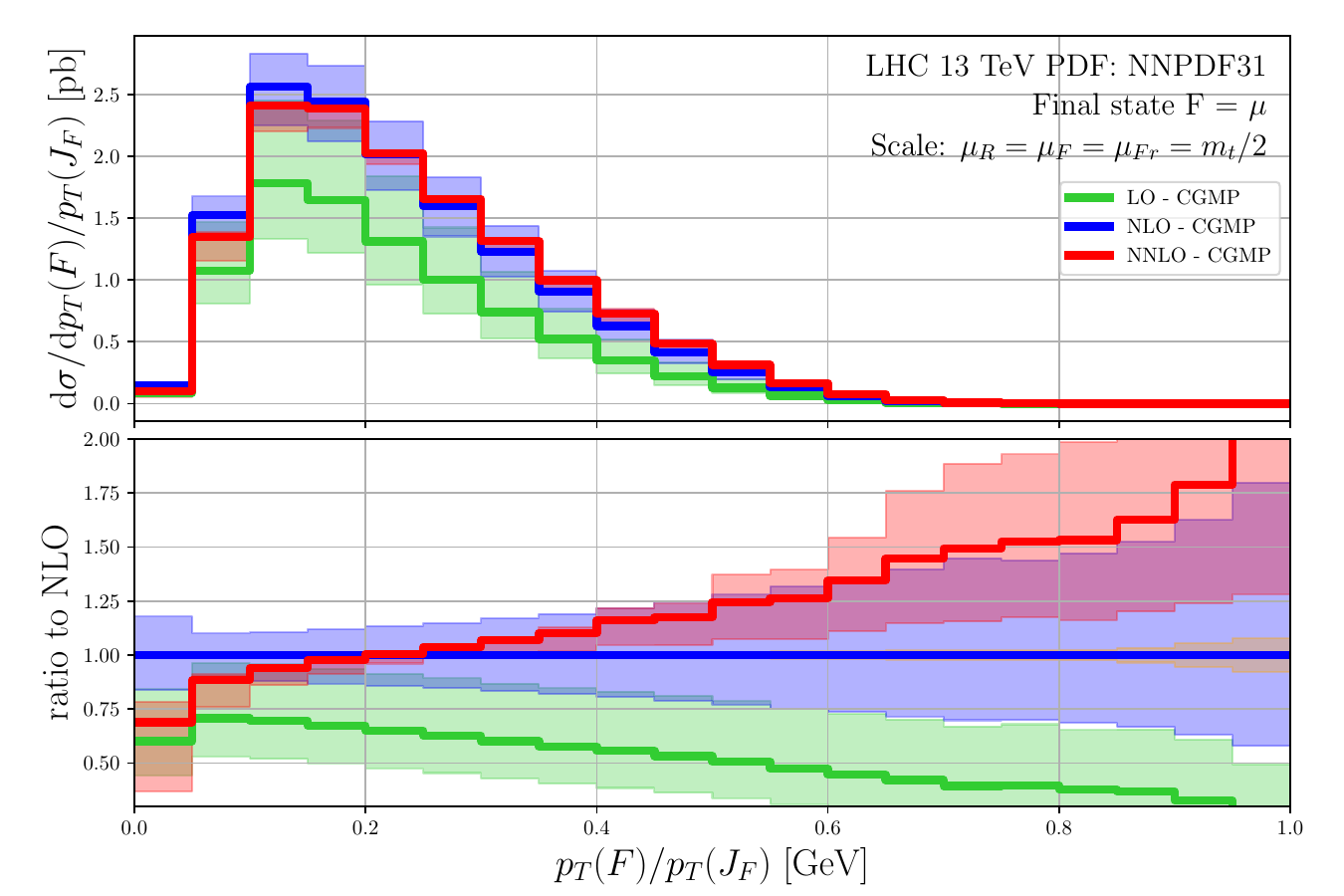}
\caption{As in fig.~\ref{fig:E_F} but for $p_T(F)$ (left column) and $p_T(F)/p_T(J_F)$ (right column).}
\label{fig:pTF}
\end{figure}

In fig.~\ref{fig:pTF} we show the predictions for $p_T(F)$. We observe that for all final states the NNLO uncertainty band is much smaller than the NLO one. The pattern of higher order corrections is, however, different for the different final states. Excluding the region of small $p_T(B)$ which is dominated by missing $m_b$ corrections and selection cuts, the $p_T(B)$ spectrum shows remarkable stability and lack of shape correction up to very high $p_T$. The lighter final states, $J/\psi$ and especially $\mu$, on the other hand show consistent change in shape between NLO and NNLO within the same $p_T$ interval. This difference between $B$, $J/\psi$ and $\mu$ is consistent with the change in the mean $z$ of the FFs for the three final states. The NLO and NNLO bands overlap in the full kinematic range considered here. 

Finally, in fig.~\ref{fig:pTF} we show the predictions for $p_T(F)/p_T(J_F)$ for all three final states. The importance of this observable is related to the fact it is sensitive to the FF, and that it can be used to extract NPFF from LHC data. It has already been extensively discussed in ref.~\cite{Czakon:2021ohs} and we refer the reader to that reference for further details.

Before concluding this section we would like to make a comment about the size of the MC uncertainty of the NNLO predictions. Overall it is small; appreciable fluctuations can only be observed in the $E(F)$ distribution and in the tails of several $\mu$ distributions. As it turns out, in order to achieve high quality predictions for all three final states $F=\{B,J/\psi,\mu\}$ considered here, one needs to use dedicated calculations with dedicated optimization for each final state. The reason for this is that the fragmentation functions for each final state is different and as a result different kinematical ranges are sampled.

\subsection{Top quark mass extraction}\label{sec:mtop}

In the following we consider the observables $E(F)$ and $m(B\ell)$ due to their sensitivity to $m_t$. We also compute the first moment of $m(B\ell)$ as a function of $m_t$. This is the first time this important observable has been derived with NNLO precision. All predictions shown in this section are subject to the following event selection requirements
\begin{itemize}
    \item $p_T(\ell) > 25$ GeV, $|\eta(\ell)| < 2.5$\,,
    \item $p_T(F) > 8$ GeV and $|\eta(F)| < 2.5$\,.
\end{itemize}
Unlike the selection requirements in sec.~\ref{sec:diff-dist}, here we do not impose any jet selection requirements.

\begin{figure}[t]
\centering
\includegraphics[width=0.49\textwidth]{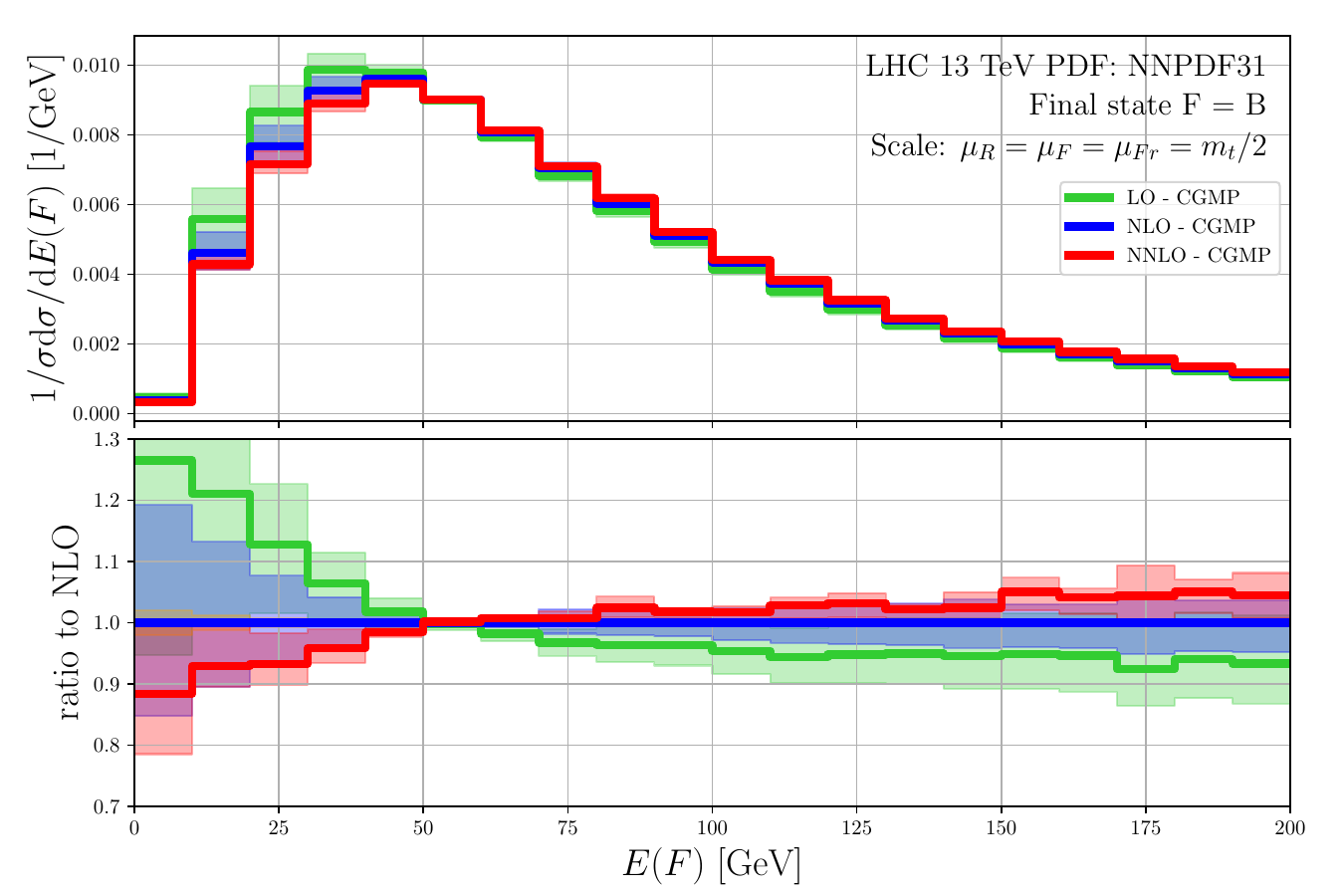}
\includegraphics[width=0.49\textwidth]{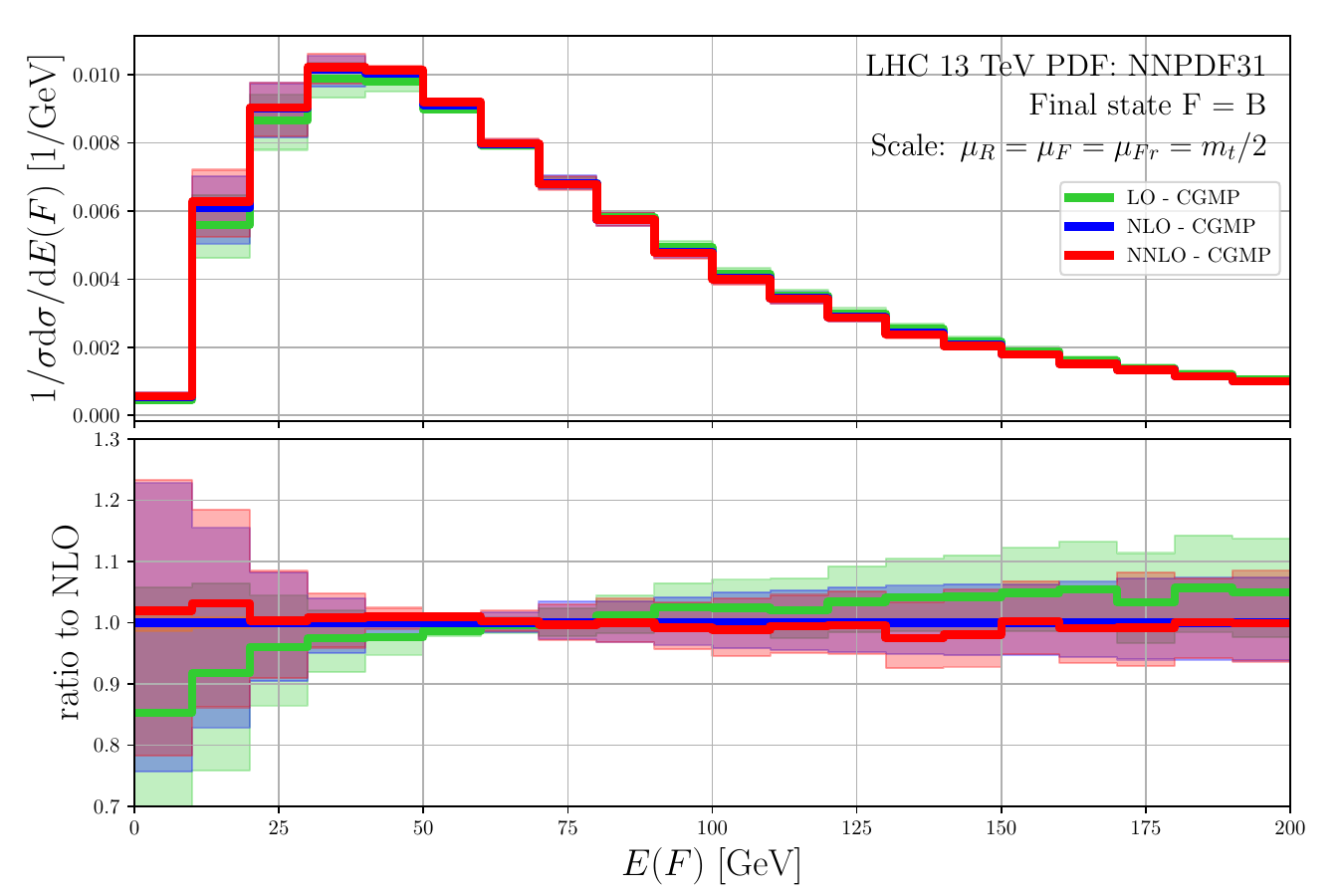}
\caption{The normalized $E(B)$ distribution for fixed scale (\ref{eq:scale-mt}). Shown are the 15 point scale variation bands for LO, NLO and NNLO as well as the NPFF r.m.s.~uncertainty band. The plot to the right is like the one to the left but with LO top decay.}
\label{fig:mt-EB}
\end{figure}

We first consider the $E(B)$ distribution. It is shown in fig.~\ref{fig:mt-EB} where we compare predictions with LO, NLO and NNLO accuracy. For reasons to be explained shortly, besides the predictions with full top quark decay (fig.~\ref{fig:mt-EB}, left), we also present predictions with top decay kept at LO (fig.~\ref{fig:mt-EB}, right). 

The $E(B)$ distribution has been proposed \cite{Agashe:2012bn,Agashe:2016bok} as a means of measuring the top quark mass relatively free of contamination due to possible physics beyond the Standard Model (BSM). The main idea is that the position of the maximum of the $E(B)$ distribution is independent of the $t\bar t$ dynamics. In particular, it is not affected by potential BSM contributions in $t\bar t$ production. Therefore, if one calculates the dependence of the position of the maximum as a function of $m_t$, one can infer the value of $m_t$ from a direct measurement of $E_{\rm max}(B)$. In practice, for values of $m_t$ close to the world average, it is sufficient to consider a linear fit 
\begin{equation}
E_{\rm max}(m_t) = a m_t + b\,.
\label{eq:E_max_fit}
\end{equation}
Our predictions for the coefficients $(a,b)$ at LO, NLO and NNLO can be found in table~\ref{tab:E_max_fit-full-dec} (with full top quark decay) and in table~\ref{tab:E_max_fit-LO-dec} (for LO top quark decay). The uncertainties on the fit parameters are explained in the following.

An important feature of the method of refs.~\cite{Agashe:2012bn,Agashe:2016bok} is that the top quark decay is a two-body one, i.e.~LO-like. Once one includes higher order QCD corrections the top decay is modified by an additional radiation which alters the top decay kinematics. This radiation affects the position of the peak of the $E(B)$ distribution and needs to be accounted for in any realistic application of this method. The NLO QCD corrections were first included in ref.~\cite{Agashe:2016bok} where a significant shift in the peak position was observed. In this work we also extend, for the first time, this observable to NNLO in QCD and observe that the NNLO QCD corrections lead to a further shift in the position of $E_{\rm max}(B)$. As can be seen in table~\ref{tab:E_max_fit-full-dec}, the inclusion of NLO corrections in top decay leads to a roughly 4 GeV shift in $E_{\rm max}(B)$ while the NNLO ones lead to a further 2 GeV shift. Clearly, the inclusion of the NNLO QCD corrections is mandatory for a reliable description of this observable. 

As a calibration of the method, in table~\ref{tab:E_max_fit-LO-dec} we have shown the same predictions but where the top quark decay has consistently been kept at LO. As can be seen there, within the uncertainty of the calculation, the position of the maximum remains stable through NNLO, as expected. 

Finally, we would like to mention that while we only show predictions for the final state with a $B$-hadron, similar predictions can also be made for final states with a $J/\psi$ or a muon.

\begin{table}
\begin{tabular}{c|c|c|c}
$m_t$ & LO & NLO & NNLO \\
\hline
$171.5$ GeV & $ 37.553~ (\pm 0.106)~ (^{+0.050}_{-0.061})$ & $ 40.994~ (\pm 0.147)~ (^{+1.178}_{-0.710})$ & $ 42.957~ (\pm 0.329)~ (^{+1.087}_{-0.818})$ \\
$172.5$ GeV & $ 37.816~ (\pm 0.109)~(^{+0.051}_{-0.062})$ & $ 41.277~ (\pm 0.158)~ (^{+1.196}_{-0.717})$ & $ 43.263~ (\pm 0.332)~ (^{+1.073}_{-0.825})$ \\
$173.5$ GeV & $ 38.093~ (\pm 0.113)~ (^{+0.051}_{-0.061})$ & $ 41.657~ (\pm 0.168)~ (^{+1.250}_{-0.745})$ & $ 43.528~ (\pm 0.222)~ (^{+1.010}_{-0.778})$ \\
\hline
Lin.~fit & LO & NLO & NNLO \\
\hline
$a = $ & $ 0.270~ (\pm 0.004)$& $ 0.329~ (\pm 0.028)$& $ 0.284~ (\pm 0.011)$\\
$b = $ & $ -8.755~ (\pm 0.708)$ GeV& $ -15.429~ (\pm 4.820)$ GeV& $ -5.666~ (\pm 1.816)$ GeV
\end{tabular}
\caption{Values of $E_{\rm max}(B)$ for the absolute differential cross section with full top quark decay at LO, NLO and NNLO and for three different values of $m_t$. Positions are fit using eq.~(\ref{eq:chi2-fit}) and 5 GeV bins. Also given are the parameters of the linear fit eq.~(\ref{eq:E_max_fit}) at LO, NLO and NNLO.}
\label{tab:E_max_fit-full-dec}
\end{table}
\begin{table}
\begin{tabular}{c|c|c|c}
$m_t$ & LO & NLO & NNLO \\
\hline
$171.5$ GeV & $ 37.553~ (\pm 0.106)~ (^{+0.050}_{-0.061})$ & $ 36.744~ (\pm 0.169)~ (^{+0.213}_{-0.313})$ & $ 36.737~ (\pm 0.311)~ (^{+0.081}_{-0.021})$ \\
$172.5$ GeV & $ 37.816~ (\pm 0.109)~ (^{+0.051}_{-0.062})$ & $ 36.981~ (\pm 0.182)~ (^{+0.223}_{-0.330})$ & $ 37.010~ (\pm 0.227)~ (^{+0.109}_{-0.019})$ \\
$173.5$ GeV & $ 38.093~ (\pm 0.113)~ (^{+0.051}_{-0.061})$ & $ 37.319~ (\pm 0.193)~ (^{+0.206}_{-0.296})$ & $ 37.292~ (\pm 0.255)~ (^{+0.113}_{-0.056})$ \\
\hline
Lin.~fit & LO & NLO & NNLO \\
\hline
$a = $ & $ 0.270~ (\pm 0.004)$& $ 0.286~ (\pm 0.029)$& $ 0.278~ (\pm 0.003)$\\
$b = $ & $ -8.755~ (\pm 0.708)$ GeV& $ -12.237~ (\pm 4.962)$ GeV& $ -10.913~ (\pm 0.556)$ GeV
\end{tabular}
\caption{As in table~\ref{tab:E_max_fit-full-dec} but for LO top quark decay.}
\label{tab:E_max_fit-LO-dec}
\end{table}

Before we conclude our discussion of the $E(B)$ distribution we would like to explain how the position of the maximum $E_{\rm max}(B)$ has been determined. This is a non-trivial task since our calculation produces not the differential spectrum but its binned version, i.e.~the position of the maximum is obscured by finite bin size effects. To this end we extract the position of the maximum with the help of a fit 
\begin{equation}
f(E) = n \exp\left( - w \left(\frac{\hat{E}}{E} + \frac{E}{\hat{E}}\right)\right)\,,
\label{eq:fit}
\end{equation}
performed at each perturbative order and for each scale combination. The fit parameters are $n, w$ and $\hat{E}$, the latter giving the curve's maximum position. The functional form in eq.~(\ref{eq:fit}) is the same one used in ref.~\cite{Agashe:2016bok}. We fit the range $E(B) \in [25,100]$ GeV using 5 GeV wide bins. The $\chi^2$ for the fit (\ref{eq:fit}) is computed as
\begin{equation}
\chi^2 = \sum_{i \in \text{Bins}} \left(f(E(B)_{i}) - \frac{\dd \sigma}{\dd E}\Big\vert_{E=E(B)_{i}}\right)^2 \left( \frac{\dd \sigma^{\rm MC\,err}}{\dd E}\Big\vert_{E=E(B)_{i}}\right)^{-2}\,,
\label{eq:chi2-fit}
\end{equation}
where $E(B)_{i}$ is the middle of each $E(B)$ bin and $\sigma^{\rm MC\,err}$ indicates that only the MC uncertainty of the calculation in each bin is used. As usual, the uncertainty on
the fit parameters is given by the interval defined by $\Delta\chi^2=1$. The uncertainties in tables~\ref{tab:E_max_fit-full-dec} and \ref{tab:E_max_fit-LO-dec} are derived as follows. The first one refers to the uncertainty due to the parameter $\hat{E}$, while the second bracket gives an estimate of the scale uncertainty which is derived by repeating the fit for each scale choice and taking the envelope of all 15 values. The uncertainties of the $a$ and $b$ parameters are the ones from the linear fit to the central scale values. We have checked that fits to absolute or normalized distributions produce the same peak position, as expected.

\begin{figure}[t]
\centering
\includegraphics[width=0.49\textwidth]{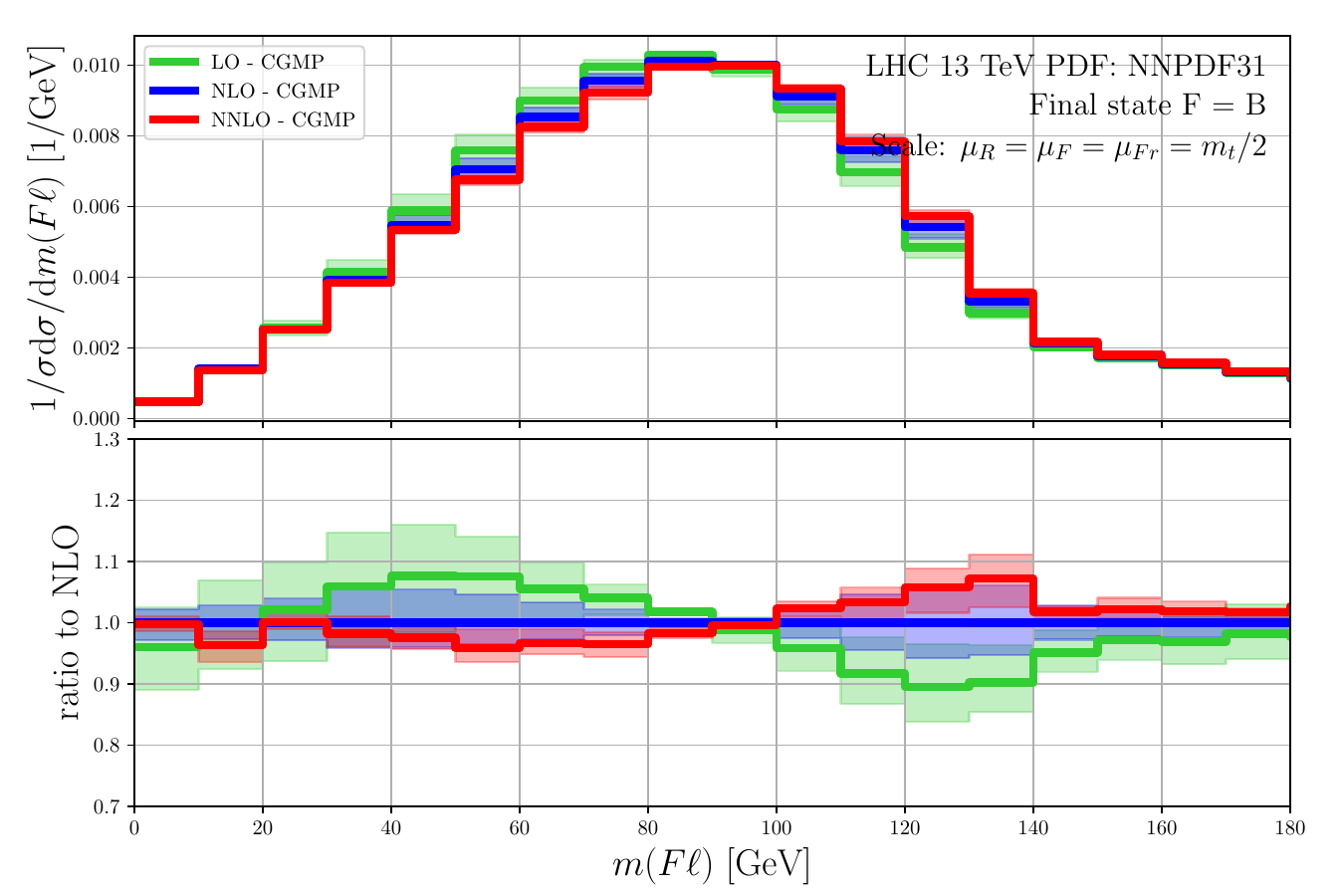}
\includegraphics[width=0.49\textwidth]{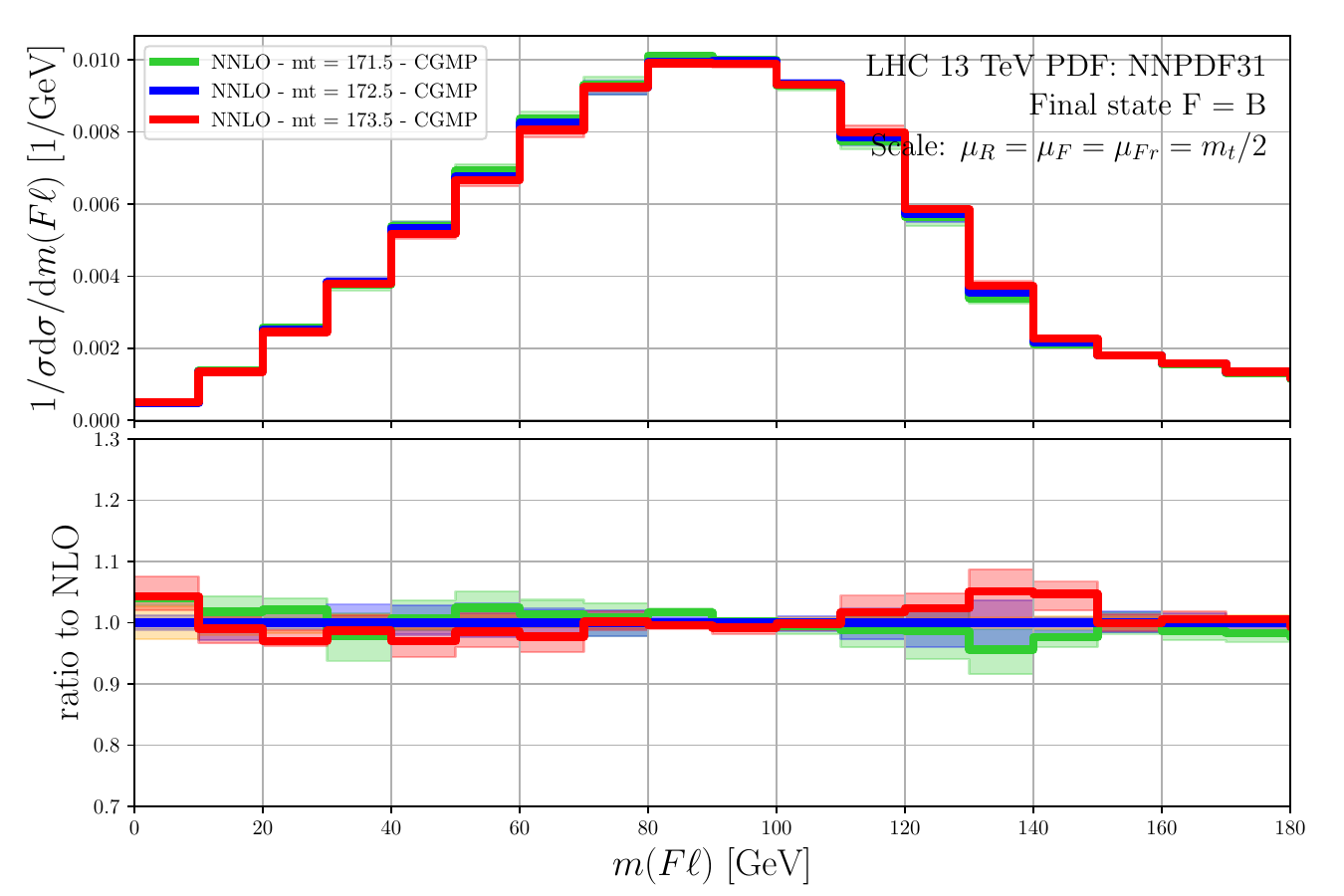}
\caption{(Left): the normalized $m(B\ell^-)$ distribution for fixed scale (\ref{eq:scale-mt}). Shown are the 15 point scale variation bands for LO, NLO and NNLO as well as the NPFF r.m.s.~uncertainty band. (Right): the NNLO prediction for 3 different values of $m_t$.}
\label{fig:mt-mBl}
\end{figure}

We next consider the $m(B\ell^-)$ distribution. We have checked that the $m(B\ell^+)$ distribution is equivalent to it within the MC error of the calculation. The prediction for the normalized $m(B\ell^-)$ distribution can be found in fig.~\ref{fig:mt-mBl}. We only show the predictions for final states with a $B$-hadron although  predictions for a $J/\psi$ or a muon can be provided. We observe the much improved perturbative description at higher orders. We also note that 1 GeV shift in $m_t$ has almost identical effect on the normalized distribution as the inclusion of the NNLO correction relative to the NLO one. This demonstrates the tight interplay between the inclusion of higher order corrections to the so-called indirect top quark mass observables and the ultimate precision of the extracted $m_t$. 

\begin{figure}[t]
	\centering
	\includegraphics[width=0.49\textwidth]{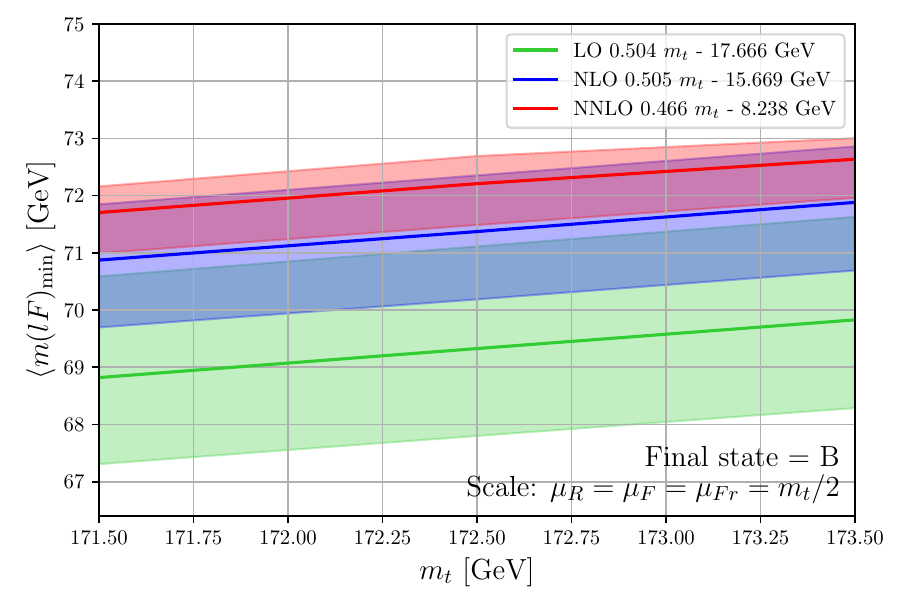}
	\includegraphics[width=0.49\textwidth]{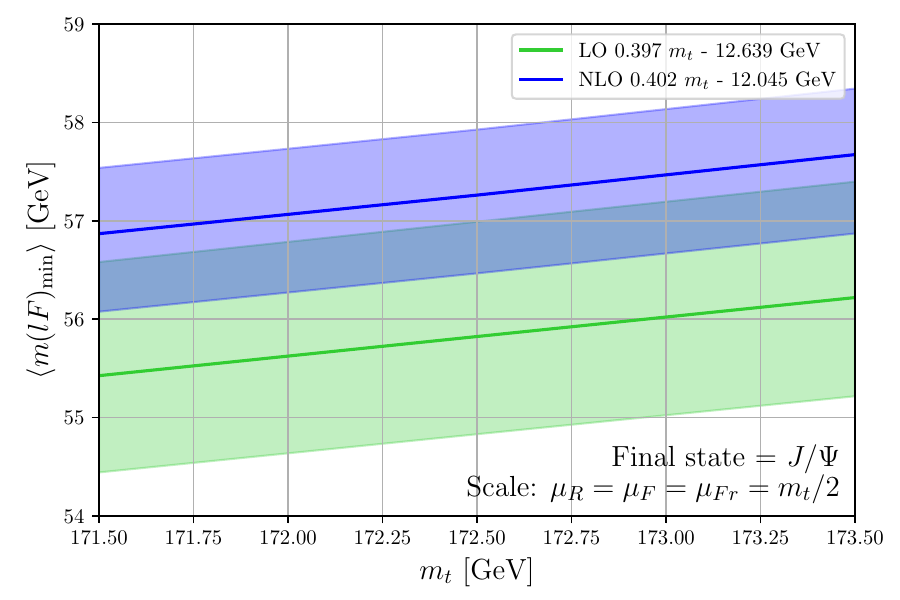}
	\includegraphics[width=0.49\textwidth]{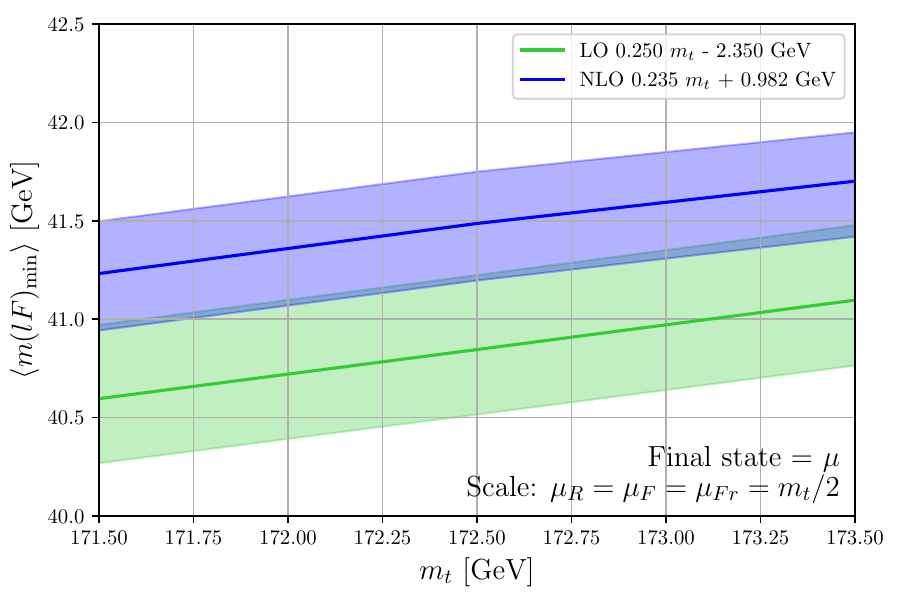}
	\includegraphics[width=0.49\textwidth]{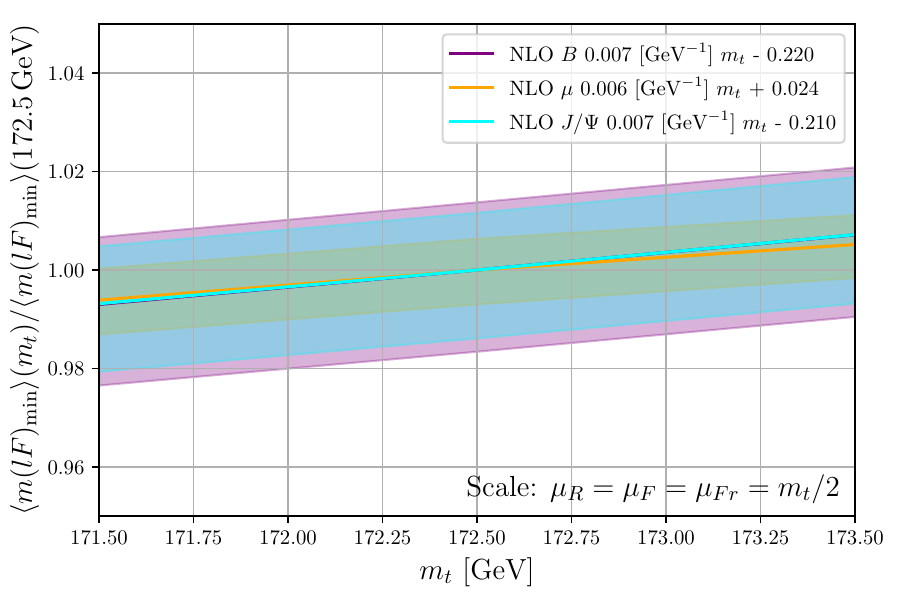}
	\caption{The first moment of the normalized $m(F\ell)_{\rm min}$ distribution for $F=B$ (upper left), $F=J/\psi$ (upper right) and $F=\mu$ (lower left). Shown are the LO, NLO and NNLO predictions as a function of $m_t$ (only the LO and NLO ones for $F=J/\psi, \mu$). Bands at each order denote 15-point scale variation for the fixed scale (\ref{eq:scale-mt}). The three NLO curves, divided by their respective central scale values at $m_t = 172.5$ GeV, are shown together in the lower-right panel.} 
\label{fig:mt-mBl-moment}
\end{figure}

A  feature of the observable $m(B\ell^-)$ is that it combines $(B, \ell)$ pairs that may originate from either the same top quark or from two different top quarks. For the purpose of $m_t$ determination, one would like to predominantly have pairs that originate from the same top quark. One way to achieve this is to consider the modified observable $m(B\ell)_{\rm min}$ where the $B$ and $\ell$ are paired not based on lepton's charge but on the requirement that their invariant mass is minimized. In fig.~\ref{fig:mt-mBl-moment} (left) we show the predicted $m_t$ dependence for the first moment of the normalized $m(B\ell)_{\rm min}$ distribution. Note that this moment is sensitive to the selection requirements listed in the beginning of this section. We have verified that for different selection requirements, for example the ones listed in sec.~\ref{sec:diff-dist}, the behavior of this observable can change significantly. 

We find the NNLO correction shifts the dependence relative to NLO, but much less than the NLO one does relative to LO. The perturbative uncertainty also decreases at higher orders which means that the inclusion of the NNLO correction will have an important impact on this observable. We also note that at NNLO the slope of the $m_t$ dependence very slightly decreases which means a slight decrease with higher orders to the $m_t$ sensitivity of this observable. This effect is nonetheless small relative to the overall theory uncertainty. While, as expected, the behavior of our prediction is broadly in line with the one in ref.~\cite{Biswas:2010sa}, a direct comparison at LO and NLO with that reference is not possible due to the differences between the NPFFs used and the event selections.

There is a longstanding interest \cite{Kharchilava:1999yj,CDF:2009mbf,CMS:2016ixg,ATLAS:2022jbw} in the $m(F\ell)$ distribution with final states other than $B$, especially $J/\psi$ or a muon. Utilizing the fragmentation functions to these final states derived in the present work, in the upper-right and lower-left panels of fig.~\ref{fig:mt-mBl-moment} we show the $m_t$ dependence of $\langle m(J/\psi\ell)_{\rm min}\rangle$ and $\langle m(\mu\ell)_{\rm min}\rangle$, respectively. Due to the high computational cost we have not computed the NNLO corrections to these observables. The LO and NLO corrections to these final states, especially when comparing with the pattern of higher order corrections for $\langle m(B\ell)_{\rm min}\rangle$, allow one to draw conclusions about the behavior of these observables. In particular, we see that through NLO $\langle m(J/\psi\ell)_{\rm min}\rangle$ has a pattern of higher order corrections very similar to $\langle m(B\ell)_{\rm min}\rangle$. We expect this trend to continue at NNLO. The $K$-factor for $\langle m(\mu\ell)_{\rm min}\rangle$ is slightly larger relative to its reduced scale uncertainties, since LO and NLO uncertainty bands barely overlap. This is due to the different shape of the muon FF which probes different kinematics relative to $B$ or $J/\psi$. 

Another feature of the three functions $\langle m(F\ell)_{\rm min}\rangle(m_t)$ is the change in slope around $m_t = 172.5$ GeV as the final state changes. The slope is largest for $F=B$ and decreases towards $F=J/\psi$ and especially for $F=\mu$. However, the relative change around $m_t = 172.5$ GeV given by $\langle m(F\ell)_{\rm min}\rangle(m_t)/\langle m(F\ell)_{\rm min}\rangle(172.5\text{ GeV})$, shown at NLO in the lower-right panel of fig.~\ref{fig:mt-mBl-moment}, is the same for $F=B$ and $F=J/\psi$ and only 20\% smaller for $F=\mu$. In fig.~\ref{fig:mt-mBl-moment}, the denominator is always evaluated at the central scale choice in order to preserve the information on the scale uncertainties. The theoretical uncertainty on a hypothetically extracted value of $m_t$ is determined by the ratio of the width of the scale uncertainty band and the slope. From fig.~\ref{fig:mt-mBl-moment} one can conclude that the theory uncertainty is slightly smaller for $F=J/\psi$ than for $F=B$ and almost a factor 2 smaller for $F=\mu$ than for $F=B$. Additionally, a corresponding measurement for $F=\mu$ would most likely also be the most precise, followed by $F=J/\psi$ and then $F=B$. 

\begin{figure}[t]
\centering
\includegraphics[width=0.49\textwidth]{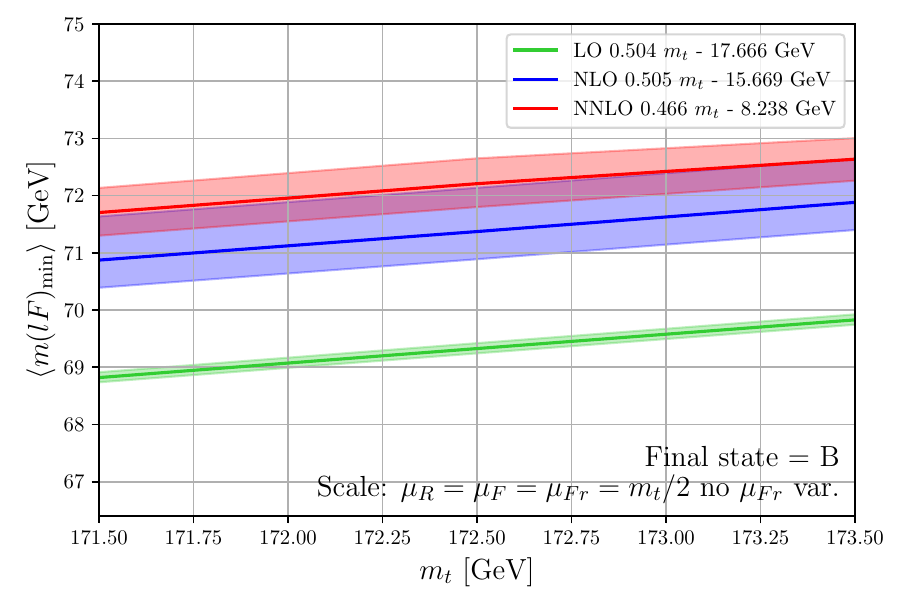}
\includegraphics[width=0.49\textwidth]{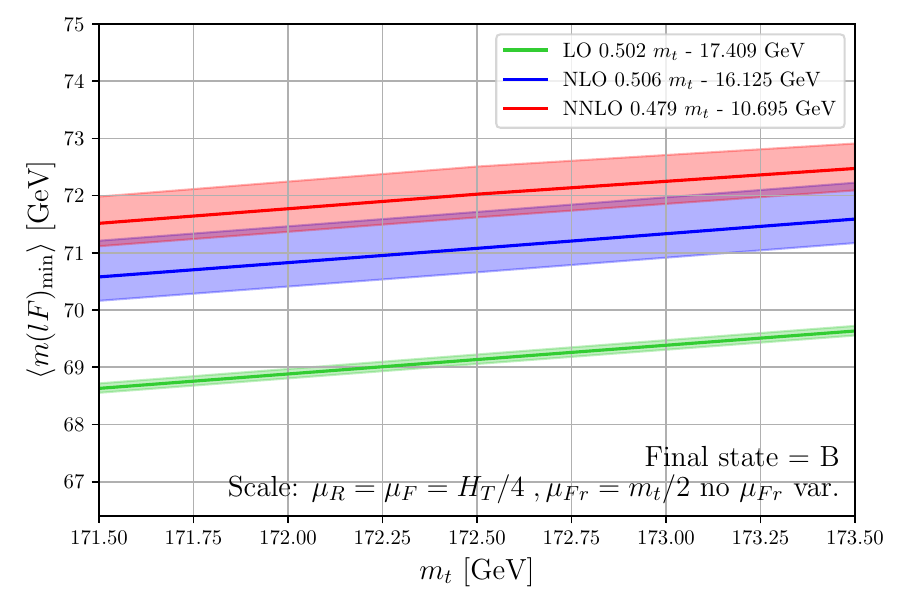}
\caption{The first moment of the normalized $m(B\ell)_{\rm min}$ distribution computed with two different scales: fixed scale (\ref{eq:scale-mt}) (left) and the dynamic scale (\ref{eq:scale-HT}) (right). In both cases bands correspond to 7-point scale variation with fixed $\mu_{Fr}$.} 
\label{fig:mt-mBl-scales}
\end{figure}

We have also investigated the sensitivity of this important observable to the choice of scale. To that end in fig.~\ref{fig:mt-mBl-scales} we have compared predictions with the fixed scale (\ref{eq:scale-mt}) (left) and with the dynamic one (\ref{eq:scale-HT}) (right). Since the calculation with dynamic scale eq.~(\ref{eq:scale-HT}) is performed with a 7-point scale variation (for fixed scale $\mu_{Fr}$), for a proper comparison of the two scales in fig.~\ref{fig:mt-mBl-scales} we show the prediction with the dynamic scale (right) and the one with the fixed scale (\ref{eq:scale-mt}) but with 7-point scale variation (left). Comparing fig.~\ref{fig:mt-mBl-scales} (left) with fig.~\ref{fig:mt-mBl-scales} (right) we conclude that the two scales lead to very similar predictions that are fully consistent within uncertainties, especially at NNLO.

\section{Conclusions}\label{sec:conclusions}

Recently, ref.~\cite{Czakon:2021ohs} presented a novel technical approach for the numerical calculation of NNLO QCD corrections to coefficient functions for identified hadron production at hadron and/or lepton colliders. The approach is applicable to both heavy and light hadron production. It utilizes numerical Monte Carlo integration techniques within the sector-improved residue subtraction scheme \cite{Czakon:2010td,Czakon:2014oma}  which implies that generic NNLO observables with complicated phase space cuts can be implemented. This sets the technique of ref.~\cite{Czakon:2021ohs} apart from the so far pursued, decades-old approach based on analytical coefficient functions applicable only to a small number of inclusive observables. 

As a first application of this computational technique, ref.~\cite{Czakon:2021ohs} presented the NNLO QCD corrections to $B$-hadron production in $t\bar t$ events at the LHC. This state of the art calculation was supplemented by an existing non-perturbative fragmentation function for $B$-hadrons. The significant improvement in the precision of theoretical predictions due to the inclusion of NNLO QCD corrections, however, is calling for a similar scrutiny at the precision and quality of the non-perturbative fragmentation contribution for heavy flavor production. We observe that, at this level of precision, the NPFF affects the ultimate precision that can be achieved in theoretical predictions of single heavy identified hadrons. 

The main goal of the present paper is to address this gap. It provides a NNLO NPFF for $B$ hadrons which is extracted from $e^+e^-$ data in a way which is {\it consistent} with the formalism used by us for state-of-the-art NNLO predictions for LHC observables. 

As a first application of the newly extracted $B$ fragmentation function, in this work we present NNLO accurate predictions for $B$-production in $t\bar t$ events at the LHC. Such application is of strong immediate interest due to the large LHC statistics on $t\bar t$ events. We observe a significant improvement from the inclusion of NNLO QCD corrections for a range of generic observables (by a factor of about four relative to NLO QCD). We have placed a particular emphasis on predictions for observables that allow precision determination of the top quark mass. Future applications of the fragmentation function extracted in this work, and of the computational formalism developed in ref.~\cite{Czakon:2021ohs}, include open $B$ production at the LHC as well as $B$ production in association with bosons. Our fragmentation function fitting procedure can also be adapted to charm production which is another application of strong interest. 

The extraction of the $B$ NPFF is based on a global fit of four $e^+e^-$ measurements, although we have studied its sensitivity to any single dataset. We have produced a set of 100 NPFFs which allows a detailed FF uncertainty analysis in any observable of interest. Our results are publicly available as LHAPDF grids.

The extraction of the $B$ NPFF is based on NNLO perturbative inputs with soft-gluon resummation through NNLL. The effect of the resummation in the perturbative initial condition for the $B$ fragmentation function is important. In analyzing the soft-gluon resummation of the FF initial condition, we have found complete consistency between its NNLO fixed order calculation, the anomalous dimensions governing that resummation and the decoupling through NNLO of the $b$-quark fragmentation function across the heavy flavor threshold. 

Another novelty of the present work is that it also derives sets of FF for single inclusive production of $J/\psi$ or soft muons originating in $B$ hadron decays. To that end we have convoluted the $B$-hadron NPFF extracted by us with precisely measured $B$ meson decays spectra. These two sets of NPFFs are, too, made public with this publication. They allow NNLO accurate predictions for observables based not only on $B$-hadrons but on their decay products. Such observables are measured very accurately at the LHC and are needed, among others, in measurements of $m_t$ with high precision or for prediction of $J/\psi$ and muons in semileptonic $B$ decays in the context of open $B$ production or in association with bosons. 

The FF sets for all final states considered in the present work are available as LHAPDF grids from the LHAPDF website (\href{https://lhapdf.hepforge.org/}{https://lhapdf.hepforge.org/}) under the names
\begin{itemize}
	\item[] \tt{CGMP\_B-hadron\_NNLO\_as\_0118\\
	                CGMP\_B-hadron\_to\_Jpsi\_NNLO\_as\_0118\\
	                CGMP\_B-hadron\_to\_muon\_NNLO\_as\_0118} .
\end{itemize}

\begin{acknowledgments}
We would like to thank Gennaro Corcella for a collaboration at an early stage of this work and for numerous useful comments and suggestions. We would also like to thank Valerio Bertone for a helpful discussion. The work of M.C.~was supported by the Deutsche Forschungsgemeinschaft under grant 396021762 – TRR 257. The work of T.G.~was supported by the DFG under grant 400140256 - GRK 2497: The physics of the heaviest particles at the Large Hadron Collider. The research of A.M.~and R.P.~has received funding from the European Research Council (ERC) under the European Union's Horizon 2020 Research and Innovation Programme (grant agreement no.~683211). A.M.~was also supported by the UK STFC grants ST/L002760/1 and ST/K004883/1. R.P.~acknowledges support from the Leverhulme Trust and the Isaac Newton Trust.
\end{acknowledgments}

\begin{appendix}

\section{Results for soft-gluon resummation}\label{app:soft-gluon}

In this appendix we collect various results used in the soft-gluon resummation of the $\ee$ coefficient functions and the initial condition of the perturbative fragmentation function. Please refer to secs.~\ref{sec:coeff-func} and \ref{sec:pert-frag} for definitions. 

The relevant coefficients of the QCD $\beta$-function in a theory with $n_f$ active flavors read
\begin{eqnarray}
b_0 &=& \frac{1}{4\pi}\bigg[\frac{11}{3} C_A-\frac{4}{3} T_F n_f\bigg]\,,\nonumber\\
b_1 &=& \frac{1}{(4\pi)^2}\bigg[\frac{34}{3} C_A^2-\frac{20}{3} C_A T_F n_f-4 C_F T_F n_f\bigg]\,,\nonumber\\
b_2 &=& \frac{1}{(4\pi)^3}\bigg[-\frac{1415}{27} C_A^2 T_F n_f+\frac{2857}{54} C_A^3-\frac{205}{9} C_A C_F T_F n_f+2 C_F^2 T_F n_f \nonumber\\
&& +\frac{158}{27} C_A T_F^2 n_f^2+\frac{44}{9} C_F T_F^2 n_f^2\bigg]\,,
\label{eq:beta}
\end{eqnarray}
where $C_F=4/3, C_A=3$ and $T_F=1/2$ are the usual QCD color factors.

\subsection{$D^{\rm ini}_{b\to b}$}\label{app:Dini}

The cusp anomalous dimension (\ref{eq:A-expand}) in a theory with $n_f$ active flavors reads though NNLL
\begin{eqnarray}
A_1 &=& C_F\,,\nonumber\\
A_2 &=& \frac{1}{2} C_F \bigg[\bigg(\frac{67}{18}-\frac{\pi^2}{6}\bigg) C_A-\frac{10}{9}T_F n_f \bigg]\,,\nonumber\\
A_3 &=& C_F \bigg[C_A^2 \bigg(\frac{11 \zeta_3}{24}+\frac{245}{96}-\frac{67 \pi ^2}{216}+\frac{11 \pi ^4}{720}\bigg)-C_A T_F n_f \bigg(\frac{7 \zeta_3}{6}+\frac{209}{216}-\frac{5 \pi ^2}{54}\bigg)\nonumber\\
&& -C_F T_F n_f \bigg(\frac{55}{48}-\zeta_3\bigg)-\frac{1}{27}T_F^2 n_f^2\bigg]\,.
\end{eqnarray}

The NNLL expansion of the wide-angle anomalous dimension $D$ in eq.~(\ref{eq:D-expand}) reads 
\begin{eqnarray}
D_1 &=& -C_F\,,\nonumber\\
D_2 &=& C_F \bigg[C_A \bigg(-\frac{9 \zeta_3}{4}+\frac{55}{108}+\frac{\pi ^2}{12}\bigg)+\frac{1}{27}T_F n_l\bigg]\;.
\end{eqnarray}

The hard matching coefficient (\ref{eq:H-expand}), expanded through NNLO, reads
\begingroup
\allowdisplaybreaks
\begin{eqnarray}
H_0^{\rm ini} &=& 1\,,\nonumber\\
H_1^{\rm ini} &=& C_F\bigg[\bigg(\frac{3}{2}-2 \gamma \bigg) \ln \bigg(\frac{\mu_{Fr,0}^2}{m_b^2}\bigg)-\frac{\pi ^2}{3}-2 \gamma ^2+2 \gamma +2\bigg]\,,\nonumber\\
H_2^{\rm ini} &=& C_F\bigg\{C_F\bigg[\frac{1}{8} (3-4 \gamma )^2 \ln ^2\bigg(\frac{\mu_{Fr,0}^2}{m_b^2}\bigg)+ \ln (\frac{\mu_{Fr,0}^2}{m_b^2})\bigg(6 \zeta_3+\frac{27}{8}-7 \gamma ^2+4 \gamma ^3-\pi ^2\nonumber\\
&&+\gamma  \bigg(\frac{2 \pi ^2}{3}-1\bigg)\bigg)-\frac{3 \zeta_3}{2}+\frac{2}{3} \gamma ^2 (\pi ^2-3)-\frac{2}{3} \gamma  (\pi ^2-6)-\frac{11 \pi ^4}{180}+\frac{\pi ^2}{4}+2 \gamma ^4-4 \gamma ^3\nonumber\\
&&+\frac{241}{32}-\pi ^2 \ln (4)\bigg]+C_A\bigg[-\frac{11}{24} (4 \gamma -3) \ln ^2\bigg(\frac{\mu_{Fr,0}^2}{m_b^2}\bigg)+\ln (\frac{\mu_{Fr,0}^2}{m_b^2}) \bigg(\frac{11}{12} (4 \gamma\nonumber\\
&& -3) \ln (\frac{\mu_{Fr,0}^2}{\mu_{R,0}^2})+\frac{1}{72} (-216 \zeta_3+315-272 \gamma -264 \gamma ^2+24 \gamma  \pi ^2)\bigg)+\frac{11}{18} (-6-6 \gamma\nonumber\\
&& +6 \gamma ^2+\pi ^2) \ln (\frac{\mu_{Fr,0}^2}{\mu_{R,0}^2})-\frac{97 \zeta_3}{18}+\gamma  \bigg(9 \zeta_3-\frac{55}{27}-\frac{14 \pi ^2}{9}\bigg)+\frac{1141}{288}-\frac{22 \gamma ^3}{9}+\frac{\pi ^4}{12}\nonumber\\
&&+\frac{1}{9} \gamma ^2 (3 \pi ^2-34)+\pi ^2 \bigg(\frac{7}{54}+\ln (2)\bigg)\bigg]+T_F n_l\bigg[\frac{1}{6} (4 \gamma -3) \ln ^2\bigg(\frac{\mu_{Fr,0}^2}{m_b^2}\bigg)\nonumber\\
&&+\bigg(\bigg(1-\frac{4 \gamma }{3}\bigg) \ln (\frac{\mu_{Fr,0}^2}{\mu_{R,0}^2})+\frac{4 \gamma ^2}{3}+\frac{8 \gamma }{9}-\frac{3}{2}\bigg) \ln (\frac{\mu_{Fr,0}^2}{m_b^2})-\frac{2}{9} (-6-6 \gamma +6 \gamma ^2\nonumber\\
&&+\pi ^2) \ln (\frac{\mu_{Fr,0}^2}{\mu_{R,0}^2})-\frac{2 \zeta_3}{9}-\frac{173}{72}-\frac{4 \gamma }{27}+\frac{8 \gamma ^2}{9}+\frac{8 \gamma ^3}{9}-\frac{4 \pi ^2}{27}+\frac{4 \gamma  \pi ^2}{9}\bigg]\nonumber\\
&&+T_F\bigg[\frac{1}{6} (4 \gamma -3) \ln ^2\bigg(\frac{\mu_{Fr,0}^2}{m_b^2}\bigg)+ \ln (\frac{\mu_{Fr,0}^2}{m_b^2})\bigg(\bigg(1-\frac{4 \gamma }{3}\bigg) \ln (\frac{\mu_{Fr,0}^2}{\mu_{R,0}^2})+\frac{4 \gamma ^2}{3}+\frac{8 \gamma }{9}\nonumber\\
&&-\frac{3}{2}\bigg)-\frac{2}{9} (-6-6 \gamma +6 \gamma ^2+\pi ^2) \ln (\frac{\mu_{Fr,0}^2}{\mu_{R,0}^2})+\frac{2 \zeta_3}{3}-\frac{\pi ^2}{3}-\frac{56 \gamma }{27}+\frac{3139}{648}\bigg]\bigg\}\,.
\end{eqnarray}
\endgroup

The coefficients of the non-truncated function $G^{\rm ini}(m,N,\alpha_S)$ read through NNLL
\begingroup
\allowdisplaybreaks
\begin{eqnarray}
g_1(\lambda) &=& -\frac{1}{2\pi b_0^2}\bigg\{2 A_1 \lambda +A_1 (1-2 \lambda ) \ln (1-2 \lambda )\bigg\}\,,\nonumber\\
g_2(\lambda) &=& \frac{1}{4\pi^2b_0^3}\bigg\{4 \lambda  \bigg[\pi  b_0 \bigg((A_1 b'_0-A'_1 b_0) \ln (\frac{\mu_{R,0}^2}{m_b^2})-A'_1 b_0 \ln (\frac{\mu_{Fr,0}^2}{\mu_{R,0}^2})\bigg)+A_2 b_0-\pi  A_1 b_1\bigg]\nonumber\\
&&+2 \ln (1-2 \lambda ) \bigg[\pi  A_1 b_0 b'_0 \ln (\frac{\mu_{R,0}^2}{m_b^2})+A_2 b_0+\pi  (2 \gamma  A_1 b_0^2-A_1 b_1+b_0^2 B_1)\bigg]\nonumber\\
&&-\pi  A_1 b_1 \ln ^2(1-2 \lambda )\bigg\}\,,\nonumber\\
g_3(\lambda) &=& \frac{1}{144\pi^3b_0^4(1-2\lambda)}\bigg\{72 \pi  b_0 \lambda  \bigg[-\pi  b_0 b'_0  \ln ^2\bigg(\frac{\mu_{R,0}^2}{m_b^2}\bigg)\big(2 A_1 b'_0 \lambda +A'_1 b_0 (1-2 \lambda )\big)\nonumber\\
&&-2 \ln (\frac{\mu_{R,0}^2}{m_b^2}) \big(b_0 (1-2 \lambda ) (A'_2 b_0-\pi  A_1 b'_1)+b'_0 (2 \lambda  (A_2 b_0-\pi  A_1 b_1)\nonumber\\
&&+\pi  b_0^2 (2 \gamma  A_1+B_1))\big)+b_0^2 (1-2 \lambda ) \ln (\frac{\mu_{Fr,0}^2}{\mu_{R,0}^2}) \bigg(\pi  A'_1 b'_0 \ln (\frac{\mu_{Fr,0}^2}{\mu_{R,0}^2})-2 A'_2\bigg)\bigg]\nonumber\\
&&-6 \lambda  \bigg[A_1 \big(24 \pi ^2 b_2 b_0 (1-\lambda)+b_0^2 (-48 \gamma  \pi ^2 b_1-14 \lambda +7)+24 \pi ^2 b_1^2 \lambda\nonumber\\
&& +8 (6 \gamma ^2 \pi ^2+\pi ^4) b_0^4\big)+24 b_0 \big(\lambda  (A_3 b_0-\pi  A_2 b_1)+\pi  (2 \gamma  b_0^2-b_1) (A_2+\pi  b_0 B_1)\nonumber\\
&&+\pi  b_0^2 B_2\big)\bigg]+3\ln (1-2 \lambda ) \bigg[24 \pi ^2 A_1 b_0 \ln (\frac{\mu_{R,0}^2}{m_b^2}) (-2 b_0 b'_1 \lambda +2 b_1 b'_0 \lambda +b_0 b'_1)\nonumber\\
&&+A_1 \big(b_0^2 (48 \gamma  \pi ^2 b_1+14 \lambda -7)-24 \pi ^2 b_2 b_0 (1-2 \lambda )-48 \pi ^2 b_1^2 \lambda \big)\nonumber\\
&&+24 \pi  b_0 b_1 (A_2+\pi  b_0 B_1)\bigg]-36 \pi ^2 A_1 b_1^2 \ln ^2(1-2 \lambda )\bigg\}\,.
\end{eqnarray}
\endgroup
Please recall that, as specified in the text following eq.~(\ref{eq:D-expand}), a prime in the above equations denotes a quantity computed with $n_f= n_l+1$ while the ones without a prime are for $n_f=n_l$.

The functions $g_1$ and $g_2$ were previously calculated in ref.~\cite{Cacciari:2001cw} for $n_f=n_l$. Substituting primed coefficients with unprimed ones in our results above should reproduce the results of ref.~\cite{Cacciari:2001cw}. This is indeed the case. The difference between the results above and the results of ref.~\cite{Cacciari:2001cw} is of $\mathcal{O}(\alpha_S^2)$, i.e.~the NLO matching implemented in ref.~\cite{Cacciari:2001cw} is insensitive to it.

The function $g_3$ was recently derived in ref.~\cite{Maltoni:2022bpy} for $n_f=n_l$. Therefore, in order to compare our result for $g_3$ with that of ref.~\cite{Maltoni:2022bpy}, we convert our expression to a coupling with $n_l$ active flavors. In practice, the primes in our result for $g_3$ need to be dropped. Once this replacement is made, the two results still disagree by the following term
\begin{equation}
    -\frac{7 A_1}{48\pi^3 b_0^2}(2 \lambda+\ln (1-2 \lambda))\,,
\end{equation}
which originates in the matching of  $\alpha_S$. After subtracting this term from our result for $g_3$, it then exactly matches the $g_3$ of ref.~\cite{Maltoni:2022bpy}. 

The result for $H^{\rm ini}$ is in full agreement with those of refs.~\cite{Cacciari:2001cw,Maltoni:2022bpy} after correcting a trivial typo for the incorrect overall sign of $H_1^{\rm ini}$ in ref.~\cite{Maltoni:2022bpy}.

Finally, the term $\sim C_F T_F \bigg(-\frac{2}{3}\ln^2\bigg(\frac{m_b^2}{\mu_\text{th}^2}\bigg)-\frac{20}{9}\ln\bigg(\frac{m_b^2}{\mu_\text{th}^2}\bigg)-\frac{56}{27}\bigg)$ which appears in the decoupling constant $C_{\rm dec}$ (\ref{eq:C-dec}) for the fragmentation function has already been discussed in ref.~\cite{Maltoni:2022bpy} but has been given a different interpretation there.

\subsection{$\ee$ coefficient functions}\label{app:ee}

The results for the cusp anomalous dimension $A_i$ can be found in appendix~\ref{app:Dini}. The anomalous dimension $B$ reads
\begin{equation}
B(a) = \sum_{i=1}\left(\frac{a}{\pi}\right)^i B_i \,, 
\label{eq:B-expand}
\end{equation}
with coefficients
\begin{eqnarray}
B_1 &=& -\frac{3}{2}C_F\,,\\
B_2 &=& C_F \bigg(C_A \bigg(5 \zeta_3-\frac{3155}{432}+\frac{11 \pi ^2}{36}\bigg)-C_F \bigg(3 \zeta_3+\frac{3}{16}-\frac{\pi ^2}{4}\bigg)+ \bigg(\frac{247}{108}-\frac{\pi ^2}{9}\bigg) T_F n_f\bigg)\,.\nonumber
\end{eqnarray}
The resummation function $G$ defined in eq.~(\ref{eq:sigma_N-matched-mult}), with expansion as in eq.~(\ref{eq:Gini-exact}), reads
\begingroup
\allowdisplaybreaks
\begin{eqnarray}
g_1(\lambda) &=& \frac{1}{\pi b_0^2}\bigg\{A_1 (\lambda +(1-\lambda ) \ln (1-\lambda ))\bigg\}\,,\nonumber\\
g_2(\lambda) &=& \frac{1}{2\pi^2b_0^3}\bigg\{2 \lambda  \bigg[\pi  A_1 b_0^2 \ln (\frac{\mu _F^2}{\mu _R^2})-A_2 b_0+\pi  A_1 b_1\bigg]+\ln (1-\lambda ) \bigg[2 \pi  A_1 b_0^2 \ln (\frac{Q^2}{\mu _R^2})\nonumber\\
&&+\pi  (-2 \gamma  A_1 b_0^2+2 A_1 b_1+b_0^2 B_1)-2 A_2 b_0\bigg]+\pi  A_1 b_1 \ln ^2(1-\lambda )\bigg\}\,,\nonumber\\
g_3(\lambda) &=& \frac{1}{12\pi^3 b_0^4 (1-\lambda)}\bigg\{\lambda  \bigg[6 \pi  b_0^2 \bigg(\big(\pi  (-2 \gamma  A_1 b_0^2+2 A_1 b_1+b_0^2 B_1)-2 A_2 b_0\big) \ln (\frac{Q^2}{\mu _R^2})\nonumber\\
&&+b_0 (1-\lambda ) \ln (\frac{\mu _F^2}{\mu _R^2}) \bigg(2 A_2-\pi  A_1 b_0 \ln (\frac{\mu _F^2}{\mu _R^2})\bigg)+\pi  A_1 b_0^2 \ln ^2\bigg(\frac{Q^2}{\mu _R^2}\bigg)\bigg)\nonumber\\
&&+\pi  b_0 \big(-6 \gamma  b_0 (-2 A_2 b_0+2 \pi  A_1 b_1+\pi  b_0^2 B_1)+6 \pi  (2 A_1 b_2+b_0 b_1 B_1)-6 (2 A_2 b_1\nonumber\\
&&+b_0^2 B_2)+\pi ^3 A_1 b_0^3+6 \gamma ^2 \pi  A_1 b_0^3\big)+6 \lambda  \big(A_3 b_0^2-\pi  (A_2 b_0 b_1+\pi  A_1 (b_0 b_2-b_1^2))\big)\bigg]\nonumber\\
&&+6 \pi  \ln (1-\lambda ) \bigg[2 \pi  A_1 (b_2 b_0 (1-\lambda )+b_1^2 \lambda -\gamma  b_1 b_0^2)+2 \pi  A_1 b_1 b_0^2 \ln (\frac{Q^2}{\mu _R^2})\nonumber\\
&&-2 A_2 b_1 b_0+\pi  b_1 b_0^2 B_1\bigg]+6 \pi ^2 A_1 b_1^2 \ln ^2(1-\lambda )\bigg\}\,.
\end{eqnarray}
\endgroup
The coefficients of the hard matching function in eq.~(\ref{eq:sigma_N-matched-mult}), in an expansion in $\alpha_S/(2\pi)$, read
\begingroup
\allowdisplaybreaks
\begin{eqnarray}
H_0 &=& 1\,,\nonumber\\
H_1 &=& C_F\bigg[\bigg(\frac{3}{2}-2 \gamma \bigg) \ln (\frac{Q^2}{\mu _F^2})+\frac{5 \pi ^2}{6}+\gamma ^2+\frac{3 \gamma }{2}-\frac{9}{2}\bigg]\,,\nonumber\\
H_2 &=& C_F\bigg\{C_F\bigg[\bigg(\frac{9}{8}-3 \gamma +2 \gamma ^2\bigg) \ln ^2\bigg(\frac{Q^2}{\mu _F^2}\bigg)+\bigg(6 \zeta_3-\frac{51}{8}+\frac{45 \gamma }{4}-\frac{3 \gamma ^2}{2}-2 \gamma ^3+\frac{3 \pi ^2}{4}\nonumber\\
&&-\frac{5 \gamma  \pi ^2}{3}\bigg) \ln (\frac{Q^2}{\mu _F^2})+6 \gamma  \zeta_3-\frac{33 \zeta_3}{2}+\frac{61 \pi ^4}{180}+\frac{5 \gamma ^2 \pi ^2}{6}+\frac{3 \gamma  \pi ^2}{4}-\frac{35 \pi ^2}{16}+\frac{\gamma ^4}{2}+\frac{3 \gamma ^3}{2}\nonumber\\
&&-\frac{27 \gamma ^2}{8}-\frac{51 \gamma }{8}+\frac{331}{32}\bigg]+C_A\bigg[\bigg(\frac{11 \gamma }{6}-\frac{11}{8}\bigg) \ln ^2\bigg(\frac{Q^2}{\mu _F^2}\bigg)+\bigg(\bigg(\frac{11}{4}-\frac{11 \gamma }{3}\bigg) \ln (\frac{Q^2}{\mu _F^2})\nonumber\\
&&+\frac{55 \pi ^2}{36}+\frac{11 \gamma ^2}{6}+\frac{11 \gamma }{4}-\frac{33}{4}\bigg) \ln (\frac{\mu _R^2}{\mu _F^2})+\bigg(-3 \zeta_3+\frac{215}{24}-\frac{367 \gamma }{36}-\frac{11 \gamma ^2}{6}-\frac{11 \pi ^2}{12}\nonumber\\
&&+\frac{\gamma  \pi ^2}{3}\bigg) \ln (\frac{Q^2}{\mu _F^2})-10 \gamma  \zeta_3+\frac{116 \zeta_3}{9}-\frac{23 \pi ^4}{240}-\frac{\gamma ^2 \pi ^2}{6}-\frac{11 \gamma  \pi ^2}{36}+\frac{1657 \pi ^2}{432}+\frac{11 \gamma ^3}{18}\nonumber\\
&&+\frac{367 \gamma ^2}{72}+\frac{3155 \gamma }{216}-\frac{5465}{288}\bigg]+T_F n_f\bigg[\bigg(\frac{1}{2}-\frac{2 \gamma }{3}\bigg) \ln ^2\bigg(\frac{Q^2}{\mu _F^2}\bigg)+\bigg(-\frac{19}{6}+\frac{29 \gamma }{9}\nonumber\\
&&+\frac{2 \gamma ^2}{3}+\frac{\pi ^2}{3}\bigg) \ln (\frac{Q^2}{\mu _F^2})+\bigg(\bigg(\frac{4 \gamma }{3}-1\bigg) \ln (\frac{Q^2}{\mu _F^2})-\frac{5 \pi ^2}{9}-\frac{2 \gamma ^2}{3}-\gamma +3\bigg) \ln (\frac{\mu _R^2}{\mu _F^2})\nonumber\\
&&+\frac{2 \zeta_3}{9}+\frac{\gamma  \pi ^2}{9}-\frac{143 \pi ^2}{108}-\frac{2 \gamma ^3}{9}-\frac{29 \gamma ^2}{18}-\frac{247 \gamma }{54}+\frac{457}{72}\bigg]\bigg\}\,.
\end{eqnarray}
\endgroup
The above results for $g_1$, $g_2$ and $g_3$ match those of ref.~\cite{Vogt:2000ci}.

\section{Fitting NPFF to one dataset at a time}\label{sec:appendix-one-dataset}

\begin{figure}[t]
\centering
\includegraphics[width=0.24\textwidth]{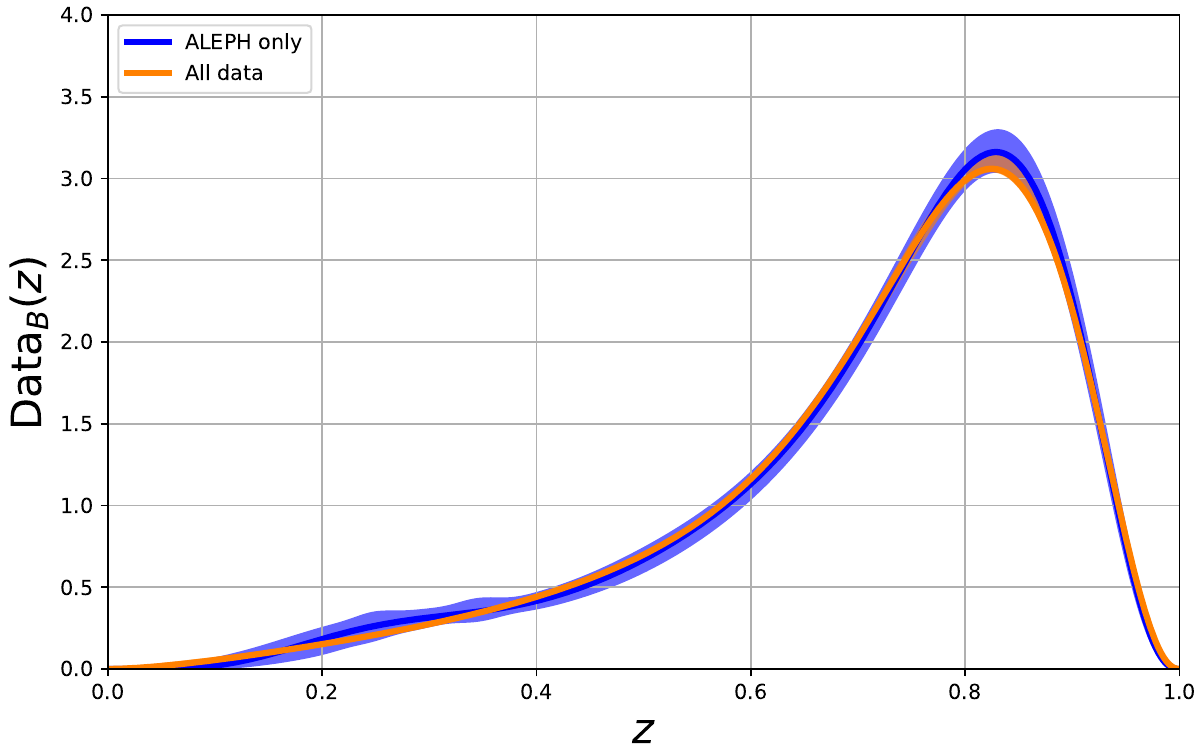}
\includegraphics[width=0.24\textwidth]{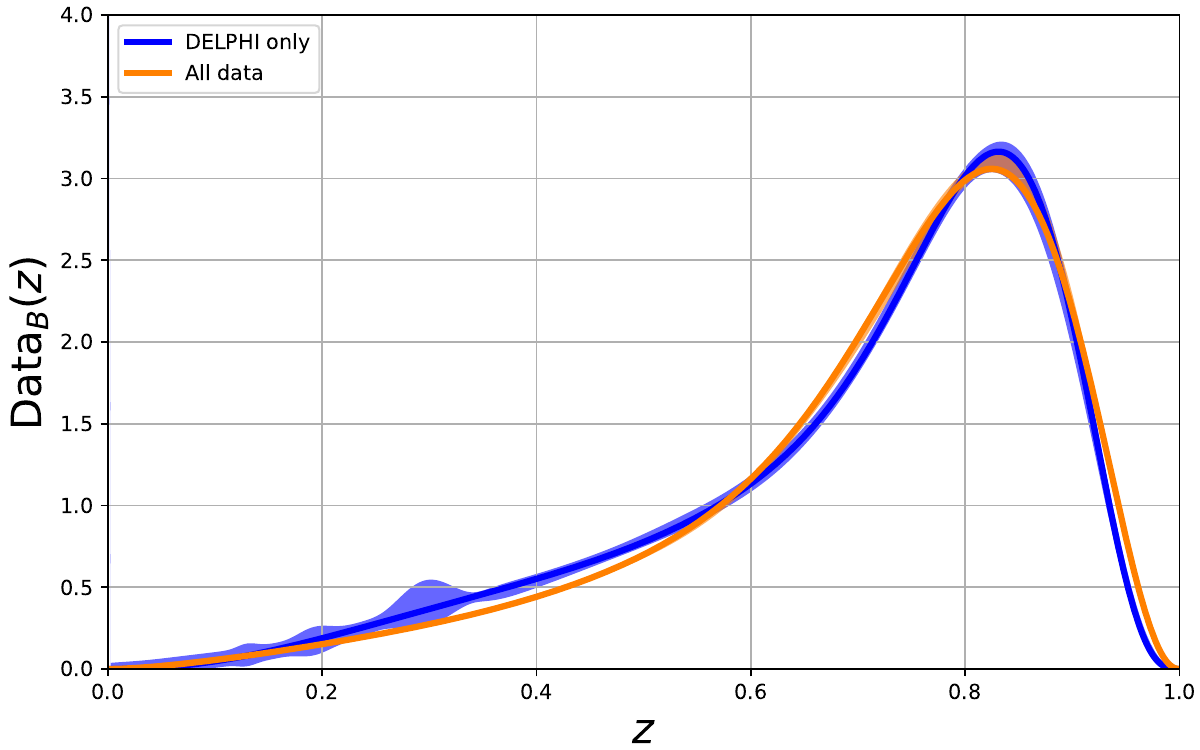}
\includegraphics[width=0.24\textwidth]{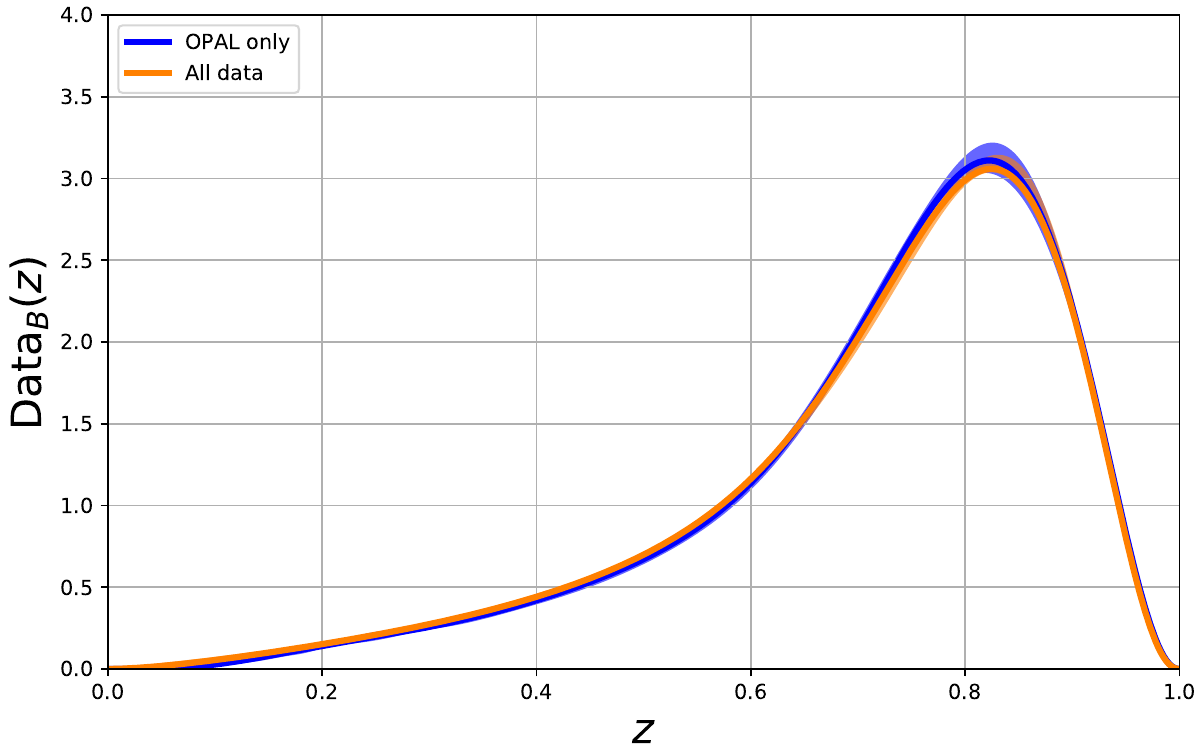}
\includegraphics[width=0.24\textwidth]{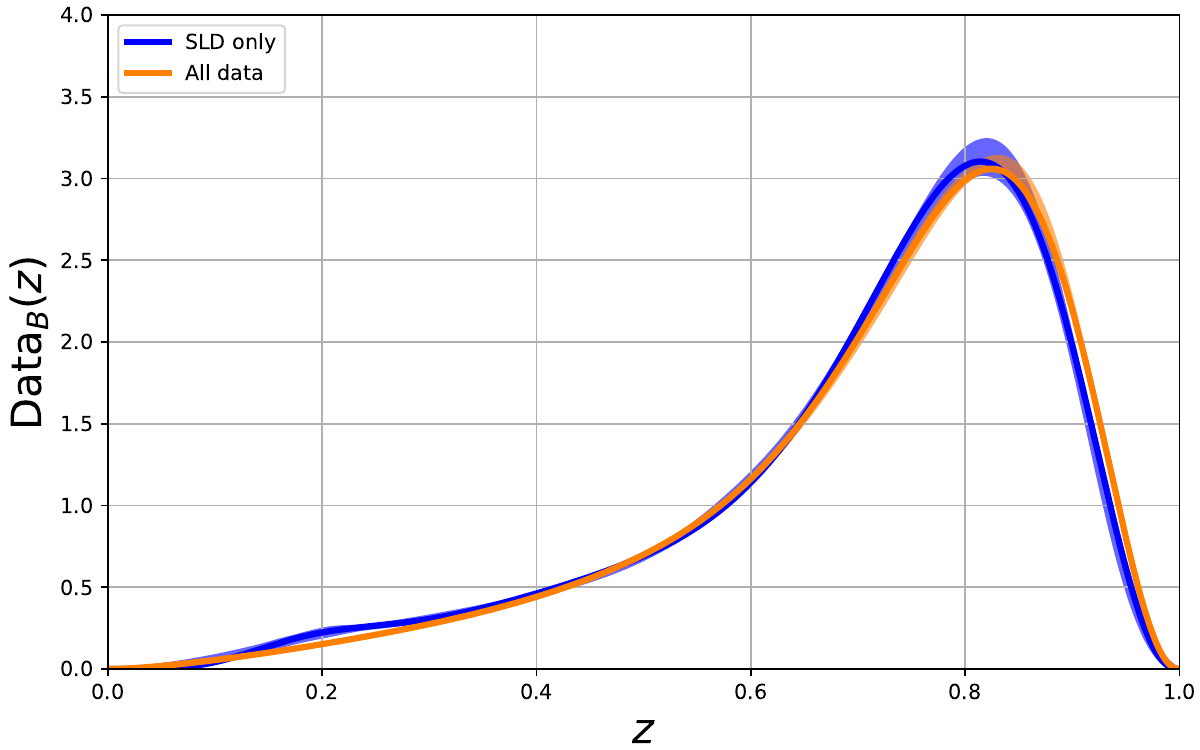}
\includegraphics[width=0.24\textwidth]{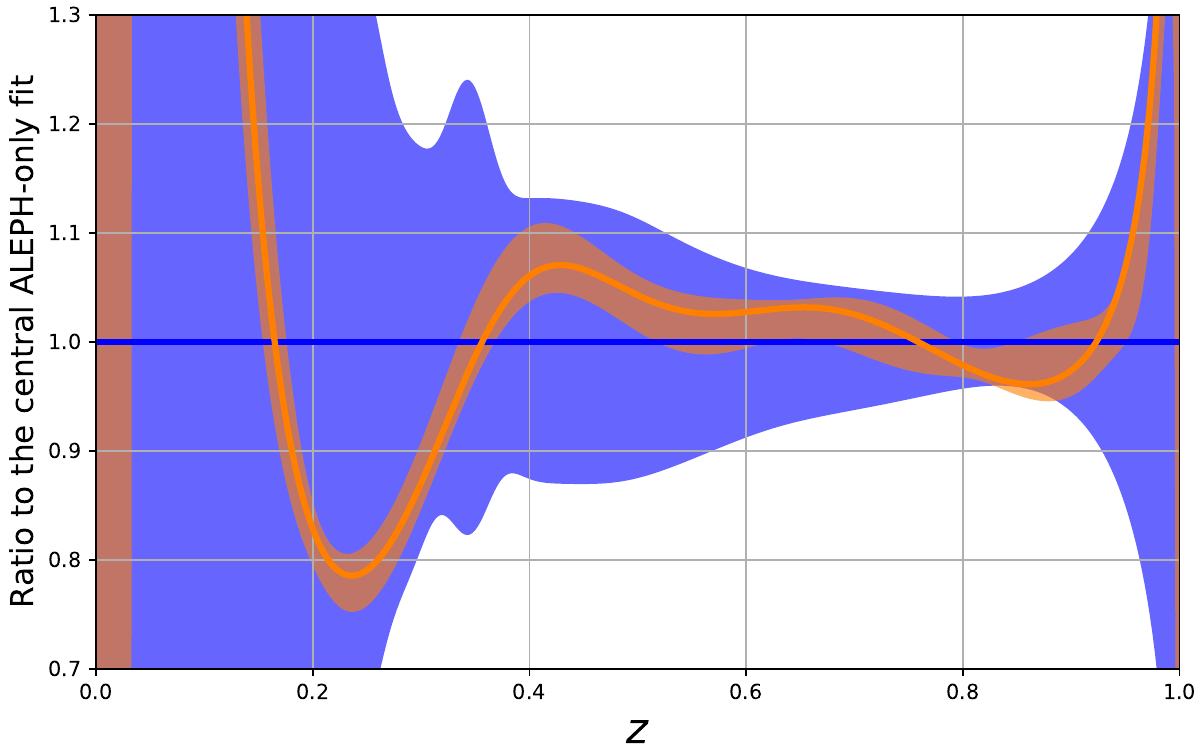}
\includegraphics[width=0.24\textwidth]{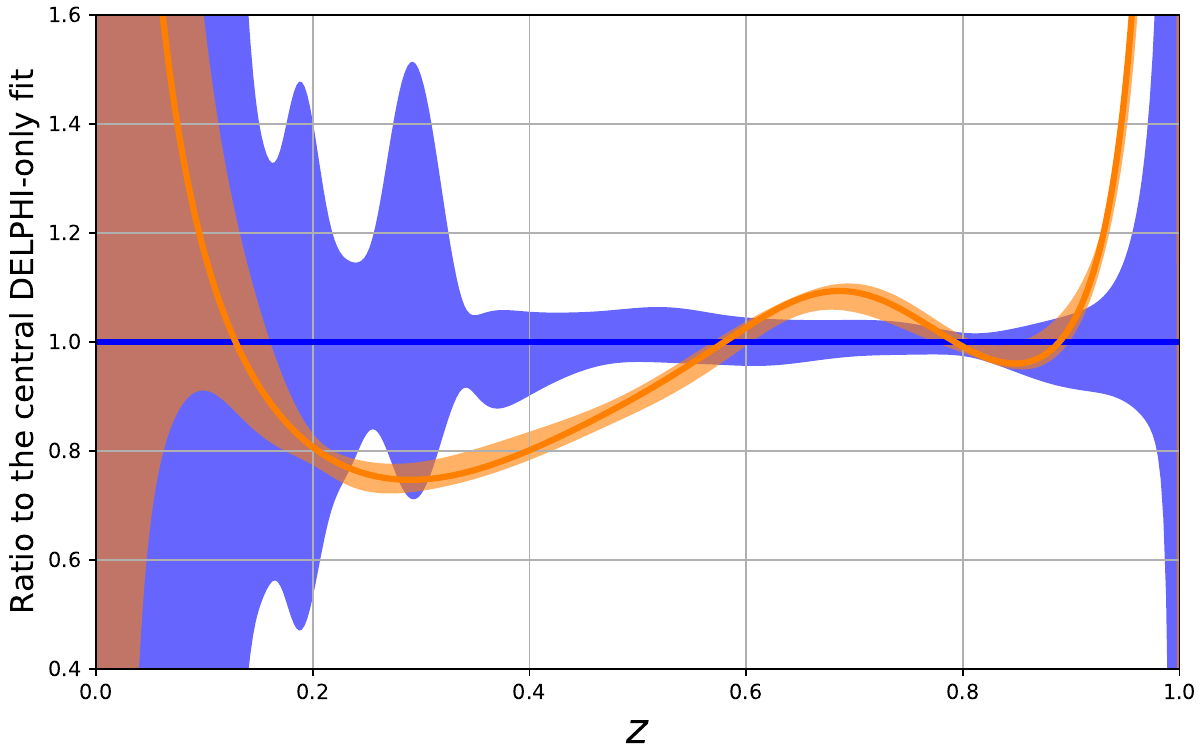}
\includegraphics[width=0.24\textwidth]{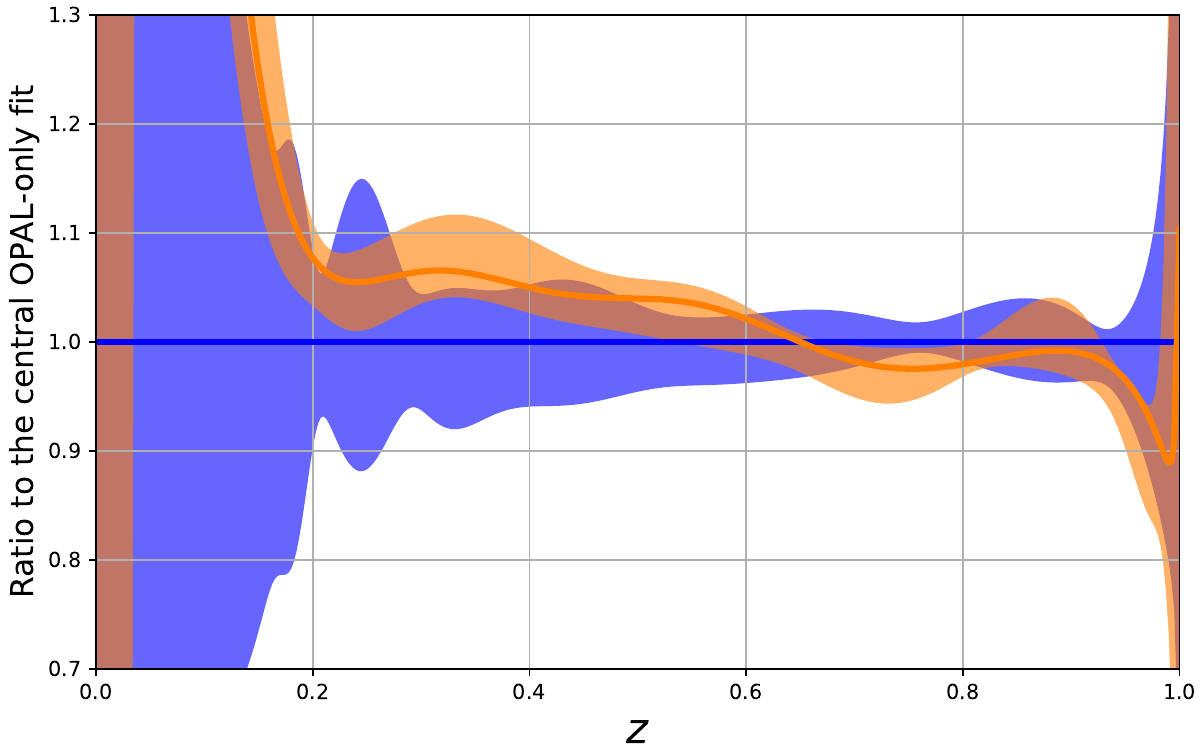}
\includegraphics[width=0.24\textwidth]{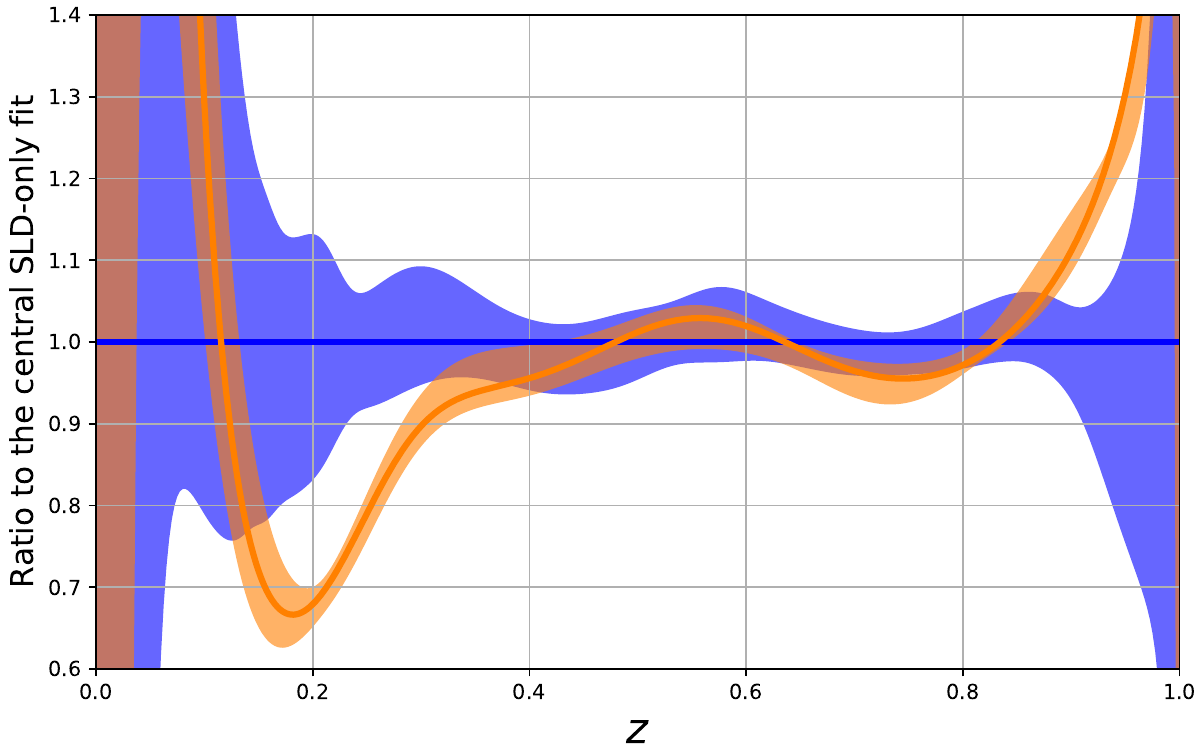}
\caption{NPFF fits from one dataset at a time (blue) versus the combined fit (orange). The four datasets are (from left to right): ALEPH, DELPHI, OPAL and SLD.}
\label{fig:onedataset}
\end{figure}

In the following we present the results for fitting NPFF to one dataset at a time. We use the four datasets ALEPH \cite{ALEPH:2001pfo}, DELPHI \cite{DELPHI:2011aa}, OPAL \cite{OPAL:2002plk} and SLD \cite{SLD:2002poq} that enter our combined fit. The extracted NPFF sets, compared to the combined extraction, can be seen in fig.~\ref{fig:onedataset}. 

Of particular interest is the fit to ALEPH data since it contains only $B$-mesons, rather than the meson-baryon mixture measured by the other collaborations. In a ALEPH-only fit one would like to check if the contribution from baryons has a visible effect on the fits. As can be seen in fig.~\ref{fig:onedataset}, when only ALEPH data is considered the uncertainty of the fit is considerably larger than the one of the combined fit. The fit to all data lies entirely within the uncertainty band of the ALEPH-only fit which suggests that it may not be possible to convincingly disentangle the effect of $B$-baryons using this measurement. 

In general, as can be seen in fig.~\ref{fig:onedataset}, there is agreement between the four individual fits. The fit showing the largest deviation with respect to the combined fit is obtained from the DELPHI dataset. Nevertheless, it is not inconsistent with the combined fit. The OPAL data appears to be the most precise. For most values of $z$ the combined fit has slightly smaller uncertainties than the OPAL-only fit. Since there are certain tensions between the four datasets one may wonder if an improvement from combining them can be expected. Upon closer inspection of fig.~\ref{fig:onedataset} it turns out that for almost all values of $z$ where the errors on the combined fit are smaller than all the individual fits, the four datasets actually do agree, which implies the consistency of the reduced uncertainties of the combined fit.

\end{appendix}

\end{document}